\documentclass[letter,11pt]{article}
 
\usepackage{jheppub}

\usepackage{graphicx}
\usepackage{dcolumn}
\usepackage{bm}
\usepackage{float}
\usepackage{hyperref}
\usepackage{subfig}
\usepackage{dsfont}
\usepackage{slashed}
\usepackage{color}
\usepackage{amsmath}
\usepackage[section]{placeins}
\usepackage{braket}
\usepackage{upgreek}
\usepackage[bottom]{footmisc}
\usepackage[normalem]{ulem}
\usepackage{xcolor}
\usepackage{mathrsfs}
\usepackage[leftcaption]{sidecap}

\newcommand{\nn}{\nonumber}
\newcommand{\sinc}{\text{sinc}}
\newcommand{\total}{\rm total}
\def\be{\begin{eqnarray*}}
\def\ee{\end{eqnarray*}}
\def\beq{\begin{eqnarray}}
\def\eeq{\end{eqnarray}}
\def\bs{\boldsymbol} 
\def\bs{\boldsymbol}

\def\rme{{\rm e}}
\def\rmd{{\rm d}}

\def\cs{{\rm cs}}

\def\out{{\rm loss}}
\def\med{{\rm med}}
\def\tr{{\rm Tr}}
\def\s{{\rm s}}

\def\bfk{{\bs k}}
\def\bfp{{\bs p}}
\def\bfx{{\bs x}}

\def\bfq{{\bs q}}
\def\bfu{{\bs u}}
\def\bfr{{\bs r}}

\def\bfx{{\bs x}}
\def\bfl{{\bs l}}
\newcommand{\bea}{\begin{eqnarray}}
\newcommand{\eea}{\end{eqnarray}}

\usepackage{marginnote}

\title{Open quantum system approach to inclusive jet production in heavy-ion collisions}

\author[1]{Yacine Mehtar-Tani,}
\author[2]{Felix Ringer,}
\author[3]{Balbeer Singh}
\author[3]{and Varun Vaidya}

\affiliation[1]{Physics Department, Brookhaven National Laboratory, Upton, NY 11973, USA}
\affiliation[2]{Department of Physics and Astronomy, Stony Brook University, NY 11794, USA}
\affiliation[3]{Department of Physics, University of South Dakota, Vermillion, SD 57069, USA}

\emailAdd{mehtartani@bnl.gov}
\emailAdd{felix.ringer@stonybrook.edu}
\emailAdd{Balbeer.Singh@usd.edu}
\emailAdd{Varun.Vaidya@usd.edu}

\abstract{We derive a factorization formula for inclusive jet production in heavy-ion collisions using the tools of Effective Field Theory (EFT). We show how physics at widely separated scales in this process can be systematically separated by matching to EFTs at successively lower virtualities. Owing to a strong scale separation, we recover a vacuum-like DGLAP evolution above the jet scale, while the additional low-energy scales induced by the medium effectively probe the internal structure of the jet. As a result, the cross section can be written as a series with an increasing number of subjets characterized by perturbative matching coefficients each of which is convolved with a {\it distinct} function. These functions encode broadening, medium-induced radiations as well as quantum interference such as the Landau-Pomeranchuk-Migdal effect and color coherence dynamics to all orders in perturbation theory. As a first application of this EFT framework, we investigate the case of an unresolved jet and show how the cross section can be factorized and fully separate the jet dynamics from the universal physics of the medium. To compare to the existing literature, we explicitly compute the medium jet function at next-to-leading order in the coupling and leading order in medium opacity.}

\begin{document}
\maketitle
\newpage

\section{Introduction}
%%%%%%%%%%%%%%%%%%%%%%%%%%%%%%%%%%%%%%%%%%%%%%

In the background of a strongly coupled medium such as the quark-gluon plasma (QGP) in heavy-ion collisions (HIC) or a nucleus in deep-inelastic scattering (DIS), hard-scattering events can create partons with energies much larger than the typical medium scales such as the temperature $T$ or $\Lambda_{\rm QCD}$. The subsequent parton cascade and fragmentation lead to collimated sprays of particles, which are known as jets. Due to the large transverse momentum $p_T$ of the observed jet, the initial production mechanism and the partonic evolution can be studied within perturbative QCD. Therefore, jet measurements are theoretically well controlled, which makes them powerful tools to probe the properties of the QGP created in HICs~\cite{ATLAS:2014ipv,ALICE:2015mjv,CMS:2016uxf,STAR:2005gfr,PHENIX:2004vcz} and nuclei in electron-nucleus collisions~\cite{AbdulKhalek:2021gbh}. Throughout this work, we primarily focus on HICs but analogous techniques can be applied to study cold nuclear matter effects in electron-nucleus collisions.

In HICs, the yield of high $p_T$ jets is suppressed compared to the vacuum case or proton-proton (pp) collisions. This phenomenon, known as ``jet quenching," has been observed experimentally at both the Relativistic Heavy Ion Collider (RHIC) and the Large Hadron Collider (LHC)~\cite{Gyulassy:1993hr,Wang:1994fx,Baier:1994bd,Baier:1996kr,Baier:1996sk,Zakharov:1996fv,Zakharov:1997uu,Gyulassy:2000er,Wiedemann:2000za,Guo:2000nz,Wang:2001ifa,Arnold:2002ja,Arnold:2002zm,Salgado:2003gb,Liu:2006ug,Qin:2007rn,Armesto:2011ht,Burke:2013yra,Kang:2017frl,He:2018xjv,Brewer:2018dfs,Qiu:2019sfj,Jia:2019qbl,Caucal:2019uvr,Casalderrey-Solana:2020rsj,Dai:2020rlu,Andres:2020vxs,JETSCAPE:2021ehl,Barata:2021byj,Clayton:2021uuv,Attems:2022ubu,Budhraja:2023rgo,Cunqueiro:2023vxl,Mehtar-Tani:2024jtd}. Jet quenching is primarily attributed to the energy loss experienced by the color charged energetic partons produced in the initial hard scattering processes. These partons lose energy through both elastic collisions with the medium constituents and medium-induced gluon radiation, where the dominant contribution is due to radiative energy loss. An important phenomenon in partonic energy loss is the Landau-Pomeranchuk-Migdal (LPM) effect, which is a coherent action of multiple scattering centers in the medium. This was first described theoretically in the 1990s~\cite{Gyulassy:1993hr,Wang:1994fx,Baier:1994bd,Baier:1996kr,Baier:1996sk,Zakharov:1996fv,Zakharov:1997uu,Gyulassy:2000er,Wiedemann:2000za,Guo:2000nz,Wang:2001ifa,Arnold:2002ja,Arnold:2002zm}. 
In addition, it has been realized that jets are extended multi-partonic systems, such that interference patterns are created by multiple fast-moving color charges in the plasma, which depend on the resolution power of the QGP. Due to the quantum nature of these dynamics, jet energy loss depends on the QGP's resolution power in the plane transverse to the jet's direction of propagation. The higher the medium's resolution scale, the greater the number of effective color charges, which, in turn, increases the radiation intensity. Conversely, if the medium does not resolve the inner structure of the jet, medium-induced radiation is sourced coherently by the total jet color charge. 
Therefore, any complete theory of jet quenching needs to systematically account for both the LPM effect and color decoherence~\cite{Mehtar-Tani:2010ebp,Mehtar-Tani:2012mfa,Casalderrey-Solana:2011ule, Casalderrey-Solana:2012evi}, which are manifestations of quantum interference in the longitudinal and transverse directions, respectively. Probing these many-body QCD phenomena is one of the key goals of ongoing heavy-ion programs. 

Several theoretical approaches have been developed to understand and improve our understanding of the jet-medium interactions. These efforts are focused on the inclusive production rate of jets, jet correlations, and jet substructures. See Refs.~\cite{Mehtar-Tani:2013pia,Blaizot:2015lma,Qin:2015srf,Cao:2020wlm,Singh:2024vwb,Singh:2024pwr} for recent reviews. A significant fraction of the theoretical efforts and Monte Carlo (MC) simulations have relied on fixed-order calculations, which have been useful for understanding the qualitative features of jet quenching. However, a comprehensive framework to improve the predictive power and accuracy to the same level as that for simpler systems like $e^+ e^-$ or pp remains a fundamental challenge.  For example, in pp collisions, fixed order calculations of inclusive jet production have been performed at next-to-next-to-leading order (NNLO) accuracy~\cite{Currie:2016bfm,Czakon:2019tmo}. Additionally, jet radius and threshold resummations have been performed~\cite{deFlorian:2007fv,Dasgupta:2014yra,Kaufmann:2015hma,Kang:2016mcy,Dai:2016hzf,Liu:2017pbb,Neill:2021std,vanBeekveld:2024jnx,Lee:2024icn}. In Ref.~\cite{Kang:2016mcy}, the leading logarithmic (LL$'$) resummation of logarithms of the jet radius was achieved by factorizing the partonic cross section into a hard function describing the production of a high-energy parton and a jet function that captures the subsequent evolution. At LL$'$, the jet function follows the Dokshitzer-Gribov-Lipatov-Altarelli-Parisi (DGLAP) evolution equations, and recently this result was extended to the two-loop level in Refs.~\cite{vanBeekveld:2024jnx,Lee:2024icn}. The understanding of jet cross-sections, particularly jet substructure observables in pp collisions, has reached an unprecedented quantitative precision in recent years with advancements in direct QCD calculations and the development of Effective Field Theory (EFT) tools. See Refs.~\cite{Larkoski:2017jix,Asquith:2018igt,Marzani:2019hun} for recent reviews. This progress has been made possible by utilizing factorization formulas that separate perturbative physics at short distances from non-perturbative but universal physics at long-distance scales~\cite{Collins:1989gx}. Identifying well-defined non-perturbative quantities, such as parton distribution functions (PDFs) and shape functions related to QCD hadronization, and extracting them from a set of reference processes and lattice data ensures the predictive power of the theoretical framework. Recently, significant progress has been made in understanding the all-order structure of jets propagating through a QCD medium. Exploiting a separation of time scales between the hard collinear shower—driven by high virtuality—and medium-induced radiation, it was shown that the inclusive jet function satisfies a non-linear DGLAP evolution equation. This equation captures the sensitivity of the observable to jet substructure fluctuations, particularly highlighting that jet energy loss is proportional to the number of resolved subjets \cite{Mehtar-Tani:2017web,Mehtar-Tani:2017ypq,Mehtar-Tani:2024smp,Mehtar-Tani:2024mvl}. This formalism has been successfully applied to describe the jet nuclear modification factor observed at the LHC \cite{Mehtar-Tani:2021fud} within a leading-logarithmic approximation. Furthermore, the medium-modified fragmentation function has also been studied within this framework \cite{Caucal:2018dla,Caucal:2019uvr}. 

The inclusive properties of the turbulent gluon cascade have been extensively investigated over the past decade \cite{Blaizot:2014ula,Blaizot:2014rla,Caucal:2019uvr,Mehtar-Tani:2018zba,Schlichting:2020lef,Mehtar-Tani:2022zwf,Soudi:2024yfy}, leading to significant theoretical advancements. Notably, substantial progress has been achieved at next-to-leading order (NLO) \cite{Arnold:2020uzm,Arnold:2021pin,Arnold:2022fku,Arnold:2022mby,Arnold:2023qwi,Arnold:2024whj,Arnold:2024bph}, including a detailed understanding of the renormalization of the jet quenching parameter \cite{Blaizot:2014bha,Caucal:2022mpp,Ghiglieri:2022gyv,Caucal:2022fhc,Arnold:2021pin,Arnold:2021mow,Caucal:2021lgf,Blaizot:2019muz,Iancu:2018trm,Mehtar-Tani:2017ypq,Wu:2014nca,Iancu:2014sha,Liou:2013qya,Iancu:2014kga}. These developments culminate in a framework that reaches next-to-leading logarithmic (NLL) accuracy \cite{Caucal:2022fhc,Caucal:2022mpp}. More recently, theoretical progress has extended to include the effects of flow and inhomogeneous QCD media \cite{Andres:2022ndd, Sadofyev:2021ohn, Barata:2022krd, Barata:2023qds}. In addition to the well-studied jet nuclear modification factor \cite{Mehtar-Tani:2021fud}, a wide range of jet substructure observables has been explored~\cite{Larkoski:2014wba,CMS:2017qlm,Chien:2016led,Mehtar-Tani:2016aco,Caucal:2021bae,Caucal:2019uvr,Barata:2023bhh,Budhraja:2025ulx}, both analytically and via Monte Carlo event generators.  To further advance the field, there is a pressing need for precision theoretical frameworks that allow for systematic, analytic calculations of jet observables in the complex environment of heavy-ion collisions.

The goal of this work is to develop factorization formulas for jets in HICs at the level of the density matrix using EFT tools and to introduce relevant non-perturbative quantities at the operator level in the phenomenologically relevant kinematic regions. In particular, we extend the factorization framework for jet production and its subsequent evolution for HICs.

The jet evolution in a thermal medium involves multiple scales such as the medium temperature, production scale $p_T$, and jet or measurement scale $p_T R$. The hierarchies between these scales can be exploited to provide an effective description of jet evolution. Hence, an EFT approach to systematically organize the physics at different energy scales provides a natural framework to address this problem. In this regard, factorization can be derived within Soft Collinear Effective Theory (SCET) \cite{Bauer:2000yr,Bauer:2001ct,Bauer:2002nz,Beneke:2002ph,Manohar:2006nz} that follows a top-down approach, starting from the QCD action, and systematically expands it out in a power counting parameter $\lambda$ for the process of interest. This leads to an \textit{effective action} given in terms of \textit{effective operators} capturing the interaction between the relevant degrees of freedom to leading power in $\lambda$.  The operator-based approach also allows for gauge invariant definitions of the relevant non-perturbative functions. In this work, we employ an extended version of the traditional SCET framework that incorporates Glauber modes~\cite{Rothstein:2016bsq}, allowing for the coupling of energetic partons to the medium.  Related work on applications of SCET to describe jet quenching in heavy-ion collisions can be found in Refs.~\cite{Idilbi:2008vm,DEramo:2010wup,Ovanesyan:2011xy}. Apart from the direct scales such as the energy of thermal partons and the medium length, the above-mentioned features of jet-medium interaction also give rise to {\it emergent} scales. In particular, this includes the mean free path of the probe $\ell_{\rm mfp}$, which is related to the jet transport parameter $\hat q$, the decoherence angle $\theta_c$ \cite{Mehtar-Tani:2012mfa,Casalderrey-Solana:2011ule, Casalderrey-Solana:2012evi,Mehtar-Tani:2017ypq}, and the quantum coherence or equivalently the formation time $t_f$ of jet partons. Therefore, to achieve a full description of the system, these emergent scales need to be incorporated into the EFT, making the complete theory of jet quenching a challenging problem. Jet evolution in a medium is characterized by non-equilibrium dynamics. Therefore, we need to treat the jet as an \textit{Open Quantum System} interacting with the thermal environment. Hence, the main question we want to address is: How do we write an EFT to describe the evolution of such an open quantum system? As a starting point, we only focus on the simplest possible observable, inclusive jet production, which can be subsequently generalized to more involved jet substructure observables and jet correlations.

The remainder of this paper is organized as follows. In section~\ref{sec:Scales}, we describe all physical scales relevant for inclusive jet production in heavy-ion collisions. This includes both manifest and emergent scales. In section~\ref{sec:modes}, we discuss the relevant degrees of freedom needed to formulate the EFT within SCET. In the next two sections~\ref{sec:StageI} and~\ref{sec:StageII}, we successively match to EFTs at lower virtualities and derive a factorization formula for our observable for both a dilute (section~\ref{sec:Dilute}) and dense (section~\ref{sec:Multi})  environment. Finally, we discuss the implications of the EFT framework derived in this work for jet transport and outline open questions that need to be addressed in future work in section~\ref{sec:JetNature}.

%%%%%%%%%%%%%%%%%%%%%%%%%%%%%%%%%%%%%%%%%%%%%%%%%%%%%%%%%%%%%%%%%%%%%%%%%%%%%%
\section{Physical scales~\label{sec:Scales}}

For inclusive jet production, the measurements on the final state jet are its transverse momentum $p_T$  with respect to the beam axis in a certain rapidity ($\eta$) bin, and the jet radius $R$. In this work, we consider the case where the jet radius is parametrically small $R\ll 1$. For $pp$ collisions, this turns out to be the phenomenologically relevant regime even for relatively large values of $R$~\cite{Aversa:1990uv,Mukherjee:2012uz}. Moreover, in HICs, smaller-sized jets are less susceptible to contamination from the background fluctuations. For $R\ll 1$, the two relevant scales $p_T$ and $p_T R$ are parametrically widely separated and are systematically factorizable. We consider the scale $p_TR$ to be perturbative, i.e., $p_T R\gg \Lambda_{\rm QCD}$ so that perturbative calculations are reliable at this scale. In the absence of the medium, these are the only relevant scales in the problem. Separating these scales, the corresponding factorization for semi-inclusive jet production in $pp$ collision has been presented in Ref.~\cite{Dasgupta:2014yra,Kang:2016mcy,Dai:2016hzf}. For recent results at higher perturbative accuracy see Ref.~\cite{vanBeekveld:2024jnx,Lee:2024tzc}. The factorization of the production cross-section is effectively an expansion in $R$, which is also one of the power counting parameters of our EFT. 

The presence of the medium introduces new scales such as the medium temperature $T$, the medium size $L$, the Debye screening mass ($m_D\sim g T$).
We assume $T\sim$ 0.5-1 GeV, which is achievable in current collider experiments and it is a non-perturbative scale in the problem. Moreover, for this range of temperatures, the coupling strength $g \sim \mathcal{O}(1)$, making $T$ and $m_D$ of same order~\cite{Kaczmarek:2003ph}. Therefore, $m_D$ is not a separate scale in the EFT. 

As a high-energy ($p_T \gg T) $ jet propagates through the medium, it can interact multiple times with the background. Due to the hierarchy in the energy, we can distinguish between the Hilbert spaces populated by the medium partons and the jet.  
Hence, it is natural to treat the jet as an open quantum system whose internal structure is being probed by an environment. This interaction introduces new dynamically emergent scales in the EFT setup:
\begin{enumerate}
\item{}
As a starting point, we assume that the medium primarily interacts with the fast-moving energetic color charge of the jet partons through small-angle forward-scattering processes with angular deflection $\delta'\sim T/E$, where $E\sim p_T$ is jet energy. In the medium, each interaction is imparting an average transverse momentum $k_{\perp}\equiv|\bfk|\sim T$. However, in a dense medium, the coherent effect of multiple scatterings between the energetic jet and thermal partons may lead to a larger momentum transfer. Therefore, the relevant momentum scale for forward scattering would be the average transverse momentum gained by the energetic parton through multiple scatterings. We denote this scale by $Q_{\rm med}$ that, in general, is parameterized by the jet transport coefficient $\hat{q}$, with $Q_{\rm med}^2=\hat{q}L$, where $L$ is medium length. This is an emergent scale that can only be computed once we have a theoretical framework in place. Whether or not the parametrization in terms of a single universal number $\hat q$ is sufficient will eventually be determined using the EFT framework.  But to set up the EFT we need some guidance about the magnitude of this scale. Based on previous leading-order calculations and models fitted to data, this scale appears to be $Q_{\med} \sim 1-3$ GeV \cite{Mehtar-Tani:2021fud}. 

\item An effect closely related to the scale of transverse momentum $Q_{\med}$ is the energy loss through medium-induced radiation. The momentum transfer $Q_{\med}$ to a fast-moving parton puts it off shell by up to $Q_{\med}$ leading to Bremsstrahlung radiation in the medium. Phenomenologically, we can model this effect of energy loss in the differential cross section as a shift of the jet $p_T$ by a certain amount $E_\out$. Then the nucleus-nucleus ($AA$)  spectrum behaves approximately as  
$\rmd \sigma_{ AA}/ \rmd p_T \propto (p_T+E_\out)^{-n}$. The nuclear modification factor, which is the ratio of the cross section $AA$ to $pp$, can be approximated by $R_{AA}\sim (1+E_\out/p_T)^{-n}\simeq 1 - n \,E_\out/p_T+\ldots\;$. Since, $R_{AA} \lesssim 1$ and $n \sim 5$, we have $E_\out \lesssim p_T/n \ll p_T$. Along with the jet radius $R$, we can define two expansion parameters 
\begin{align}
\beta \equiv  \frac{E_\out}{p_T}\,, \quad \quad \quad \delta \equiv \frac{Q_\med} {E_\out}\,.
\end{align}
However, we also know that  any loss of energy through radiation can only occur at angles larger than $R$, implying $E_{\out} \sim Q_{\rm med} /R$ such that
\begin{align}
\delta \sim R\,.
\end{align}
 
\item{} 
Due to the possibility of multiple interactions, an important question is that of quantum interference between successive interactions. Does the parton go on-shell between successive interactions with the medium thereby suppressing any interference effects? The scale that determines this is known as coherence or formation time $t_f=\omega/\bfq^2$, where $\omega$ is the energy of the emitted gluon and $|\bfq|$ is its transverse momentum measured from the jet axis. For partons with $t_f\gg L$ interaction with the medium throughout its evolution is completely coherent leading to large quantum interference effects. This effectively leads to a large suppression for emitting partons with a very large formation time. Therefore, the formation time $t_f\sim L$ sets another phase space boundary for emissions inside the jet~\cite{Mehtar-Tani:2017web,Caucal:2018dla}. 
 
\item{}
The medium can resolve color of the partons separated by a transverse distance $1/Q_{\med}$, which is the transverse resolution power of the QGP. In particular, partons emitted with relative angle $\theta$ are resolved if $\theta L >  1/Q_{\rm med}$, where $\theta L$ is the transverse distance accumulated between two partons over the length of the medium. This defines the phase space beyond which the resolved partons are independent sources of radiation and sets the corresponding scale, characterized by the (color) decoherence angle $\theta_c\sim 1/{Q_{\rm med} L}$~\cite{Mehtar-Tani:2011ezl}. Therefore, the greater the resolution power of the QGP, the larger the number of independent sources for medium-induced radiation. For most phenomenological purposes, $\theta_c \in \{0.02, 0.2\}$ \cite{Mehtar-Tani:2021fud}. Therefore we have another transverse or virtuality scale in the problem, $\sim p_T\theta_c$. For the purposes of this paper, to simplify the discussion, we are going to assume $\theta_c \sim R$, so that we do not have a separate scale. We will see that the presence of color decoherence will lead to factorization formula that effectively probes the substructure of the jet. For a discussion on the case where $\theta_c \ll R $ see Refs.~\cite{Mehtar-Tani:2021fud,Mehtar-Tani:2024mvl}. 
\end{enumerate}

The evolution of the jet in the medium involves multiple scales associated with both the jet and the medium dynamics. While the jet dynamics is governed by scale $p_TR$, the medium dynamics is primarily encoded in the medium scale $Q_{\rm med}$. We therefore consider two possible hierarchies based on the virtuality of the jet $p_TR$ and the intrinsic medium scale $Q_{\rm med}$, which will determine the structure of our EFT. 
\begin{enumerate}
\item{$p_TR \sim  Q_{\text{med}}$:} 
Here, the hierarchy of all relevant scales is $p_T \gg p_TR \sim Q_{\text{med}} \geq m_D$. 
Therefore, the EFT only requires the separation between the two scales, i.e., $p_T$ and $p_T R$ that involves integrating out modes with virtuality $\geq p_T R$. 
In this case $E_{\text{loss}} \sim p_T$ so that $\beta \sim 1$. Hence $\delta \sim R$ is the only expansion parameter. This allows us to factorize the hard-scattering matrix elements, where the jet is initiated at time scales $p_T^{-1}$, from its subsequent collinear evolution in the vacuum and medium at times larger than $p_T/Q^2\sim (p_T R^2)^{-1}$. This regime is relevant when considering jets of low energy (few 10s of GeVs) or for very narrow jets ($R \ll 0.1$). We also note that the formation time for medium-induced radiation is $t_f \sim E_{\text{loss}}/Q_{\med}^2 \sim 1/(Q_{\med} R)$. So even though $E_{\text{loss}} \sim p_T$, for the regime of very narrow jets, the formation time can be substantially larger than $L$ leading to a large LPM suppression of medium-induced radiation. The corresponding EFT was discussed in Refs.~\cite{Vaidya:2020cyi,Vaidya:2020lih,Vaidya:2021mly,Vaidya:2021vxu} \footnote{For $R \ll 0.1$, $\theta_c$ is always greater than $R$ and hence the medium cannot resolve jet substructure. For low energy jets with larger radius, $\theta_c$ will be relevant but this effect was not considered in these papers.}. 
\item{$p_TR \gg Q_{\text{med}}$:} In this case, the three scales $p_T, p_T R$ and $Q_{\text{med}}$ are well separated from each other and hierarchy in the scales is $p_T\gg p_T R\gtrsim E_{\out}\gg Q_{\rm med}\gtrsim T\sim \Lambda_{\rm QCD}$. 
We therefore have at least two expansion parameters $\delta \sim R$ and $\beta$. 
This requires a two-stage EFT where we first integrate out the hard jet production at scale $p_T$ and then a second stage where we further integrate out physics at $p_T R$. This regime will be the focus of this paper. 
\end{enumerate}

Since the energy loss depends on the resolution power of the QGP, we further consider two possible hierarchies between jet radius $R$ and the color decoherence angle $\theta_c$. First, in the case where $R\ll \theta_c$, the medium cannot resolve the color charge of the energetic partons inside the jet, causing the jet to behave coherently as a single color source for medium-induced radiation. We call this scenario the \textit{unresolved jet}. Second, in the case where $R \gtrsim \theta_c$, the medium can resolve the color charges inside the jet. Each resolved charge inside the jet acts as an independent source of radiation. We define these distinct radiation sources as \textit{subjets} inside the jet. In this paper, we develop an EFT for both cases. However, for doing perturbative calculations, we will limit ourselves to the case of the unresolved jet. We will leave explicit perturbative calculations for $R \gtrsim \theta_c$ for the future. For this scenario, we will however, not assume a wide separation between $\theta_c$ and $R$ and therefore $p_T\theta_c \sim p_TR$ will be the same scale in our hierarchy.

%%%%%%%%%%%%%%%%%%%%%%%%%%%%%%%%%%%%%%%%%%%%%%%%%%%%%%%%%%%%%%%%%%%%%%%%%

\section{The EFT landscape}
\label{sec:modes}

\begin{figure}
\centering
\includegraphics[width=0.6\linewidth]{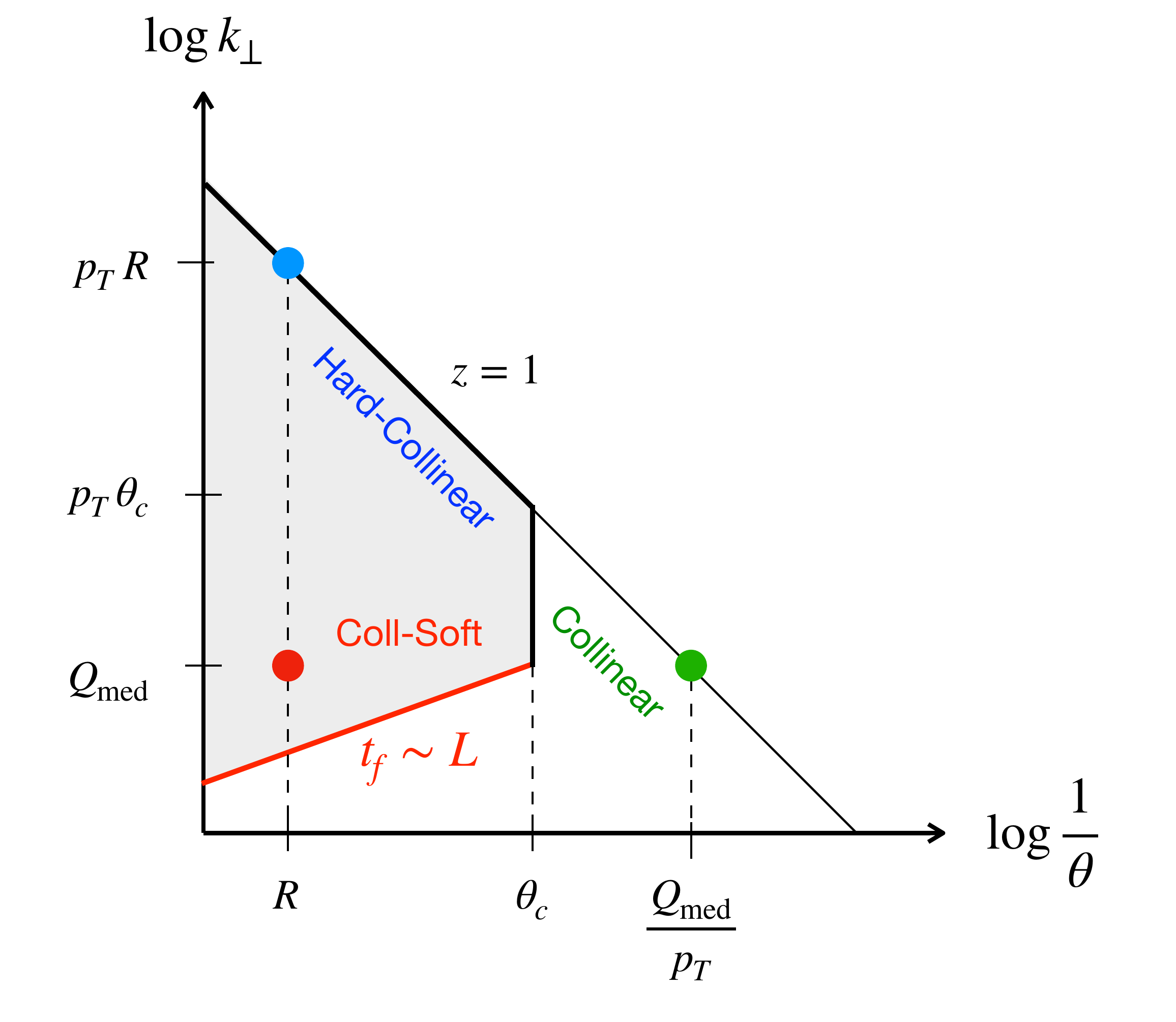}
\caption{Lund diagram for the factorization of the inclusive jet cross section in heavy-ion collisions. The different measurements are illustrated by straight lines in the $k_{\perp}\equiv|\bfk|,\theta$ plane, and the intersections of those lines are associated with modes of the corresponding EFT.~\label{fig:LJ}}
\end{figure}

To understand the phase space regions and the corresponding kinematic and measurement constraints on the multiple scales involved in the problem, we illustrate them using the Lund plane~\cite{Andersson:1988gp,Dreyer:2018nbf} in Figure~\ref{fig:LJ}. Here, for illustration purposes, we consider a jet with $p_TR\sim 10-100$~GeV, and medium temperature $T\sim 0.5$~GeV, which is also a typical scenario at current experiments. Furthermore, from phenomenological studies we have $\hat{q}\sim 1-2$~\text{GeV}$^2/\text{fm}$ and medium length $L\sim 2-5$~fm, so that the intrinsic medium scale $Q_{\rm med}\sim 1-3$~GeV. These values are an example of the hierarchy considered here 
\begin{equation}
p_T\,\,\gg\,\, p_T R\,\,\gtrsim \,\,E_{\out}\,\,\gg\,\, Q_{\rm med}\,\,\gtrsim\,\, T\,\,\sim\,\, \Lambda_{\rm QCD}\,.
\end{equation}
The Lund plane in Figure~\ref{fig:LJ} displays the transverse momentum $k_{\perp}=z\theta p_T$ of emissions with respect to the jet direction as a function of their angle $\theta$. Here, $z$ is the momentum fraction of the initial parton. The two vertical dotted lines represent the angular phase space introduced by the measurement $R$ and decoherence angle $\theta_c$. Emissions with angular separation $\theta_c<\theta< R$ are resolved by the medium and they contribute to the measurement of the final state jet. Further, the dotted horizontal line at $Q_{\rm med}$ denotes the transverse momentum of induced emission after the average momentum is imparted from the medium. The contour at $t_f\sim L$ sets another phase space boundary for determining the impact of LPM suppression on radiation. Without loss of generality, we assume that the jet initiating parton is moving in the light-like direction $n^{\mu}=(1,0,0,1)$. The light-like vector in the $-z$ direction is $\bar{n}^{\mu}=(1,0,0,-1)$ so that $n^2=\bar{n}^2=0$ and $n\cdot\bar{n}=2$. For any four-vector $p^{\mu}=(p^{-},p^{+},p_{\perp})$ with $p^{-}=\bar{n}\cdot p$ and $p^{+}=n\cdot p$ we adopt  the light-cone decomposition 
\begin{equation}
p^{\mu}=\bar{n}\cdot p\frac{n^{\mu}}{2}+n\cdot p\frac{\bar{n}^{\mu}}{2}+p_{\perp}^{\mu}\,.    
\end{equation}
    
The philosophy of the EFT is that we need to split the phase space available for radiation into various regions that contribute at leading power in our expansion parameters. We can then formulate an effective gauge invariant action, keeping only these relevant degrees of freedom and their interactions. In the Lund plane, these special regions correspond to the intersections of the kinematical constraints with the measurement ones. We will work in the framework of SCET, where this has been implemented systematically for describing the dynamics of high-energy partons in gauge theories. We will introduce various components of the SCET Lagrangian as we need them in our framework. Below we systematically describe all the relevant modes and their constraints. 

\begin{figure}
\centering
\includegraphics[width=\linewidth]{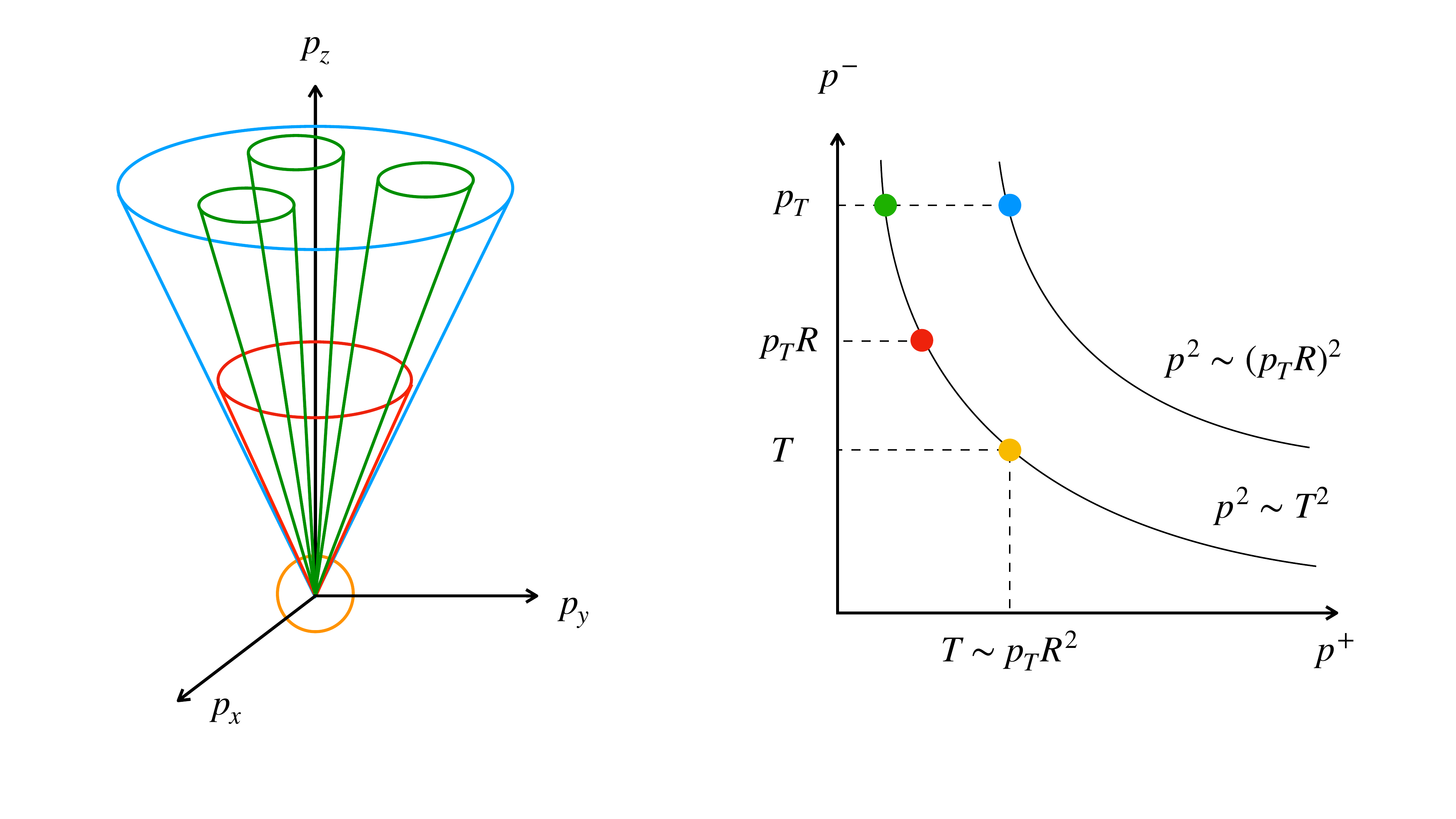}
\caption{Illustration of the relevant momentum modes for inclusive jet production in heavy-ion collisions and their scaling in SCET. See the main text for details.~\label{Mm}}
\end{figure}

In the absence of the medium, we only have two constraints arising from the measurements on the final state jet. First, the transverse momentum or $p_T$ of the jet, which is represented by the line $ z\sim 1$ in the Lund plane. Second, the radius of the jet $R$ introduces an angular scale in the phase space of gluon radiation. From the intersection of the corresponding two lines, we identify a hard-collinear (hc) mode, which is represented by the blue dot in the Lund plane. The hc mode has an energy of $\sim p_T$ with a transverse momentum spread or virtuality of $|\bfk| \sim p_TR$. We can therefore read off the scaling of this mode in momentum space using light-cone coordinates
\bea
p_{\rm hc}^{\mu} \equiv (p^-, p^+, p^{\perp}) \sim p_T\left( 1,  R^2, R \right)\,.
\label{eq:Collinear}
\eea
This includes the cone-shaped region in momentum space around the $z$-axis and angular size $R$ as indicated by the blue cone in Figure~\ref{Mm} (left panel). The factorization for the cross section in the process $p+p \rightarrow \text{jet}(p_T,R)+X$ in vacuum can be written in terms of a hard function and a collinear jet function, which we will discuss in the next section. 
As discussed in the previous section, in this hierarchy we have two expansion parameters $\delta \sim R$ and $\beta$. Given the phenomenological regime that we are working in, we also have $ \beta \sim R$ and we will work with this scaling for the rest of the paper. In a dilute medium, the typical transverse momentum transfer in a single interaction from the medium to the energetic jet parton is $|\bfk| \sim T $. However, in a dense medium, the average transverse momentum $Q_{\rm med}\geq T$ is transferred from the medium to the jet parton due to multiple scatterings. 
Therefore, at this virtuality we can describe the medium partons by a soft mode (s) with  scaling 
\bea
p^{\mu}_{\rm s} \sim p_T( \delta \beta, \delta \beta, \delta \beta) \sim p_T( R^2,R^2,R^2)\,.
\eea
Since we are working in the rest frame\footnote{This holds for the analysis of this paper where we do not assume $Q_{\rm med}$ to be a well-separated scale from T.} of the medium, this mode scales uniformly in all its components. We observe that the line $|\bfk| \sim Q_{\med}$ intersects the measurement of the jet radius $R$ and the jet energy $p_T$ ($ z\sim 1$) at two distinct points. Therefore, we define two distinct radiation modes corresponding to possible medium-induced radiation that can contribute to our measurement. First, the collinear-soft (cs) mode, which is represented by the red dot in Figure~\ref{fig:LJ}. The momentum scaling is given by
\bea
p^{\mu}_{\rm cs}	\sim \left(\frac{Q_{\text{med}}}{R},  Q_{\text{med}} R, Q_{\text{med}} \right). 
\eea 
Given our hierarchy and power counting, we can also write the momentum scaling of cs mode as 
\bea 
p^{\mu}_{\rm cs}	\sim p_T\beta(1, \delta^2 ,\delta) \sim p_TR\left(1, R^2,R \right). 
\eea
By comparing the center-of-mass energy of the interaction between collinear and soft modes to the typical momentum transfer $Q_{\text{med}}$, we find
\bea 
s\, \sim\, p^-_{\rm cs} p_{\rm s}^+  \,\sim p^2_T R^3 \,\,\gg \,\ t\,\sim\, Q^2_{\text{med}}\,\sim \,p_T^2 R^4\,,
\eea
which indicates that the interaction is dominated by forward scattering. Consequently, the dynamics of the collinear-soft interaction are governed by Regge kinematics. This justifies the use of Glauber SCET to describe the interactions between collinear-soft and soft modes.

Note that the cs mode has the same angular scaling as the hc mode, see Eq.~(\ref{eq:Collinear}), but with a lower energy. As a result, the contribution to the final state jet energy and the jet energy loss will be power suppressed. Below, we will show this with an explicit one-loop calculation. A natural question is whether we should also work at next to leading power in $R$ for vacuum evolution of the jet. However, these corrections are sub-leading compared to those from the cs radiation and will therefore be ignored. This is because the suppression for the cs mode is only linear in the jet radius ${\cal O}(\beta \sim R)$, compared to the ${\cal O}(R^2)$ corrections in the vacuum. At the same time, despite the $\mathcal{O}(\beta)$ suppression, the cs radiation can still contribute at $\mathcal{O}(1)$ to the $R_{AA}$ when $\beta\sim 1/n$, as described in the previous section. 

The vacuum evolution of the jet creates high-energy partons of energy $\sim p_T$ and angular fluctuations of $\mathcal{O}(R)$. As these energetic color charges or partons move through the medium, the subsequent interactions with the thermal partons cause angular fluctuations of order $Q_{\text{med}}/p_T \sim R^2$. We account for these with a collinear mode (c), which is represented by the green dot in Figure~\ref{fig:LJ}. The momentum scaling  for this mode in $n$- directions is given by
\bea
p^{\mu}_{\rm c}	\sim p_T(1,\beta^2 \delta^2,\beta \delta )  \sim p_T(1, R^4, R^2)	\,.
\eea
Each high-energy parton created in vacuum with angular separation $\theta_c$ or larger contains an independent phase space for the evolution. As a result, the jet is populated by several ``subjets'' as shown by the green cones in Figure~\ref{Mm}, each with angular size $\sim Q_{\text{med}}/p_T \sim R^2$. The contribution to the jet radius measurement by medium-induced radiation inside any of the green cones is power suppressed by ${\cal O}(R^2)$ and it is suppressed compared to the contribution of the cs mode (red) which populates all angles. Therefore, we only need to include the tree-level contribution of the collinear mode as a color source for the $cs$ radiation\footnote{This will change if additional jet substructure measurements are performed.}. The modes can also be visualized on fixed virtuality hyperbolas in a $p^+p^-$ plane as shown in the right panel of Figure.\ref{Mm}. 
Finally, we consider the coherence or formation time $t_f$, which decides whether the partons in the jet interact with the medium coherently. The corresponding phase-space boundary is shown as the red line in Figure~\ref{fig:LJ}. From Figure~\ref{fig:LJ}, we also observe that the cs mode is right on the border of this phase-space boundary so that this mode will be sensitive to the medium length.  On the other hand, the collinear mode has a lifetime that is much larger than $L$ so that it will always interact with the medium coherently, and therefore collinear radiative corrections (high-energy radiations inside any of the green cones in Figure~\ref{Mm}) are suppressed due to the LPM effect. Moreover, emissions at much larger virtualities $\sim p_TR$ and high energy, i.e. the hard collinear mode, have a much smaller lifetime than $L$. These fluctuations do not see the medium at all since their interaction with the medium through an exchange of transverse momentum ($Q_{\rm med}$) is power suppressed by a factor of $R^2$. With these modes in place, we can now visualize all the modes of our EFT in momentum space and on contours of constant virtuality as shown in Figure~\ref{Mm}. 

This mode analysis allows us to conclude that the leading contribution to the cross section from the medium is through the $cs$ mode, which contributes to the jet energy loss at ${\cal O}(\beta \sim R)$. 

We can visualize the evolution of the jet as follows. First, the hard interaction creates an initial parton with large energy($p_T$) and virtuality $\gg p_TR$. This parton splits into multiple collinear partons ($ z\sim 1$) at angles $\theta \geq R$. This splitting process is not seen or affected by the medium due to the large transverse momentum transfer $ \geq p_T\theta_c \sim p_TR$ such that they are essentially vacuum splittings. Therefore, this evolution obeys the vacuum DGLAP equation. As the system evolves to lower virtuality $Q_{\text{med}}$, each of the high-energy partons with $z \sim 1$, well separated ($\geq R^2$) in angle, now define a subjet inside the jet in terms of a collinear mode and act as a source of collinear soft radiation. The collinear and $cs$ modes interact with the medium through forward scattering. 
%%%%%%%%%%%%%%%%%%%%%%%%%%%%%%%%%%%%%%%%%%%%%%%%%%%%%%%%%%%%%%%%%%%%%%%%%%%%%%%
\section{Hard to hard-collinear matching (Stage I EFT)}
\label{sec:StageI}

We first discuss the EFT at the virtuality $p_TR$ integrating out all the physics above this scale. This allows us to factorize the cross section into a hard function, which describes the production of jet-initiating parton. At this stage, we identify the hard-collinear $p^{\mu}_{\rm hc}$ mode, which fluctuates in virtuality from $p_TR$ down to $T$. As a result, all three modes - the collinear, collinear-soft, and soft modes are contained in the hard-collinear mode, which can be observed in Figure~\ref{Mm}. Therefore, all the IR physics at this stage of the EFT is described through the hc mode. For an EFT with a single mode, this is equivalent to a full QCD calculation which is generally challenging. To simplify our calculations, we use the mode analysis discussed in the previous section to anticipate that the interaction of the jet with the medium is primarily through forward scattering between collinear and soft as well as collinear-soft and soft modes. Therefore, we raise the virtuality\footnote{This can be understood as temporarily raising the medium temperature to $p_TR$ or, equivalently, the scale of transverse momentum transfer $|\bfk|$ from the medium to the jet partons from $Q_{\rm med}\geq T$ to $p_TR$.} of the soft mode to $p_T R$. We will eventually expand it in $T/(p_T R)$ and keep only leading terms with $|\bfk|\ll p_T R$ when matching it to virtuality $T$.
Hence, at this stage, the jet evolves with the effective Lagrangian that contains the hc modes, which interacts with the soft medium modes that scale as $p_T(R,R,R)$. The corresponding SCET Lagrangian density is given by~\cite{Bauer:2000yr,Bauer:2001ct,Bauer:2002nz} 
\bea
\mathcal{L}_{\text{SCET}} =  \mathcal{L}_n +\mathcal{L}_{\s}+ \mathcal{L}_{G}^{n\s},
\eea
where $ \mathcal{L}_n$ is the standard SCET collinear Lagrangian \cite{Becher:2014oda} that describes the partons that make up the jet.  Moreover, $\mathcal{L}_{s}$ is the soft SCET Lagrangian, which is equivalent in form to the full QCD Lagrangian since all the components scale uniformly. Finally, the interaction between the energetic jet partons and the medium is mediated by the Glauber mode, which scales as $p^{\mu}_G \sim p_T(R, R^2, R)$ and the corresponding Glauber interaction Lagrangian $\mathcal{L}_{G}^{ns}$ was derived in Ref.~\cite{Rothstein:2016bsq}. We do not make any further assumptions about the medium. The EFT, in general, allows for inhomogeneities in the medium over a length scale greater than $1/T$, which will eventually allow us to account for the medium evolution. In terms of SCET operators the Glauber Lagrangian density is given by
\bea
\mathcal{L}_G^{n \s}&=& C_G \sum_{i,j \in \{q,g\}} \mathcal{O}_{n\s}^{ij} , \  \ \  \text{with} \ \  \ \mathcal{O}_{ns}^{ij}  =\mathcal{O}_{ n}^{ib}\frac{1}{\mathcal{P}_{\perp}^2}\mathcal{O}_{\s}^{jb},
\label{eq:EFTOp}
\eea
where $b$ is a color index, $C_G(\mu)=8\pi\alpha_s(\mu)$ and ${\cal P}_\perp^\mu$ is a derivative operator in the $\perp$ direction, which pulls out the transverse Glauber momentum from the medium soft operators. For this form, the Glauber mode does not appear explicitly since it is not a propagating degree of freedom. Instead, its effect appears through the propagator $1/\mathcal{P}_{\perp}^2$. The gauge invariant $n$-collinear and soft operators are 
\bea
\mathcal{O}_{n}^{qb}&=&\bar{\chi}_{n}t^b\frac{\slashed{n}}{2}\chi_{n}, \ \ \ 
\mathcal{O}_{n}^{gb}=\frac{i}{2} f^{bcd}\mathcal{B}_{n\perp\mu}^c \frac{\bar{n}}{2} \cdot (\mathcal{P}+\mathcal{P}^{\dagger}) \mathcal{B}_{n\perp}^{d\mu}, \\
\mathcal{O}_{\s}^{qb}&=&\bar{\chi}_{\s}t^b\frac{\slashed{n}}{2}\chi_{\s}, \   \   \    
\mathcal{O}_{\s}^{gb}=\frac{i}{2}f^{bcd}\mathcal{B}_{\s \perp \mu}^c \frac{n}{2} \cdot (\mathcal{P}+\mathcal{P}^{\dagger})\mathcal{B}_{\s \perp}^{d\mu},
\label{eq:Opn}
\eea
 where the operator $\mathcal{P}$ pulls out the momentum from the operator it acts on. The soft and collinear operators are built out of the gauge invariant SCET building blocks
\bea
\label{eq:Opbuild}
&& \chi_{n} = W_{ n}^{\dagger}\xi_{ n}, 
  \   \   \   \    
  W_{n} = \text{FT} \  {\bf P} \exp\Big\{ig\int_{-\infty}^0 ds\, \bar n\cdot A_{ n}(x+\bar n s)\Big\},\nonumber\\
 &&\chi_{\s} = S_{n}^{\dagger}\xi_{\s},  
   \   \   \   \    
   S_{ n} = \text{FT} \ {\bf P} \exp\Big\{ig\int_{-\infty}^0 ds\,n\cdot A_{\s}(x+sn )\Big\},
    \nonumber \\
 && \mathcal{B}_{n \perp}^{c\mu} t^c = \frac{1}{g}\Big[W_{n}^{\dagger}iD_{n \perp}^{\mu}W_{ n}\Big],    \  \  
 \mathcal{B}_{S\perp}^{c\mu} t^C = \frac{1}{g}\Big[S_{ n}^{\dagger}iD_{S \perp}^{\mu} S_{ n}\Big].
 \eea
where $W_n$ and $S_n$ are collinear and soft Wilson lines. FT denotes the Fourier transform from position to momentum space. These operators encode bare quarks and gluons dressed by Wilson lines.

%%%%%%%%%%%%%%%%%%%%%%%%%%%%%%%%%%%%%%%%%%%%%%%%%%%%%%%%%%%%%%%%%%%%%%%%%%%%%%%%%%%

\subsection{Factorization}
\label{sec:fact1}
We now use the effective action, discussed in the previous section, to derive a factorization formula for inclusive jet production. To simplify the calculation, we start with an initial state $e^+e^-$. We note that the factorization derived here is sufficiently general such that it also applies to $pp$ and heavy-ion collisions by modifying the initial production of the jet in terms of hard matching coefficients at the scale $p_T$. For heavy-ion collision, the corresponding factorization can be obtained by replacing the $e^+e^-$ hard coefficient function with the appropriate quark and gluon expressions integrated against nuclear PDFs. Moreover, the subsequent jet evolution in the vacuum and medium will be identical in both cases. Since the jet evolution is the main focus here, we will work here with an $e^+e^-$ initial state to avoid a cluttered notation. We derive the form of the factorization for a quark jet and the corresponding result for a gluon-initiated jet can be obtained analogously. 

The basic vertex that we need  to create an initial high-energy quark from the lepton pair is $L_{\mu}J^{\mu}$, where $L_{\mu}$ is the leptonic tensor and 
\bea
J^{\mu}(x) = \bar{\chi} \gamma^{\mu} \chi(x),
\eea
is the hadronic current. Without loss of generality, we can assume the initial quark from the hard interaction to be along the $z$ direction. At tree level, we have an energetic anti-quark in the $\bar n$ direction but this will eventually be integrated out via an operator product expansion (OPE). Therefore, the total Hamiltonian is given by 
\bea
H_{\total}(t)= H(t)+\int d^3{\bm x} \, C(q^2)\, L^{\mu} \, J_{\mu}(x),
\eea
where $H$ is obtained from the SCET Lagrangian defined in the previous section and contains collinear $H_n$, soft $H_s$ and their coupling in terms of Glauber modes $H^{nG}_{\rm int}$ so that the entire SCET Hamiltonian is given by $H=H_n+H_s+H^{nG}_{\rm int}$. Throughout this paper, we will work in Feynman gauge. Moreover,
$C(q^2)$ is the coefficient of the effective hard vertex that creates the jet from the incoming leptons. $q= p_{e^+}+p_{e^-}$ is the energy of the photon in the center of mass frame that creates the initial $q \bar q$ pair. At leading order this is just the photon propagator $\sim 1/q^2$. At tree level, this is just the photon propagator. To proceed further, we consider an initial density matrix for the $e^+ e^-$ state in the background of a medium and evolve it with the effective Hamiltonian 
\bea
\Sigma=\lim_{t \rightarrow \infty} \text{Tr}\Big[ \rme^{-i\int_0^{t}\,  d \hat t\, H_{\total}(\hat{t})}\rho(0)\,\rme^{i\int_0^{t}\, d \hat t\, H_{\total}(\hat{t})} \,\mathcal{M}\Big].
\eea
The trace runs over all final states. We are ignoring any radiative corrections from QED hence we do not consider it here. Here, $\mathcal{M}$ is the measurement imposed on the jet at the detector level, which in the present case is its transverse momentum $p_T$, rapidity $\eta$, and radius $R$. From here on, we leave the limit $t \rightarrow \infty$ implicit. We make one insertion of the hard current on each side of the cut to create the jet from the lepton pair to obtain
\bea
\Sigma =|C(q^2)|^2\int d^4x \int d^4y  \text{Tr} \Big[\rme^{-i   \int d \hat{t}\, H} L_{\mu}\,J^{\mu}(x) \,\rho(0) \,L_{\nu\,}J^{\nu}(y) \,\rme^{i  \int d \hat{t}\, H}\,\mathcal{M}\Big].
\label{eq:Sigma0}
\eea
We assume that the initial density matrix is factorized in terms of the $e^+e^-$ initial state and the background medium density matrix $\rho_M$. The factorization is agnostic to the detailed form of the density matrix, requiring only that it describes a (pure or mixed) state of soft partons with energy much smaller than  $p_T$. Therefore, the total initial density matrix can be written as
\bea
\rho(0) = |e^+e^- \rangle \langle e^+ e^-| \otimes \rho_{M} \,.
\eea 
Plugging this density matrix back into Eq.~(\ref{eq:Sigma0}) and factoring out the leptonic tensor, we obtain
\bea
\Sigma =|C(q^2)|^2L_{\mu \nu}\int d^4x \int d^4y e^{i q\cdot (y-x)} \text{Tr} \Big[e^{i \int d\hat{t}\, H } J^{\mu}(x) \rho_M J^{\nu}(y) e^{-i  \int d\hat{t}\, H }\mathcal{M}\Big],
\eea
where $q= p_{e^+}+p_{e^-}$ is the c.o.m energy. The action of the lepton tensor $L_{\mu\nu}$ annihilates the $e^+e^-$ state leading to a phase factor with q. We can now use the translational invariance of space-time to factor out a 4D volume
\bea
\Sigma =V_4|C(q^2)|^2L_{\mu \nu}\int d^4r \,\rme^{i q\cdot r}\, \text{Tr} \Big[\,\rme^{-i   \int d\hat{t}\, H} J^{\mu}(0) \,\rho_M\, J^{\nu}(r) \,\rme^{i  \int d\hat{t}\, H }\mathcal{M}\Big],
\label{eq:sig}
\eea
after which we can perform an OPE as $ q \rightarrow \infty$ to match onto the collinear jet function following ~\cite{Kang:2016mcy}. This involves integrating out the final state with invariant mass much greater than the jet scale $p_T R$, defined through the measurement imposed on the jet leaving behind modes with virtuality $p_T R$ and lower. We obtain the following result for the inclusive jet production cross section
\begin{align}
\label{eq:factI}
\frac{d\sigma}{dp_T d\eta}&=|C(q^2)|^2L_{\mu \nu}\sum_{i\in q,\bar{q},g}\int  d\omega H^{\mu \nu}_{i}(\omega, \mu)\int \frac{dz}{z}\delta\left( z-\frac{\omega_J}{\omega}\right) J_i(z,\omega_J, \mu),\nn \\
& \equiv  \sum_{i\in q,\bar{q},g}\int \frac{dz}{z}\, H_i(\omega, \mu)\, J_i(z, \omega_J, \mu),
\end{align}
where $\mu$ is factorization scale and $\omega=\omega_J/z$. Here, $H_i(\omega)$ is the hard function that describes the hard-scattering matrix elements and the production of the jet initiating the parton $i$ with energy $\omega$, while the jet function $J_i$ captures its subsequent evolution in the vacuum as well as in the medium with the SCET Hamiltonian. Here, we will focus on the quark jet function, which is given by
% \small
\begin{align}
J_q(z, \omega_J,\mu)= \frac{z}{2N_c}\tr\Big[\,\rho_M\, \frac{\slashed{\bar n}}{2} \chi_{n} (0)\,\rme^{i  \int d\hat{t} H }\mathcal{M} \delta^2(\mathcal{P}_{\perp})\delta(\omega-\bar{n}\cdot\mathcal{P})|J X\rangle \langle J X| \rme^{-i  \int d\hat{t} H }\bar{\chi}_{n}(0)\Big],
\label{eq:Sigma}    
\end{align}
where we have left integrals over the phase space of the final state particles implicit. Here, the delta functions impose momentum conservation along the transverse directions with respect to the jet axis as well as along the direction of the large momentum component $\bar{n}\cdot p$. This also fixes the initial quark produced in the hard interaction along $n^{\mu}$ direction.  Further, $\chi_n$ is the collinear quark field in SCET dressed with collinear Wilson lines composed of the collinear modes in our EFT. Furthermore,  $|J\rangle$ denotes the Hilbert space of all the (collinear) partons that make up the jet while $|X\rangle$ are all the collinear partons that are emitted outside the observed jet.

For semi-inclusive jet production, the measurements in $\mathcal{M}$ are given by  
\begin{equation}
\mathcal{M} = \delta(z - \omega_J/\omega)\Theta_{\text{alg}}\,.
\end{equation}
Here, $\omega_J$ is the $\bar{n}\cdot p$ component of the momentum of all the particles inside the jet defined by the jet algorithm $\Theta_{\text{alg}}$. 
$z$ is the energy fraction of the initial $\bar{n}\cdot p$ component of the hard parton that ends up in the observed jet. We can relate $\omega_J$ to the jet $p_T$ and  rapidity $\eta$ through $\omega_J = 2p_T \cosh \eta$. At NLO, the jet algorithm constrains the relative angle between two partons to be smaller than the jet radius~\cite{Ellis:2010rwa}. Specifically, for $k_T$-type algorithms~\cite{Catani:1993hr,Cacciari:2008gp}, for single emission it is given by
\bea\label{eq:jetconstraint}
\Theta_{\text{alg}}=\Theta\Big(\omega_J x(1-x)\tan\frac{R}{2}-|\bfq|\Big),
\eea 
where $x$ is the longitudinal momentum fraction of final state parton. In principle, Eq.~\eqref{eq:factI} is the factorization formula at this stage of the matching procedure. It is given in terms of a convolution of a hard function that describes the physics at the scale $p_T$ with a collinear jet function that describes the evolution of the jet both in the vacuum and medium at scales below $p_TR$. However, as stated earlier, to make the calculations at this stage more tractable and to explicitly separate the medium-induced evolution from the vacuum, we can further expand out the Hamiltonian order by order in the number of Glauber exchanges between the energetic color partons of the jet and the soft partons of the medium.  The interaction of the hard-collinear and the soft mode is given by $H_{\rm int}^{nG}$, which does not allow for a direct factorization of the jet and the medium to all orders in $\alpha_s$. Hence, we need to expand the above result order by order in $H_{\rm int}^{nG}$ and factorize the soft physics from the hard-collinear sector at each order. This is also useful for doing an explicit calculation, to eventually match this EFT to the one at the scale $T$. To facilitate this, we rewrite Eq.~(\ref{eq:Sigma}) as
%\small
\begin{align}
J_q(z, \omega_J,\mu) &= \frac{z}{2N_c}\text{Tr}\Bigg[ e^{-i\int d\hat{t}\,(H_n+H_{\s})}\mathcal{T}\Big[e^{-i\int_0^t dt_l H^{nG}_{\rm int, I}(t_l)}\bar{\chi}_{n,{\rm I}}(0)\Big] \rho_M\frac{\slashed{\bar n}}{2}\nn\\
&\times \mathcal{\bar T}\Big[e^{-i\int_0^t dt_r H^{nG}_{\rm int, I}(t_r)} \delta^2(\mathcal{P}_{\perp})\delta(\omega-\bar{n}\cdot\mathcal{P})\chi_{n,{\rm I}}(0)\Big] e^{i\int d\hat{t}\,(H_n+H_{\s} )} \mathcal{M}\Bigg]. 
\label{eq:fact}  
\end{align}
where $\mathcal{T}$ and $\bar{\mathcal{T}}$ represents and time and anti-time ordering of the operators. The series of steps that lead from Eq.~\ref{eq:Sigma} to Eq.~\ref{eq:fact} have been explained in detail in Ref~.\cite{Vaidya:2020lih}. Here the subscript I indicates that the operators are now dressed with the Hamiltonian $H_{n}+H_{s}$, i.e.
\bea
  \mathcal{O}_{\rm I} = e^{i (H_n+H_{\s} )t} \mathcal{O}  e^{-i (H_n+H_{\s} )t}.
\eea
 We can now expand out the jet function in terms of the number of jet-medium interactions
 \bea
 \label{eq:Jexp}
 J_q(z, \omega_J, \mu) = \sum_{i=1}^{\infty} J_q^{(i)}.
 \eea
%%%%%%%%%%%%%%%%%%%%%%%%%%%%%%%%%%%%%%%%%%%%%%%%%%%%%%%%%%%%%%%%%%%%%%%%%%%

\subsubsection{Vacuum evolution}
\label{sec:vacI}
In the vacuum, there are no interactions between the energetic jet partons and the soft partons of the medium at leading power. Therefore, we need to recover the vacuum factorization involving the semi-inclusive jet function as discussed in Ref.~\cite{Kang:2016mcy}. In our framework, this is achieved by expanding out the time and anti-time ordered exponentials containing the jet-medium interaction term in Eq.~\eqref{eq:fact}, which reduces to the identity at leading order. We can write the cross section as
\begin{equation}
\frac{d\sigma^{(0)}}{dp_T d\eta}=\int \frac{dz}{z} H_q(\omega) \tr\Big[ e^{-i\int dt(H_n+H_{\s} )}\bar{\chi}_{n,{\rm I}}(0) \rho_M\frac{\slashed{\bar n}}{2}\delta^2(\mathcal{P}_{\perp})\delta(\omega-\bar{n}\cdot\mathcal{P})\chi_{n,{\rm I}}(0) e^{i\int dt (H_n+H_{\s} )} \mathcal{M}\Big].
\end{equation}
Here, the superscript $d\sigma^{(0)}$ indicates that there are no insertions of $H_{\rm int,I}^{nG}$. Further, since the Hilbert space of the collinear and the soft modes  are factorized as $|X \rangle = |X_n\rangle |X_{\s} \rangle$, we can use the commutativity of $H_n$ and $H_{\s}$ to separate the term containing $\rho_M$ as 
\begin{equation}
\frac{d\sigma^{(0)}}{dp_Td\eta}\!=\!\!\int\! \frac{dz}{z} H_q(\omega) \tr\Big[ e^{-i\int dt H_n }\bar{\chi}_{n,{\rm I}}(0) \frac{\slashed{\bar n}}{2}\delta^2(\mathcal{P}_{\perp})\delta(\omega-\bar{n}\cdot\mathcal{P})\chi_{n,{\rm I}}(0) \mathcal{M} e^{i\int dt H_n }\Big] 
\tr\Big[e^{-iH_{\s}t}\rho_M e^{iH_{\s} t}\Big].
\label{eq:vacfact}
\end{equation}
Since the medium partons are not measured the second term containing medium density matrix trivially reduces to one. We can now write everything in Eq.~\eqref{eq:vacfact} in the interaction picture of the free theory by writing the collinear Hamiltonian in terms of free and interacting Hamiltonian $H_n = H_n^{(0)} +H_{n}^{\text{int}}$. Following some simplifications, we can write the cross section as
\begin{equation}
\frac{d\sigma^{(0)}}{dp_Td\eta}=\int \frac{dz}{z} H_q(\omega) \text{Tr}\Big[ \mathcal{T}\Big[e^{-i\int dt_l H_{n,\text{I}}^{\text{int}}}\bar{\chi}_{n,\text{I}}(0)\Big] \frac{\slashed{\bar n}}{2}\mathcal{\bar T}\Big[e^{-i\int dt_r H_{n, \text{I}}^{\text{int}}}\delta^2(\mathcal{P}_{\perp})\delta(\omega-\bar{n}\cdot\mathcal{P})\chi_{n,\text{I}}(0)\Big] \mathcal{M}\Big]. 
\label{eq:vacfact1}
\end{equation}
It is worth mentioning that in the absence of soft and Glauber modes, all operators in the above equation are now dressed with the free theory collinear Hamiltonian. The operators of the form $\mathcal{O}_I$ are given by
\bea
\mathcal{O}_\text{I} = e^{iH_n^{(0)}t} \mathcal{O}  e^{-iH_n^{(0)}t},
\eea
which is in the usual interaction picture. For brevity, we are going to drop the explicit factors of $H^{\text{int}}$,  which are always understood to be included in a time-ordered product. Therefore, we can write the jet production cross section in the standard factorized form
\bea
\frac{d\sigma^{(0)}}{dp_Td\eta} &=& \int \frac{dz}{z} H_q(\omega)\otimes J_q^{(0)}(z,\omega_J,\mu),
\eea
where the vacuum jet function $J_q^{(0)}$ is defined as
\bea
J^{(0)}(z,\omega_J,\mu) =\frac{z}{2N_c}\text{Tr}\Big[\bar{\chi}_{n,\text{I}}(0)\frac{\slashed{\bar n}}{2}\delta^2(\mathcal{P}_{\perp})\delta(\omega-\bar{n}\cdot\mathcal{P})\chi_{n,\text{I}}(0) \mathcal{M}\Big].
\eea
Note that the vacuum semi-inclusive jet function is identical to the one in \cite{Kang:2016mcy} where the tree and one-loop results for quark and gluon-initiated jets were computed. For completeness, we provide the corresponding jet function expressions in Appendix~\ref{app:JVOneLoop}.

%%%%%%%%%%%%%%%%%%%%%%%%%%%%%%%%%%%%%%%%%%%%%%%%%%%%%%%%%%%%%%%%%%%%%%%%%%%%%%%

\subsubsection{Factorization for a single medium interaction}
\label{sec:JMedI}
Now we consider higher order terms in the Glauber Hamiltonian expansion in Eq.~\eqref{eq:fact}. The first term, which is linear in $H^{nG}_{\text{int}}$, does not contribute since it contains a single soft operator. This follows from the assumption that the medium is color singlet state so that the expectation value of a color density operator $\langle \bar \chi_s \Gamma T^A \chi_s \rangle$ vanishes in the medium. Therefore, the leading non-vanishing contribution involves two Glauber insertions  $\mathcal{O}(H^{nG}_{\text{int}})^2$. This is essentially the single scattering regime of the jet interacting with the medium. At this order, there are two contributions -- one Glauber insertion on each side of the cut and both insertions on the same side of the cut. We refer to these as real and virtual contributions, respectively. For the real contribution, the differential cross section takes the form 
\begin{align}
\frac{d\sigma_R^{(2)}}{dp_Td\eta} &=|C_G|^2\int \frac{dz}{z} H_q(\omega)\int d^4x \int d^4y
\sum_{i=q,\bar{q},g} \tr\bigg[ e^{-i\int dt\, (H_n+H_{\s})}\mathcal{T}\bigg[\bigg(\mathcal{O}_{n}^{qa}\frac{1}{\mathcal{P}_{\perp}^2}\mathcal{O}_{\s}^{ib} \bigg)(x)\bar{\chi}_{n,{\rm I}}(0)\bigg] \rho_M\nn\\
&\qquad\qquad\times  
\frac{\slashed{\bar n}}{2}\mathcal{\bar T}\bigg[\bigg(\mathcal{O}_{n}^{qb}\frac{1}{\mathcal{P}_{\perp}^2}\mathcal{O}_{\s}^{ib}\bigg)(y)\delta^2(\mathcal{P}_{\perp})\delta(\omega-\bar{n}\cdot\mathcal{P})\chi_{n,{\rm I}}(0)\bigg]e^{i\int dt\, (H_n+H_{\s})} \mathcal{M}\bigg],   
\end{align}
where $a$ and $b$ are color indices and Glauber Wilson coefficient is given by $C_G=8\pi \alpha_s$. There is also an implicit summation over i, which is the flavor $i \in \{q, \bar q ,g \}$ of partons that make up the medium soft partons. 
Similar to vacuum case, the delta functions represent transverse momentum and energy conservation. Following the same procedure as in the vacuum case, we can now separate the collinear and soft functions 
\begin{align}
&\frac{d\sigma_R^{(2)}}{dp_Td\eta}=|C_G|^2\int \frac{dz}{z} H_q(\omega)\int d^4x \int d^4y\sum_{i=q,\bar{q},g}\tr \Bigg[e^{-i\int\,dt H_n}\mathcal{T}\Big[\mathcal{O}_{n}^{qa}(x)\bar{\chi}_{n,{\rm I}}(0)\Big]\frac{\slashed{\bar n}}{2}\mathcal{\bar T}\Big[\mathcal{O}_{n}^{qb}(y)\nn\\
\,\,&\delta^2(\mathcal{P}_{\perp})\delta(\omega-\bar{n}\cdot\mathcal{P})\chi_{n,{\rm I}}(0)\Big]e^{i\int dt\,H_n} \mathcal{M}\Bigg] \tr\Bigg[ e^{-i\int dt\, H_{\s}}\frac{1}{\mathcal{P}_{\perp}^2} \mathcal{O}_{\s}^{ia}(x) \rho_M(0)\frac{1}{\mathcal{P}_{\perp}^2}\mathcal{O}_{\s}^{ib} (y)e^{i\int dt\,H_{\s}}\Bigg].
\label{eq:CollGlauber}
\end{align}
Here, the second term is a trace over soft operators computed in the background medium. We note that there is no measurement on this correlator. 
We can now express everything in terms of the free theory interaction picture using $H_n = H_n^{(0)}+H_{n}^{\text{int}}$ for the collinear Hamiltonian and $H_{\s} = H_{\s}^{(0)}+H_{\s}^{\text{int}}$ for the soft Hamiltonian. We obtain
\begin{align}
\frac{d\sigma_R^{(2)}}{dp_Td\eta}&=|C_G|^2\int \frac{dz}{z} H_q(\omega)\int d^4x \int d^4y \text{Tr}\Bigg[\mathcal{T}\Big[e^{-i\int dt_l H^{\text{int}}_{n, \text{I}}}\mathcal{O}_{n,{\rm I}}^{a}(x)\bar{\chi}_{n,\text{I}}(0)\Big]\frac{\slashed{\bar n}}{2}\nn\\
&\mathcal{\bar T}\Big[e^{-i\int dt_r H^{\text{int}}_{n, \text{I}}}\mathcal{O}_{n,{\rm I}}^{b}(y)\delta^2(\mathcal{P}_{\perp})\delta(\omega-\bar{n}\cdot\mathcal{P})\chi_{n,{\rm I}}(0)\Big] \mathcal{M}\Bigg] S^{ab}(x,y),  
\end{align}
where the medium correlator in terms of the soft operators is given as 
\bea
\label{eq:Medcorr1}
S^{ab}(x,y)=\sum_{i=q,\bar{q},g}\tr\Bigg[\mathcal{T}\Big[e^{-i\int dt_l H^{\text{int}}_{\s,\text{I}}}\frac{1}{\mathcal{P}_{\perp}^2}\mathcal{O}^a_{\s,{\rm I}}(x)\Big]\rho_M(0)\mathcal{\bar T}\Big[e^{-i\int dt_r H^{\text{int}}_{\s,\text{I}}}\frac{1}{\mathcal{P}_{\perp}^2}\mathcal{O}^b_{\s,{\rm I}}(y)\Big]\Bigg].
\eea
Assuming that the medium has no net color, the correlator will be proportional to $\delta^{ab}$.
In a homogeneous medium, the correlator only depends on the relative distance between the soft operators. For a medium that is evolving in time and space, this is no longer true. However, in general, the non-homogeneity in the medium is only visible over length scales much greater than those probed by a single medium interaction. We can therefore make a change of variables, defining $\hat x = x- y$ and $\bar x = x+ y$. The coordinate $\hat x$ tracks the short distance physics of the medium while $\bar{x}$ tracks its slow evolution over space and time.
We can express the medium correlator in the momentum space conjugate to $\hat x $ as
\bea
\label{eq:Medcorr2}
S^{ab}(\hat x, \bar x) = \delta^{ab} \int \frac{d^4k}{(2\pi)^4} \, e^{i k \cdot \hat x} S(k,\bar x)\,,
\eea
where $k$ scales as the Glauber momentum. From power counting arguments, we find that $S(k, \bar x)$ does not depend on $k^+$ since the contribution from the Glauber exchange is sub-leading. We can now write the real contribution (i.e. for Glauber insertions on opposite sides of the cut) to the cross section as
\begin{align}
\frac{d\sigma_R^{(2)}}{dp_Td\eta} &= |C_G|^2\int \frac{dz}{z} H_q(\omega)\int d^4 \hat x  \int d^4 \bar x \int \frac{d^4k}{(2\pi)^4} e^{i k \cdot \hat x} \tr\Bigg[\mathcal{T}\Big[e^{-i\int dt_l H^{\text{int}}_{n, \text{I}}}\mathcal{O}_{n,{\rm I}}^{a}(x)\bar{\chi}_{n,\text{I}}(0)\Big]\frac{\slashed{\bar n}}{2}\,\nn\\
& \times\mathcal{\bar T}\Big[e^{-i\int dt_r H^{\text{int}}_{n, \text{I}}}\mathcal{O}_{n,{\rm I}}^{b}(y)\delta^2(\mathcal{P}_{\perp})\delta(\omega-\bar{n}\cdot\mathcal{P})\chi_{n,{\rm I}}(0)\Big] \mathcal{M}\Bigg]S(k^-,\bfk,\bar x )\delta^{ab},
\end{align} 
where $\bfk$ is the transverse component of the Glauber momentum.
Given the power counting of the $p^-$ component of the momentum in the collinear sector, we can drop the $k^-$ contribution from the Glauber term. In other words, the Glauber interaction does not change the energy of the jet, i.e., there is no collisional energy loss. 
The $\bar x $ dependence in the function S encodes the spatio-temporal inhomogeneity in the medium. Assuming that the medium is inhomogeneous over much longer scales than those probed in a single jet-medium interaction,  we can drop the $\bar x_{\perp}$ dependence from the medium correlator $S$. Likewise, we can also drop $\bar x^+$ dependence from $S$ since it is sub-leading to the contribution from the jet function which has a very large $p^-$ momentum.  Therefore, we can write   
\begin{align}
&\frac{d\sigma_R^{(2)}}{dp_Td\eta}  =|C_G|^2\int \frac{dz}{z} H_q(\omega)\int d^4 \hat x \int d^4 \bar x\int \frac{d^2\bfk dk^+ }{(2\pi)^3}e^{i \bfk \cdot \bf \hat x+i k^+ \hat x^-}\Bigg[\mathcal{T}\Big[e^{-i\int dt_l H^{\text{int}}_{n, \text{I}}}\mathcal{O}_{n,{\rm I}}^{a}(x)\nn\\
&\bar{\chi}_{n,\text{I}}(0)\Big]\frac{\slashed{\bar n}}{2}\mathcal{\bar T}\Big[e^{-i\int dt_r H^{\text{int}}_{n, \text{I}}}\mathcal{O}_{n,{\rm I}}^{b}(y)\delta^2(\mathcal{P}_{\perp})\delta(\omega-\bar{n}\cdot\mathcal{P})\chi_{n,{\rm I}}(0)\Big] \mathcal{M}\Bigg]\int \frac{dk^-}{2\pi}S(k^-, \bfk, \bar x^-),
\end{align}
Now doing the integral over $k^+$, which explicitly sets $\hat x^- =0$, we find that the interaction with the medium is instantaneous. We can explicitly carry out the integrals over the variables $\hat x^+, \bar x^+,  \bf \hat x, \bf \bar x$. This yields energy and transverse momentum conservation at each vertex. We find
\begin{align}
\frac{d\sigma_R^{(2)}}{dp_Td\eta}&=|C_G|^2\int \frac{dz}{z} H_q(\omega)\int d\bar x^- \int \frac{d^2\bfk }{(2\pi)^2}\tr\Bigg[\mathcal{T}\Big[e^{-i\int dt_l H^{\text{int}}_{n, \text{I}}}\Big(\delta(\bar{n}\cdot\mathcal{P})\nn\\
&\delta^2( \bfk+\mathcal{P}_{\perp})\mathcal{O}_{n,{\rm I}}^{a}(\frac{\bar x^-}{2})\Big)\bar{\chi}_{n,\text{I}}(0)\Big]\frac{\slashed{\bar n}}{2}\mathcal{\bar T}\Big[e^{-i\int dt_r H^{\text{int}}_{n, \text{I}}}\Big(\delta(\bar{n}\cdot\mathcal{P})\delta^2(\bfk-\mathcal{P}_{\perp})\mathcal{O}_{n,\text{I}}^{b}(-\frac{\bar x^-}{2})\Big)\nn\\
&\delta^2(\mathcal{P}_{\perp})\delta(\omega-\bar{n}\cdot\mathcal{P})\chi_{n,{\rm I}}(0)\Big] \mathcal{M}\Bigg]  \int \frac{dk^-}{2\pi}S(k^-, \bfk, \,\bar x^-), \nn\\
& \equiv  |C_G|^2\int_{\in QGP} d \bar x^- \int \frac{dz}{z} H_q(\omega) \int \frac{d^2 \bfk}{(2\pi)^2}J_r^{(1)}(z, \omega_J, R; \bfk, \bar x^-)\varphi(\bfk, \bar x^-)\,,
\end{align}
where we defined 
\begin{align}
&J_{r}^{(1)}(z,\omega_J,R ; \bfk,\bar x^-)=\frac{e^{i(p^+_a-p^+_b)\frac{\bar x^-}{2}}}{\bfk^2}\tr\Bigg[\mathcal{T}\Big[e^{-i\int dt_l H^{\text{int}}_{n, \text{I}}}\Big(\delta(\bar{n}\cdot\mathcal{P})\delta^2( \bfk+\mathcal{P}_{\perp})\mathcal{O}_{n,{\rm I}}^{a}(0)\Big)\nn\\
&\bar{\chi}_{n,\text{I}}(0)\Big]\frac{\slashed{\bar n}}{2}\mathcal{\bar T}\Big[e^{-i\int dt_r H^{\text{int}}_{n, \text{I}}}\Big(\delta(\bar{n}\cdot\mathcal{P})\delta^2(\bfk-\mathcal{P}_{\perp})\mathcal{O}_{n,\text{I}}^{b}(0)\Big)\delta^2(\mathcal{P}_{\perp})\delta(\omega-\bar{n}\cdot\mathcal{P})\chi_{n,{\rm I}}(0)\Big] \mathcal{M}\Bigg]\frac{z}{4\pi N_c} \,.
\label{eq:Injetrn}
\end{align}
Here $p_a^+, p_b^+$ are the plus momentum components of the operators $\mathcal{O}_n^a$ and $\mathcal{O}_n^b$, respectively. 
The medium correlator is defined as  
\bea
\label{eq:Medcorr3}
\varphi(\bfk, \bar x^-)= \frac{1}{\bfk^2}\frac{\delta_{ab}}{N_c^2-1}\int \frac{dk^-}{2\pi}\int d^4r e^{i(k^-\frac{r^+}{2}-\bfk \cdot \bfr)}\text{Tr}\Bigg[e^{-i\int d\hat{t} H_{\s} }\mathcal{O}^{ia}_{\s}(r^+, \bfr , r^-+\bar x^-)\rho_M \mathcal{O}^{ib}_{\s}(0)e^{i\int d\hat{t} H_{\s} }\Bigg], \nn\\   
\eea
We can treat the case of two Glauber insertions on the same side of the cut in the same way. We denote the corresponding part of the medium jet function as $J_v^{(1)}$ and define it as 
\begin{align}
&J_v^{(1)}(z,\omega_J,R;\bfk,\bar x^-)=e^{i(p^+_a-p^+_b)\frac{\bar x^-}{2}}
\tr\Bigg[\mathcal{T}\Big[e^{-i\int dt_{1l} H^{\text{int}}_{n, \text{I}}}\Big(\delta(\bar{n}\cdot\mathcal{P})\delta^2( \bfk+\mathcal{P}_{\perp})\mathcal{O}_{n,{\rm I}}^{a}(0) \Big)\nn\\
&e^{-i\int dt_{2l} H^{\text{int}}_{n, \text{I}}}\left(  \delta(\bar{n}\cdot\mathcal{P})\delta^2(\bfk-\mathcal{P}_{\perp})\mathcal{O}_{n,{\rm I}}^{b}(0)\right)\bar{\chi}_{n,\text{I}}(0)\Big]\frac{\slashed{\bar{n}}}{2}\mathcal{\bar T}\Big[\delta^2(\mathcal{P}_{\perp})\delta(\omega-\bar{n}\cdot\mathcal{P})\chi_{n,{\rm I}}(0)\Big] \mathcal{M}\Bigg]\frac{z}{8\pi N_c}. 
\end{align}
The full result can be written as
\begin{equation}
\frac{d\sigma^{(2)}}{dp_Td\eta}  =|C_G|^2\,\int 
 d\bar x^-\int \frac{dz}{z} H_q(\omega)\int \frac{d^2\bfk}{(2\pi)^2 }\left( J_{r}^{(1)}-J_{v}^{(1)}\right)(z,\omega_J,R; \bfk, \bar x^-)\varphi(\bfk ,\bar x^-), \label{eq:realminusvirtual} 
\end{equation}
If the medium is homogeneous, which is the brick model of the medium, and extends in the $x^-$ direction over a length $L$, then we can drop the $\bar x^-$ dependence from the correlator $\varphi$ and explicitly do the integral over $\bar x^-$, which yields an explicit enhancement by the factor $L$ 
\begin{align}
\frac{d\sigma_R^{(2)}}{dp_Td\eta} =|C_G|^2\,L\int \frac{dz}{z} H_q(\omega)\int \frac{d^2\bfk}{(2\pi)^2}\, (J_{r}^{(1)} -J_{v}^{(1)})(z,\omega_J,R; \bfk,L) \varphi(\bfk)\,.
\end{align}
The medium jet function now becomes
\begin{align}
&J_{r}^{(1)}(z,\omega_J,R ; \bfk,L)=\frac{e^{i(p^+_a-p^+_b)\frac{L}{4}}}{\bfk^2}\text{sinc} \Big[ (p^+_a-p^+_b)\frac{L}{4}\Big]\tr\Bigg[\mathcal{T}\Big[e^{-i\int dt_l H^{\text{int}}_{n, \text{I}}}\Big(\delta(\bar{n}\cdot\mathcal{P})\delta^2( \bfk+\mathcal{P}_{\perp})\mathcal{O}_{n,{\rm I}}^{a}(0)\Big)\nn\\
&\bar{\chi}_{n,\text{I}}(0)\Big]\frac{\slashed{\bar n}}{2}\mathcal{\bar T}\Big[e^{-i\int dt_r H^{\text{int}}_{n, \text{I}}}\Big(\delta(\bar{n}\cdot\mathcal{P})\delta^2(\bfk-\mathcal{P}_{\perp})\mathcal{O}_{n,\text{I}}^{b}(0)\Big)\delta^2(\mathcal{P}_{\perp})\delta(\omega-\bar{n}\cdot\mathcal{P})\chi_{n,{\rm I}}(0)\Big] \mathcal{M}\Bigg]\frac{z}{4\pi N_c} \,.
\label{eq:jetrn}
\end{align}
with an identical modification to $J_v^{(1)}$.
%%%%%%%%%%%%%%%%%%%%%%%%%%%%%%%%%%%%%%%%%%%%%%%%%%%%%%%%%%%%%%%%%%%%%

\subsection{The single-interaction medium jet function}\label{sec:cal1loop}

For the purpose of matching the EFT at the medium scale $Q_{\med}$, we explicitly compute the medium jet function $J_r^{(1)}-J_v^{(1)}$ up to the one-loop level. In the factorization derived so far, we have assumed that the medium partons have energy ranging from $p_TR$ down to the medium temperature $T$ and that $|\bfk|$ has support over this whole momentum range. However, we know that the density matrix of the medium $\rho_M$ consist of partons of energy $T$, which source the Glauber exchange while higher energy partons are suppressed. Likewise, due to the $1/\bfk^4$ propagator of the Glauber exchange that is cut off in the IR at the scale $m_D$, the medium mostly exchanges $|\bfk| \sim m_D$ in one interaction with a strong suppression for higher momenta. Therefore, we are interested in matching the EFT at $p_T R$ to the one at the scale $m_D$ with $|\bfk|  \ll p_TR$. There are two equivalent ways to perform the calculation. Either we carry out the full calculation with the current setup and expand out the final result in $m_D/(p_TR)$. Alternatively, we can expand out the Feynman integrands assuming $|\bfk| \ll  p_TR$  before performing integrations explicitly. Here we adopt the latter approach as it significantly simplifies the calculation. In the following, we derive the results at the tree and one-loop level, keeping on the results at leading power in $m_D/(p_TR)$ with the goal of matching this to an EFT at a lower scale later on.
%%%%%%%%%%%%%%%%%%%%%%%%%%%%%%%%%%%%%%%%%%%%%%%%%%%%%%%%%%%%%%%%%%%%%%%%%%%%%%%%%

\begin{figure}
\centering
\includegraphics[width=0.8\linewidth]{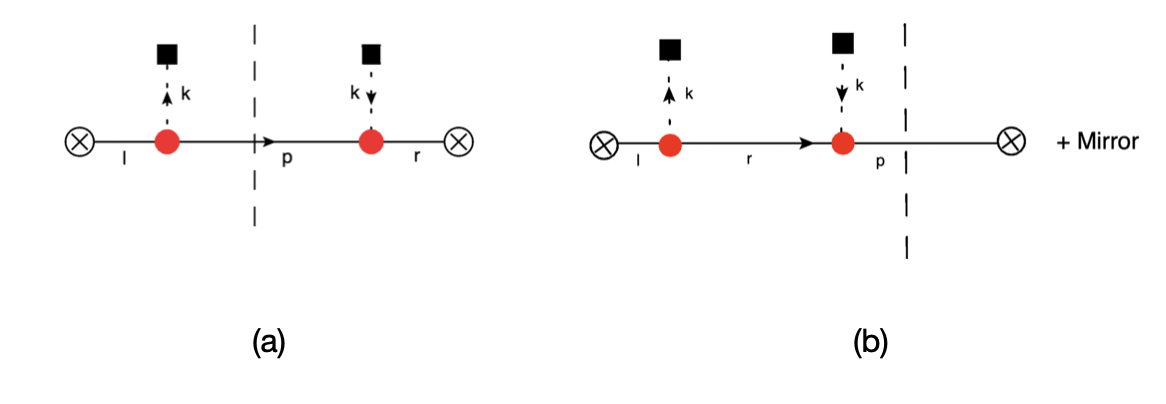}
\caption{Feynman diagrams contributing to the in-medium scattering of a single high-energy quark.~\label{Treeh}}
\end{figure}

\noindent 
\textbf{Tree level result:~\label{sssec:treeJ}} At tree level, we have a single high-energy light quark that makes up the measured jet. Without loss of generality, we assume that this quark has zero transverse momentum so that before any interaction with the medium it moves along $n =(1,0,0,1)$. The transverse momentum of this quark can fluctuate by $|\bfk|$ acquired during forward scattering with the medium. 
To account for the interaction with the medium, we need to consider two types of diagrams associated with $J_{r}^{(1)}$ and $J_{v}^{(1)}$, see Fig.~\ref{Treeh}.
The result for the real contribution, shown as the left diagram($J_{r}^{(1)}$) in Fig.~\ref{Treeh}, can be written as
\begin{align}
3a =&\,\delta(1-z)\int \frac{d^4p}{(2\pi)^4} \delta(p^2)\int \frac{dl^+}{2\pi} \int \frac{dr^+}{2\pi} \delta(\omega-\bar{n}\cdot p)\delta^2(\bfp-\bfk) \frac{ p^-}{ l^+ +i\epsilon}  \frac{1}{r^+- i\epsilon}\nn\\
&\qquad\qquad\,\times e^{i \frac{L}{4}(l^+-r^+)}\sinc\Big[\frac{L}{4}(l^+-r^+)\Big],\nn\\
=&\,\delta(1-z).    
\end{align}
Here we obtain contour integrals over $l^+$  and $r^+$ since the $+$ component of momentum is not conserved at the Glauber vertex (red dot) due to the finite medium size in the $x^-$ direction. This results in the sinc function in the $+$ component of the momentum in the definition of the jet function in Eq.~\eqref{eq:jetrn}.
Similarly, we can compute the tree-level term for insertions on the same side of the cut ($J_{v}^{(1)}$) 
\bea
3b + \text{mirror}=\delta(1-z).
\eea
We observe that the combination $J_r^{(1)}-J_v^{(1)}$ vanishes at the tree level due to the relative minus sign in Eq.~(\ref{eq:realminusvirtual}). We thus need at least two partons for the medium to modify the vacuum jet function for this observable.

\begin{figure}
\centering
\includegraphics[width=0.75\linewidth]{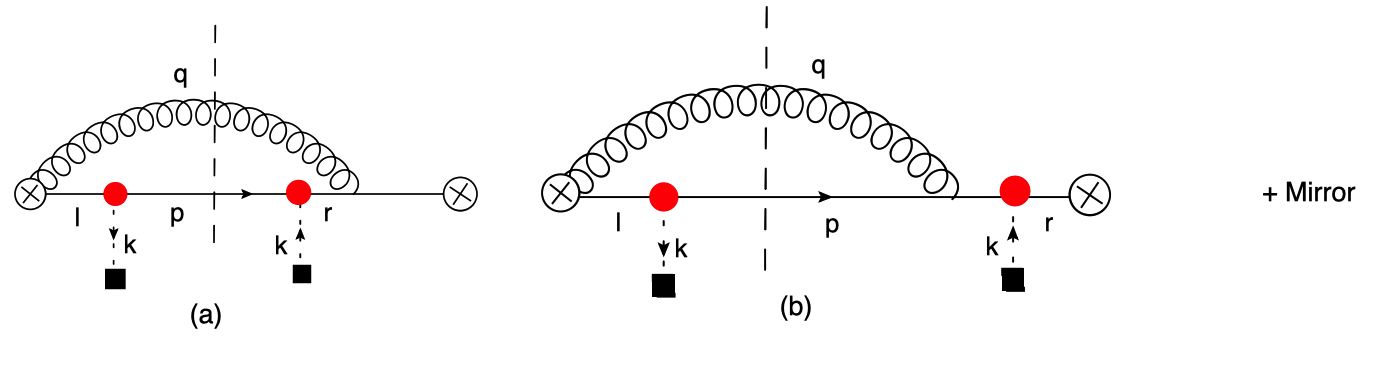}
\caption{One-loop Feynman diagrams contributing to the medium-modified inclusive jet cross section.~\label{LW1}}
\end{figure}

%%%%%%%%%%%%%%%%%%%%%%%%%%%%%%%%%%%%%%%%%%%%%%%%%%%%%%%%%%%
\noindent
\textbf{One-loop result:~\label{sssec:OneloopJ}}We now consider a single gluon emitted from the initial high-energy quark followed by a single interaction of either parton with the medium. As mentioned earlier, we will only keep the leading non-zero contributions in $|\bfk| \ll p_TR$ while evaluating each diagram. There are various Feynman diagrams that need to be taken into account. We first look at the diagrams shown in Fig.~\ref{LW1} to understand which regions of phase space contribute at leading power. The remainder of the calculation is presented in Appendix \ref{app:Loop}. From the scaling analysis outlined in section \ref{sec:modes} we note that the relevant contributions come from the collinear-soft and collinear momentum modes that interact with the medium. To see how this works in practice, we start by considering the left diagram in Fig.~\ref{LW1}, where we have a gluon emission with momentum $q$ off the quark, which retains momentum $p$. In this case, the quark is interacting with the medium.
When the gluon emission is inside the jet, we get
\begin{align}
&4a =g^2\text{Tr}[t^ct^at^at^c]\int \frac{d^4p}{(2\pi)^4} \delta(p^2) \int \frac{d^4q}{(2\pi)^3} \delta(q^2)\delta^2(\bfp+\bfq-\bfk) \delta(\omega-p^--q^-)  \int \frac{dl^+}{2\pi} \int \frac{dr^+}{2\pi}\frac{1}{q^-}\nn\\
&\qquad\,\frac{(p^-)^3}{l^+p^- - q_{\perp}^2+i\epsilon} \frac{1}{p^-r^+ -q_{\perp}^2 -i\epsilon}\frac{1}{\omega^- (q^++r^+)-i\epsilon}  e^{-i \frac{\bar x^-}{2} (l^+-r^+)}\delta(1-z)\Theta_{\rm alg},
\end{align}
where $\Theta_{\rm alg}=\left(R/2-|\bfq|\omega/(q^-(\omega-q^{-}))\right)$, see Eq.~(\ref{eq:jetconstraint}) above. We first carry out the contour integration over $l^+$ by closing the contour in the lower half plane. Next, we perform the contour integration over $r^+$, which has two poles. We note that the pole $r^+ = \bfq^2/p^-$ corresponds to the quark going on-shell before interacting with the medium. In this case, $L$ dependence vanishes. For the contribution from the first pole, we find
\begin{equation}
4a_{1}=4g^2C_F\int \frac{d^2\bfq}{(2\pi)^2}\frac{1}{\bfq^2} \int \frac{dx}{2\pi} \frac{x}{1-x}\,\delta(1-z)\Theta_{\rm alg} ,    
\end{equation}
where we have used $q^-=(1-x)\omega$. For the second pole contribution, we have
\begin{equation}
4a_{2}=-4g^2C_F\int \frac{d^2\bfq}{(2\pi)^3}\frac{1}{\bfq^2} \int\frac{dx}{2\pi}\frac{x}{1-x} e^{-i \frac{\bar x^-}{2} \frac{\bfq^2}{x(1-x)\omega}}\delta(1-z)\Theta_{\rm alg}.     
\end{equation}
We can now consider several distinct momentum scalings of the emitted gluon:
\begin{enumerate}
\item{Hard-collinear emission $q^{\mu} \sim p_T( 1, R^2, R)$:}  Since $ |\bfk| \ll |\bfq|$, we can drop the $|\bfk|$ dependence at leading power. Then this diagram is just the vacuum contribution and will cancel exactly with the corresponding contribution from the Glauber insertions on the same side of the cut, i.e. $J_v^{(1)}$. Therefore, when keeping the first non-zero contribution, this gives a correction that is suppressed by $R^2$.

\item{Collinear emission $ q^{\mu} \sim p_T( 1, R^4, R^2)$:} We can verify again that the first non-zero contribution is suppressed by $R^2$. In this case, this is due to the emission angle for this scaling $\ll R$.
 
\item{Collinear-soft emission $q^{\mu} \sim p_TR( 1, R^2, R)$:} We find that this contribution is also power suppressed but only by a factor linear in the jet radius $R$. This can be understood from the fact that this type of radiation contributes to the jet radius $R$ at leading power but to the jet energy only at ${\cal O}(\beta \sim R)$. We will show this explicitly in the complete one-loop result. Therefore, we conclude that the main contribution from the medium for this observable comes from the collinear-soft mode scaling of the emitted gluon.
\end{enumerate}

Keeping only the collinear-soft contribution, we arrive at the following result where we also include the complex conjugate contribution
\begin{align}
4a=&8g^2C_F\int \frac{d\bfq}{(2\pi)^2}\frac{1}{\bfq^2} \int\frac{dx}{2\pi}\frac{1}{1-x}  \bigg[1-\cos \left(\frac{\bar x^-}{2} \frac{\bfq^2}{(1-x)\omega}\right)\bigg]\delta(1-z)\Theta_{\rm alg}  \,.
\label{eq:verdiag}
\end{align}

We note that the location in the medium $\bar x^-$ always appears through the ratio $\frac{\bar x^-}{2}\frac{\bfq^2}{(1-x)\omega} \equiv \bar x^-/t_f$ where $t_f$ is the formation time of the emitted gluon defined in Section \ref{sec:Scales}. The function $ F\left(\frac{\bar x^-}{t_f} \right)= \cos \left(\frac{\bar x^-}{t_f} \right) $ which appears in the result above encodes the quantum interference between the vacuum and medium evolution. If we use the brick model in the limit where the medium length $L \gg t_f$, the $\bar x^-$ integral over  $F\rightarrow 0$ and the interference term vanishes. For the collinear soft radiation considered here, based on the Lund diagram Fig.\ref{fig:LJ}, for current colliders  $t_f \sim L$ so we need to retain this factor.  Given the analysis of the different gluon momentum regions at one loop, from here on we will only consider the contributions where the emitted medium-induced gluon is collinear soft.  We have four sets of diagrams for the real and virtual corrections to $J_r^{(1)}$ and $J_v^{(1)}$. The full set of diagrams along with the one loop integrals with the explicit medium length dependence is given in Appendix \ref{app:Loop}.
The jet function at one loop can be written as 
\bea
J^{(1)}(z, \omega, R; \bfk) = J^{(1)}_{qg}(z,\bfk)+ J^{(1)}_{q}(z,\bfk),
\eea
where jet function for the gluon emission inside the identified jet is given by
\begin{align}
J^{(1)}_{qg}(\omega,z,R;\bfk, \bar x^-)&= \frac{\alpha_s(N_c^2-1)}{4\pi^2\bfk^2}\int_0^1\frac{dx }{1-x}\int d^2\bfq\Bigg[    \Big\{\frac{\bfk^2}{\bfq^2(\bfq+\bfk)^2}+\frac{1}{(\bfq+\bfk)^2}-\frac{1}{\bfq^2}\Big\}\nn\\
& \times \Big\{1- \cos\Big[\frac{(\bfq+\bfk)^2 \bar x^-}{2\omega(1-x)}\Big] \Big\}\left(\Theta_{\rm alg}-1 \right)\Bigg] \delta(1-z) \,.
\end{align}
where we have set $q^- = \omega(1-x)$. The $ \cos x $ term  is a result of interference betwee vacuum and medium-induced emissions due to a finite medium size. Here, $\Theta_{\rm alg}=\Theta(\omega(1-x)R/2 -|\bfq|)$. For the case where the gluon emission is outside the cone, the quark jet function is
\begin{align}
J^{(1)}_{q}(z,\omega,R;\bfk, \bar x^-)&=\frac{\alpha_s(N_c^2-1)}{4\pi^2\bfk^2}\frac{z}{1-z} \int d^2\bfq\Bigg[  \frac{\bfk^2}{\bfq^2(\bfq+\bfk)^2}+\frac{1}{(\bfq+\bfk)^2}-\frac{1}{\bfq^2}\Bigg] \nn\\
&\times \Big\{1- \cos\Big[\frac{(\bfq+\bfk)^2 \bar x^-}{2\omega(1-z)}\Big] \Big\}\Theta'_{\rm alg},
\end{align}
where $\Theta'_{\rm alg}=\Theta(|\bfq|-R(1-z)\omega/2)$. We can combine the terms to write the result in a more compact form as
\begin{eqnarray}
J^{(1)}_{qg}(z,\omega,R;\bfk , \bar x^-)&=& \frac{\alpha_s(N_c^2-1)}{4\pi^2 \bfk^2}\,\delta(1-z)\, \int d^2\bfq\int_0^1\frac{dx }{1-x} \frac{2\bfk \cdot \bfq}{\bfq^2(\bfq+\bfk)^2}\Theta\Big(|\bfq|-\frac{(1-x)\omega R}{2}\Big)\nn\\ 
& \times &  \Big\{1- \cos\Big[\frac{(\bfq+\bfk)^2 \bar x^-}{2\omega(1-x)}\Big] \Big\}.
\label{eq:StageInloop}     
\end{eqnarray}
Similarly, for the gluon emission outside the jet, we get
\begin{eqnarray}
J^{(1)}_{q}(z,\omega,R;\bfk,\bar x^-) &= &  -\frac{\alpha_s(N_c^2-1)}{4\pi^2 \bfk^2} \frac{z}{1-z} \int d^2\bfq \frac{2\bfk\cdot \bfq}{\bfq^2(\bfq+\bfk)^2} \Theta\Big(|\bfq|-\frac{(1-z)\omega R}{2}\Big)\nn\\
& \times &  \Big\{1- \cos\Big[\frac{(\bfq+\bfk)^2 \bar x^-}{2\omega(1-z)}\Big] \Big\},
\label{eq:StageOloop}    
\end{eqnarray}
This result agrees exactly with the one-loop result of the GLV formalism Refs.~\cite{Gyulassy:2000er,Wiedemann:2000za}, which is an important cross check on our framework. We compute this one loop result explicitly in the limit $L \rightarrow \infty$ so that the interference term $\cos$ vanishes.  
Carrying out the remaining integrals and expanding the result in terms of plus distributions, the one-loop result of the jet function evaluates to 
\begin{align}
J^{(1)}(z,\omega,R;\bfk)&=\frac{1}{\bfk^2}\frac{(N_c^2-1)\alpha_s}{2\pi} \Bigg\{ \left( \Big[\frac{\ln (1-z)}{1-z}\Big]_+-\Big[\frac{\ln (1-z-t)}{1-z}\Big]_+\right) \Theta (1-z-t)+\Bigg[-\frac{1}{\epsilon}\left(\frac{\mu}{|\bfk|}\right)^{2\epsilon}\nn\\
&\hspace{-1.5cm}\times\Big[\frac{1}{1-z}\Big]_{+}-\left( \left(\frac{\mu}{|\bfk|}\right)^{2\epsilon}\frac{1}{2\epsilon} +2 \ln t \right) \frac{1}{[1-z]_+} + \Big[\frac{ \ln [(1-z) (t-1+z)]}{1-z}\Big]_+ \Bigg] \Theta( t-1+z)\Bigg\},
\end{align}
where $ t = |\bfk|/(\omega R)$, which scales as $R$ in our hierarchy of scales. 
The plus distribution is defined in a standard way with the appropriate modification by the $t$-dependent phase space constraint. For example, we have
\bea
\int_0^1 dz \Big[\frac{f(z)}{1-z}\Big]_{+}\Theta(t-1+z) &=& \int_{1-t}^1 dz\, \frac{f(1)-f(z)}{1-z}.
\eea
We see that we do not have a pure $\ln t$ term since in the limit $t \rightarrow 0$, the terms in the first line vanish. In the second line, the $\Theta$ function forces $z \rightarrow 1$ so that the corresponding terms also vanish as the expressions appear in the form of plus distributions. Therefore, when we expand out our result for $t \ll 1-z $, the leading non-zero term is proportional to $t$. We also notice that this expression is not UV divergent. All divergences regulated by dimensional regularization are IR divergences which will be cut off by screening effects ($m_D$) in the medium. 

We find that the medium jet function only starts at ${\cal O}(\alpha_s)$. As a result, the one-loop expression does not contain any UV or rapidity divergences. However, this will not be true when we consider the two-loop result. Even without doing any explicit calculation, we can say that the two-loop result for the jet function will have a rapidity divergence and the corresponding Renormalization Group (RG) equation will be the negative of the BFKL evolution equation. We can see this in two ways. First, the anomalous dimension is a property of the operator irrespective of the state for which it is evaluated. Hence, we expect a BFKL evolution given the earlier computation of the same operator in Ref.~\cite{Vaidya:2021vxu}. In this case, a $k_T$ measurement led to a non-zero result at tree level and a BFKL rapidity divergence at one loop. Second, as we will see in the next section, the other function that appears in our factorization formula, the medium correlator $\varphi(\bfk)$, obeys the BFKL equation. From the RG consistency of the factorization, the jet function has to follow the negative of the BFKL evolution. We can therefore write the RG equation for $J^{(1)}(z,\bfk)$ as 
\begin{align}
\frac{dJ^{(1)}(z,\bfk;\mu,\nu)}{d\ln \nu} &= - \int d^2\bfu\, \mathcal{K}_{\text{BFKL}}(\bfk,\bfu)\,J^{(1)}(z,\bfu;\mu,\nu),\nn\\
&= -\frac{\alpha_s N_c}{\pi^2} \int d^2\bfu\, \Bigg[\frac{J^{(1)}(z,\bfu;\mu,\nu)}{(\bfu-\bfk)^2} -\frac{\bfk^2J^{(1)}(z,\bfk;\mu,\nu) }{2\bfu^2(\bfu-\bfk)^2}\Bigg].
\end{align}
where we have suppresses the $\omega, R$ arguments. The natural scale $\nu$ for the jet function is the $-$ component of the gluon momentum, which is here the scale $\omega(1-z)$. The corresponding scale for the $\varphi$ function is $|\bfk| \sim m_D$ so that the solution of the BFKL equation resums logarithms of the form $\alpha_s\ln (m_D/(\omega(1-z))$.
Likewise, by demanding $\mu$ independence of the cross section, we can also infer that just like the vacuum jet function, the single interaction medium jet function $J^{(1)}(z,\bfk;\mu,\nu)$ also obeys the DGLAP type equation in the renormalization scale $\mu$. As with the rapidity logs, the DGLAP type logs are also only visible at two loops due to the vanishing of the tree level result.
%%%%%%%%%%%%%%%%%%%%%%%%%%%%%%%%%%%%%%%%%%%%%%%%%%%%%%%%%%%%%%%%%%%%%%%%%%%%

\subsection{The medium correlator}

We now consider the medium correlator, which was defined in Eq.~(\ref{eq:Medcorr3}) as a correlation function written in terms of the soft SCET operators 
\begin{align}
\varphi(\bfk)= \frac{1}{\bfk^2}\frac{\delta_{ab}}{N_c^2-1}\int \frac{dk^-}{2\pi}\int d^4r e^{i(k^-r^+/2-\bfk \cdot \bfr)}\text{Tr}\Bigg[e^{-i\int d\hat{t} H_{\s} }\mathcal{O}^{ia}_{\s}(r)\rho_M \mathcal{O}^{ib}_{\s}(0)e^{i\int d\hat{t} H_{\s} }\Bigg].
\end{align} 
The medium is non-perturbative and, hence, this function cannot be computed order by order in $\alpha_s$. Nonetheless, we can compute its anomalous dimension for the evolution between two perturbative renormalization scales analogous to a parton distribution function. As with the jet function, the anomalous dimension is a property of the operator irrespective of the state for which it is evaluated. The one-loop computation of this correlator in a thermal medium was carried out in Ref.~\cite{Vaidya:2021vxu}. Here we quote the RG evolution equations for $\varphi$, which can be written as
\begin{align}
\frac{d  \varphi(\bfk;\nu;\mu)}{d\ln \nu} &= \int d^2\bfu\, \mathcal{K}_{\text{BFKL}}(\bfu,\bfk)\,\varphi(\bfk;\nu;\mu),\nn\\
&= \frac{\alpha_s N_c}{\pi^2} \int d^2\bfu \Bigg[\frac{\varphi(\bfu;\nu;\mu)}{(\bfu-\bfk)^2}-\frac{\bfk^2\varphi(\bfk;\nu;\mu) }{2\bfu^2(\bfu-\bfk)^2}\Bigg],\nn\\
\frac{d  \varphi(\bfk;\nu;\mu)}{d\ln \mu}& =  -\frac{\alpha_s \beta_0}{\pi} \varphi(\bfk;\nu;\mu). 
\label{eq:BRG}
\end{align}
The medium correlator has both a rapidity and a UV divergence. The rapidity divergence leads to a BFKL equation while the UV anomalous dimension is the one-loop QCD beta function, which leads to the running of the coupling $\alpha_s$. Solving this RG equation leads to $\alpha_s$ being evaluated at the scale $\mu = |\bfk|$, the transverse momentum exchanged in a single interaction with the medium.

%%%%%%%%%%%%%%%%%%%%%%%%%%%%%%%%%%%%%%%%%%%%%%%%%%%%%%%%%%%%%%%%%%
\section{Hard-collinear to soft-collinear matching (Stage II EFT)~\label{sec:StageII}}

In this section, we refactorize the medium jet function shown in Eq.~\eqref{eq:realminusvirtual} and match the EFT to a lower virtuality $Q_{\med} \gtrsim T$ where the jet energy loss is entirely associated with collinear-soft modes. In the following, we discuss the factorization and matching procedure in more detail.  

%%%%%%%%%%%%%%%%%%%%%%%%%%%%%%%%%%%%%%%%%%%%%%%%%%%%%%%%%%%%%%%%%%%%%
\subsection{Factorization}

The factorization formula discussed in Eq.~\eqref{eq:realminusvirtual} separates the physics at the hard scale $p_T$ from the jet scale $p_TR$ and below by effectively performing an expansion in the power counting parameter $R$. The one-loop calculation of the jet function in Section~\ref{sec:cal1loop} therefore accounts for contributions from modes of virtuality $p_T R$ down to the medium scale $T$. Since the transverse momentum $\bfk$ of the exchanged Glauber mode between energetic jet and medium partons in a single interaction is of the order of $m_D$, we can further expand out the jet function in $|\bfk|/(p_T R) \ll 1$. As a result, we can match the EFT discussed in the previous section to a lower virtuality by integrating out the contributions from modes of virtuality $p_TR$ in terms of matching coefficients, which we will describe below. As discussed in section~\ref{sec:modes}, the EFT at virtuality $Q_{\med}$ contains three modes -- the collinear, collinear-soft, and soft modes which are separated in rapidity as shown in Fig.~\ref{Mm}. The collinear modes account for high-energy partons with momentum fractions $z\sim 1$, created during vacuum splittings with time scale $p_T^{-1}$ and the corresponding shower down to the virtuality $Q_{\med}$. These modes then interact with the soft modes of the medium, sourcing medium-induced collinear-soft radiation that contributes to the jet energy loss at leading order. Further, if the angular separation between the collinear modes is larger than the decoherence angle $\theta_c$, they are resolved by the medium and behave as independent sources of medium-induced collinear-soft radiation. In that case, the phase space is populated by multiple collinear partons, each of which constitute a subjet with an angular size $\ll R$ inside the jet of radius $R$. On the other hand, if $R\ll \theta_c$, the medium cannot resolve the color of the subjets, and the jet acts as a single coherent source of medium-induced collinear soft radiation. We first focus on the case $R\geq \theta_c$, where the phase space is sufficiently large to allow for multiple subjets each of size $Q_{\med}/p_T \sim R^2$ that are separated from each other by an angle $\theta \gg Q_{\med}/p_T$. As the jet evolves, both the collinear and collinear-soft modes from each subjet interact with the medium through forward scatterings mediated by Glauber modes. The EFT and corresponding operators, which include all three sectors coupled via Glauber modes, were derived in Ref.~\cite{Rothstein:2016bsq}. We can write the Lagrangian density that mediates the forward scattering between the three modes in terms of explicit SCET operators as follows:
\begin{equation}
\mathcal{L}_G^{\text{int}}(x)= \sum_{n'} \sum_{i,j= q, \bar q, g}\mathcal{O}^{ia}_{c\,n', \cs}\frac{1}{\mathcal{P}_{\perp}^2}\mathcal{O}^{ja}_{\s,n'}+  \sum_{i,j=q,\bar q, g}\mathcal{O}^{ia}_{\cs,n}\frac{1}{\mathcal{P}_{\perp}^2}\mathcal{O}^{ja}_{\s,n} \equiv \mathcal{L}^{\text{int}}_{c-\cs-\s}+ \mathcal{L}^{\text{int}}_{\cs-\s},
\label{eq:LG}       
\end{equation}
While the Lagrangian describes both quark and gluon jets, we will focus here on quark jets $i=q$ in the collinear sector.  $1/\mathcal{P}_{\perp}^2$ is the Glauber propagator, which extracts the corresponding momentum from the soft operators. The first term in Eq.~(\ref{eq:LG}) describes the effective interactions between all three modes. $\mathcal{O}_{\s,n'}^{ja}$ is the soft operator,  and $\mathcal{O}_{c\,n',\cs}^{ia}$ describes collinear-soft emission from a collinear parton. Here, $c\,n'$ denotes different collinear modes, and we sum over different directions inside the jet labeled by distinct light-light vectors $n'$. The second term describes the forward scattering of collinear-soft radiation with the soft modes of the medium. Here, $n$ denotes the direction of the entire jet. The collinear-soft radiation occupies the full angular phase space of the jet, allowing interactions between distinct subjets. The operators $\mathcal{O}^{j,a}_{\s,n}$ and $\mathcal{O}^{i,a}_{\cs,n}$ are identical to the soft $\mathcal{O}_{\s}^{i,a}$ and collinear $\mathcal{O}_n^{ia}$ operators, respectively, as given in Eq.~(\ref{eq:Opn}). The operator $\mathcal{O}^{i,a}_{c\,n,\cs}$ that describes the interaction between collinear and collinear-soft modes reads 
\begin{equation}
\mathcal{O}^{ia}_{c\,n,\cs}= \mathcal{O}^{ib}_n \frac{1}{\mathcal{P}_{\perp}^2}\mathcal{O}_{ \cs}^{ba} \,,    
\end{equation}
where $\mathcal{O}^{ib}_n $ is again the collinear current given in Eq.~(\ref{eq:Opn}). Lastly, the collinear-soft operator $\mathcal{O}_{ \cs}^{ba}$ which encodes medium induced gluon emission is given as
\begin{align}
\mathcal{O}_{cs}^{ba}&= 8\pi \alpha_s\Bigg[\mathcal{P}_{\perp}^{\mu}S_n^TW_{n}\mathcal{P}_{\perp \mu} -\mathcal{P}_{\mu}^{\perp}g \mathcal{\tilde B}_{S\perp}^{n\mu}S_n^TW_n-S_n^TW_ng \mathcal{\tilde B}^{n \mu}_{\perp}\mathcal{P}_{\mu}^{\perp}-g \mathcal{\tilde B}_{S \perp}^{n\mu}S_n^TW_ng\mathcal{\tilde B}^{n}_{\perp \mu}\nn \\
&\qquad\,\qquad -\frac{n_{\mu}\bar n_{\nu}}{2}S_n^Tig  G^{\mu\nu}W_n\Bigg]^{ba},
\label{eq:csop}
\end{align}
which is defined in terms of SCET building blocks, see Eq.~(\ref{eq:Opbuild}), and $G^{\mu \nu}$ is the gauge field tensor. The gauge field that appears in all operators in Eq.~\eqref{eq:csop} is $A^{\mu}_{cs}$. Further, the operators $\tilde B$ and $ G$  are evaluated in the adjoint representation with  
\bea
\mathcal{\tilde B}_{n\perp}^{ab}= -if^{abc}\mathcal{B}_{n\perp}^c\,.
\eea
Since the jet is populated by multiple subjets, we expect that the matching procedure leads to a series in terms of an increasing number of subjets. We now begin with the jet function in Eq.~\eqref{eq:realminusvirtual} and refactorize it in terms of matching functions convolved with subjet functions at virtuality $Q_{\med}$ that are resolved by the medium. 
\begin{align}
J_{i}(z,\omega_J,\mu)&=\int_{0}^{1} \rmd z' \int_0^{\infty}\rmd \epsilon_L \, \delta(\omega_J'-\omega_J-\epsilon_L)\sum_{m}\prod_{j=2}^m\int \frac{{\rm d}\Omega(n_j)}{4\pi} \langle \mathcal{C}_{i\rightarrow m}(\{\underline{n}\},z', \omega_J', \mu)\,{\cal S}_{m} (\{\underline{n}\} , \epsilon_L,\mu)\rangle\,,
 \label{eq:mdfactorization}
\end{align}
which is valid up to power corrections of $\mathcal{O}(Q_{\rm med}/p_TR)$. We have $\omega_J'= z'\omega_J/z$, and the term $\{\underline{n}\}\equiv\{n_1,n_2,...,n_m\} $ denotes the direction of $m$ subjets with $n_i\cdot n_j\sim \theta_c$. Analogous to the hard function that describes the production of jet initiating parton at scale $p_T$ in Eq.~\eqref{eq:realminusvirtual}, the fist term $\mathcal{C}_{i\to m}$ describes the production of $m$ collinear partons from an initial parton $i$ at virtuality $p_TR$. The second term $\mathcal{S}_m$ describes the production and evolution of collinear-soft radiation from the $m$ resolved collinear partons and $\epsilon_L$ is the energy loss out of the jet cone due to emission of collinear-soft radiation. $\langle ... \rangle $ denotes color trace. The function $\mathcal{S}_{m}$ depends on the radius $R$ through the jet algorithm measurement. The variable $\Omega(n_j)$ denotes the angular phase space of $m$ subjets inside the jet. The convolution between the matching function and ${\cal S}_m$ is given in terms of the angle $\Omega(n_j)$ and energy loss $\epsilon_L$.  While phase space is in principle available for multiple subjets for $R\gg \theta_c$, in practice it may be sufficient to truncate the series in Eq.~\eqref{eq:mdfactorization} to a finite order and consider only the first few terms. Since the subjets are well separated from one another, each of them will act as an independent source of collinear-soft radiation through a Wilson line in the direction of the subjet. As a result, ${\cal S}_m$ can be written as a correlator of $m$ distinct collinear-soft Wilson lines dressing the $m$ collinear fields along the directions $n_j$ that source them. For a single subjet with only one collinear parton in the $n$ direction and one collinear-soft Wilson line, the function ${\cal S}_1$ is given by 
\begin{align}
\mathcal{S}_{1}(\epsilon_L) = &\,\int d^2\bfr \int dr^+e^{-i\omega r^+}\tr\Big[ \mathcal{T}\Big[e^{-i\int dt \bar H}[U(n)U(\bar n)\bar{\chi}_{cn}](0)\Big]\frac{\slashed{\bar n}}{2} \rho_M \nn \\
&\times\, \mathcal{\bar T}\Big[e^{-i\int dt \bar H}[U(n)^{\dagger}U(\bar n)^{\dagger}\chi_{cn}](\bfr, r^+)\Big]\mathcal{M}\Big],
\label{eq:subjet1}
\end{align}
where we have suppressed all other arguments in $\mathcal{S}_1$ except for the energy loss $\epsilon_L$. This function evolves with the total Hamiltonian $\bar H$
\begin{equation}
\label{eq:Ham2}
\bar H = H_{c}+H_{\cs}+H_{\s}+H^{\rm int}_{G}.
\end{equation}
Here, $H^{\rm int}_{G}= H_{c-\cs-\s}^{\rm int}+ H^{\rm int}_{\cs-s}$ is the Glauber Hamiltonian which can be obtained from the Lagrangian in Eq.~\eqref{eq:LG}. Further, $H_{c}$ and $H_{cs}$ are two copies of the collinear SCET Hamiltonian, while $H_{s}$ is the soft SCET Hamiltonian which is identical to the full QCD Hamiltonian. The collinear-soft Wilson line $U(n)$ appearing in Eq.~\ref{eq:subjet1} is defined as
\begin{equation}
U(n)= {\bf P} \exp\left[ig \int_0^{\infty} \rmd s  \, n\cdot A_{\rm cs}(s n)\right].
\end{equation}
Lastly, $\chi_{cn}$ in Eq.~\eqref{eq:subjet1} is the SCET collinear quark field operator defined in Eq.~\eqref{eq:Opbuild}. 

%%%%%%%%%%%%%%%%%%%%%%%%%%%%%%%%%%%%%%%%%%%%%%%%%%%%%%%%%%%%%%%%%%%%%%%%%

\subsection{Integrating out the collinear sector}

As discussed in the previous section, the leading contribution to the jet energy loss comes from collinear-soft radiation while the contribution from collinear emissions is subleading. Therefore, at leading power, we can treat the collinear Hamiltonian ($H_c$ in Eq.~\eqref{eq:subjet1}) as a free theory Hamiltonian, which we denote as $H_c^{(0)}$. As a result, the physical process involves the propagation of a highly energetic collinear parton that interacts with the medium and induces collinear-soft radiation. Consequently, in the Hamiltonian $\bar{H}$, the collinear quark field $\chi_{cn}$ appears either through a kinetic term $H_c^{(0)}$ or in the Glauber interaction operator $H_G^{\rm int}$ via the three-sector operator
\begin{equation}
\sum_{i,j}\mathcal{O}_n^{ib}\frac{1}{\mathcal{P}_{\perp}^2}\mathcal{O}_{\rm cs}^{ba} \frac{1}{\mathcal{P}_{\perp}^2} \mathcal{O}_{\rm s}^{ja},    
\end{equation}
which is the first term in Eq.~\eqref{eq:LG}. For a jet initiated by a quark $i=q$, the operator $\mathcal{O}_n^{qb}$ is the same as the collinear quark current operator defined in Eq.~\eqref{eq:Opn} where the collinear Wilson line is now set to the identity since we are ignoring all radiative corrections to the collinear sector. Therefore, the Hamiltonian is quadratic in the collinear degrees of freedom and can be integrated out. With these simplifications, the result for the subjet function reduces to 
\begin{equation}
\mathcal{S}_{1}(\epsilon_L)= \tr\Big[ \mathcal{T}\Big[e^{-i\int dt\,\tilde H}U_{\bar n}(0)U_n(0)\Big]\rho_M \mathcal{\bar T}\Big[e^{-i\int dt\, \tilde H}U_{n}^{\dagger}(0)U_{\bar n}^{\dagger}(0)\Big]{\mathcal{M'}}\Big].
\label{eq:1subjet}
\end{equation}
Here, the measurement function $\mathcal{M'}$ is defined as   
\begin{equation}
\mathcal{M'} =\Theta_{\text{alg}}\times \delta\left(\epsilon_L- \bar{n}\cdot \mathcal{P}\right), 
\end{equation}
where $\bar{n}\cdot\mathcal{P}$ is the energy of the collinear-soft radiation that flows out of the jet. The function ${\cal S}_1$ now evolves with the Hamiltonian 
\begin{equation}
\label{eq:H2}
\int\! {\rm d}t\, \tilde H(t)\!=\!\int {\rm d}t \left(H_{\rm cs}(t)+H_{\rm s}(t)+ H_{\rm cs\text{-}s}(t)\right) + \int {\rm d} s\, \mathbf{O}_{\rm cs\text{-}s}(sn).
\end{equation}
As before, $H_{\text{cs}}$ is standard collinear SCET Hamiltonian, and $H_{\text{s}}$ describes the dynamics of the soft field. Moreover, $H_{\text{cs-s}}$ describes the forward scattering of the collinear-soft gluon off a soft medium parton. We have now completely eliminated the collinear sector and, as a result, we obtained the operator $\mathbf{O}_{\text{cs-s}}$ accounting for medium-induced radiation to all orders, and it is defined along the world line of the collinear parton that sources it. The only information about the collinear sector that is retained is the direction of the resolved collinear parton. The Hamiltonian interaction density for the forward scattering of collinear-soft gluons with the soft partons of the medium is given by 
\begin{align}
\label{EFTOp}
\mathcal{H}_{\rm cs\text{-}s}
 &= C_G\frac{i}{2}f^{abc}\mathcal{B}_{n \perp\mu}^b\frac{\bar n}{2}\cdot(\mathcal{P}+\mathcal{P}^{\dagger})\mathcal{B}_{n \perp}^{c\mu}\frac{1}{\mathcal{P}_{\perp}^2}\mathcal{O}_{\rm s}^a. 
\end{align}
The operator along the world line is given by
\begin{align}\label{eq:Oc-s}
\mathbf{O}_{\text{cs-s}}(sn) = \int {\rm d}^2\bfq \frac{1}{\bfq^2}\Big[\mathcal{O}^{ab}_{\text{cs}}\frac{1}{\mathcal{P}_{\perp}^2}\mathcal{O}_{\text{s}}^b\Big](sn,\bfq) t^a,
\end{align} 
which is expressed in terms of $\mathcal{O}_{\text{cs}}$, see Eq.
~\eqref{eq:csop}, and the soft operator $\mathcal{O}_{\text{s}}$ as defined earlier. A detailed derivation of this simplified formulation starting from Eq. \eqref{eq:subjet1} and Eq.~\eqref{eq:Ham2} is given in Appendix \ref{sec:collint}.  
We can further generalize this result to the $m$ subjet configuration and define the ${\cal S}_m$ subjet function in Eq.~\eqref{eq:mdfactorization} as
\begin{align}
\mathcal{S}_{m}(\epsilon_L)= \tr\Big[ \mathcal{T}\Big[e^{-i\int dt\,\tilde H_m}U_m(n_m)...U_{1}(n_1)U_0(\bar n)\Big]\rho_M \mathcal{\bar T}\Big[e^{-i\int dt\, \tilde H_m}U^\dag_0(\bar n)U_1^\dag(n_1)...U_m^\dag(n_m)\Big]{\mathcal{M'}}\Big],
\label{eq:msubjet}
\end{align}
which is given by a correlator of $m$ collinear soft Wilson lines along the directions of the $m$ subjets. Note that we have suppressed color indices on the Wilson lines which can be either fundamental or adjoint based on the parton species that sourced them. As stated earlier in Eq.~\eqref{eq:mdfactorization} there is an implicit contraction of these color indices with the coefficient $\mathcal{C}_{i\rightarrow m}$ and an overall color trace. This function encodes the interference between the collinear soft radiation from different subjets which includes the LPM and color decoherence effects.
What is remarkable is that the refactorization of the jet function in Eq.~\eqref{eq:mdfactorization} takes the exact same form as encountered in the context of non-global logarithms in Ref.~\cite{Becher:2015hka}. See also ~\cite{Dasgupta:2001sh,Larkoski:2015zka}.
However, in comparison to the case of non-global logarithms, the Hamiltonian with which it evolves is different and is given as 
\begin{equation}
\label{eq:Hm}
\int\! {\rm d}t\, \tilde H_m(t)\!=\!\int {\rm d}t \left(H_{\rm cs}(t)+H_{\rm s}(t)+ H_{\rm cs\text{-}s}(t)\right) + \sum_{i=i}^m\int {\rm d} s_i\, \mathbf{O}_{\rm cs\text{-}s}(s_i n_i),
\end{equation}
with an $\mathbf{O}_{\rm cs\text{-}s}$ operator along the word line of each subjet.
The matching functions ${\cal C}_{i\to m}$ in Eq.~\eqref{eq:mdfactorization} start at $\mathcal{O}(\alpha_s^{m-1}(p_TR))$, allowing us to express the refactorized jet function as a perturbative series. The interference between the collinear-soft radiation off distinct collinear partons leads to the emergence of the decoherence angle $\theta_c$. To explicitly see this scale we require a two-loop calculation, which is beyond the scope of this work. We refer the reader to~\cite{Mehtar-Tani:2011lic}, where an explicit calculation for a dipole antenna was performed. Note that the function $\mathcal{S}_{m}$  depends on the jet radius $R$. It contributes to the measurement of the jet radius as well as out-of-jet radiation, i.e., energy loss. In a medium of finite size, it explicitly depends on the medium length $L$ through the enhancement by an overall factor $L$ and the LPM effect.

%%%%%%%%%%%%%%%%%%%%%%%%%%%%%%%%%%%%%%%%%%%%%%%%%%%%%%
\subsection{One-loop matching from stage I to stage II}

In this paper, we focus on the regime $\theta_c \geq R$ so that only the first term of the series in Eq.~\eqref{eq:mdfactorization} is retained. In this case, we can simplify the factorization as
\begin{align}
J_{i}(z,\omega_J,\mu)&=\int_{0}^{1} \rmd z' \int_0^{\infty}\rmd\epsilon_L \, \delta(\omega_J'-\omega_J-\epsilon_L)\,\mathcal{C}_{i\rightarrow 1}(\{\underline{n}\},z', \omega_J', \mu)\,{\cal S}_{1} (\{\underline{n}\} , \epsilon_L,\mu).
 \label{eq:Onejet}
\end{align}
For this case of a single subjet, the color structure for the coefficient $\mathcal{C}_{i\rightarrow 1}$ is trivial. 
For stage I, we expanded out the jet function order by order in the number of interactions with the medium Eq.\ref{eq:Jexp}.
\bea
J_{i}(z,\omega_J,\mu) = \sum_{j=0}^{\infty} J_i^{(j)}
& \equiv & J_i^{(0)} + |C_G|^2L\int \frac{d^2 \bfk}{(2\pi)^2}(J_{r}^{(1)}-J_{v}^{(1)})(\bfk) \,\varphi(\bfk) +  \ldots\,.
\eea
The tree and one-loop results for the stage I vacuum quark jet function $J_q^{(0)}$ are given in Appendix \ref{app:JVOneLoop}. Results up to one loop for the single interaction medium jet function  $(J_{r,n}^{(1)}-J_{v,n}^{(1)})$) were calculated in section \ref{sec:cal1loop}. Our goal in this section is to match the one-loop results from stage I to the stage II EFT and obtain the matching coefficient $\mathcal{C}_{i \rightarrow m}$. 

To achieve this, we explicitly compute the function ${\cal S}_1$ in Eq.~\eqref{eq:mdfactorization}. In this case, the jet acts as a coherent color source for the energy loss leading to a single subjet that we need to take into account. As a result, at NLO, we expect $\mathcal{S}_1$ to capture both medium-induced effects as well as all the vacuum physics below the scale $p_TR$, expanded in $T/(p_TR)$. As for Stage I, we will also expand this function order by order in the number of jet-medium interactions.
\begin{equation}
\mathcal{S}_{1} = \sum_{i=0}^{\infty}\mathcal{S}_{1}^{(i)}.
\label{eq:seriess}
\end{equation}
Since the factorization should remain valid even when the medium is turned off, comparing the vacuum result is sufficient to extract the matching coefficient $\mathcal{C}_{q\rightarrow 1}$. However, as a consistency check of our factorization, we will also compute the result for a single medium interaction to one loop for the stage II factorization and verify that the EFT reproduces the stage I result expanded out to leading power in  $T/p_TR$.  The function ${\cal S}_1$ encodes the physics of both collinear soft modes of the jet and the soft modes that populate the medium. To further separate out the physics of these two modes, we need to expand out the Glauber interaction Hamiltonian order by order to factorize the medium soft physics from measurement-dependent collinear-soft dynamics in the jet. The procedure we follow is the same as the one we employed for the stage I EFT.  Therefore, starting from Eq.~\eqref{eq:1subjet}, we rewrite ${\cal S}_1$ in the interaction picture as    
\begin{align}
\mathcal{S}_{1}(\epsilon_L) &=  \tr\Big[e^{-i(H_{\cs}+H_{\s})t} \mathcal{T}\Big[e^{-i\int dt {H}^{\text{int}}_{G,{\rm I}}}U_{\bar n}(0)U_n(0)\Big]\nn\\
 &\times \rho_M \mathcal{\bar T}\Big[e^{-i\int dt {H}^{\text{int}}_{G,{\rm I}}}U_n^{\dagger}(0)U^{\dagger}_{\bar n}(0)\Big]e^{i(H_{\cs}+H_{\s})t}\mathcal{M'}\Big],
\end{align}
where ${H}^{\text{int}}_{G,{\rm I}}$ is the Glauber interaction for a single subjet and reads as
\begin{equation}
\int dt H^{\text{int}}_{G,{\rm I}}= \!\int {\rm d}t  H_{\rm cs\text{-}s,I}(t) + \int {\rm d} s\, \mathbf{O}_{\rm cs\text{-}s,I}(sn).
\end{equation}
Here, the subscript I refers to dressing the operator with $H'=H_{\cs}+H_{\s}$ so that $\mathcal{O}_{\rm I} = e^{iH't}\mathcal{O} e^{-iH't}$.
We can now expand out ${\cal S}_1$ order by order in ${H}^{\text{int}}_{G,{\rm I}}$, which corresponds to an expansion in the number of collinear-soft--soft interactions and prove factorization at each order.

\subsubsection{Matching for the vacuum evolution}

\begin{figure}
\centering
\includegraphics[width=0.7\linewidth]{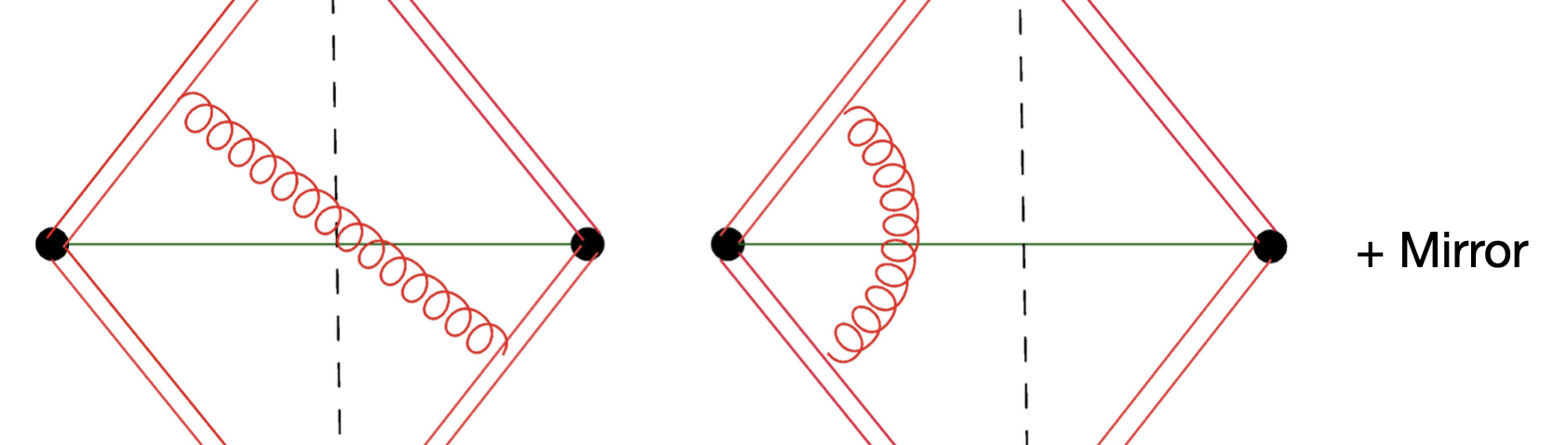}
\caption{One-loop diagrams for the single-subjet function in the vacuum. The straight red lines correspond to collinear soft Wilson lines along n and $\bar n $ directions.}
\label{vac2}
\end{figure}

The matching for the vacuum results is already known and appears in the factorization for threshold resummation~\cite{deFlorian:2007fv,Dai:2017dpc,Liu:2017pbb,Neill:2021std}. We reproduce it here in the context of jets in heavy-ion collisions.
At $\mathcal{O}({H}_{G,{\rm I}}^{\text{int}(0)})$, the function $\mathcal{S}_{1}^{(0)}$ is given by a correlator of collinear-soft Wilson lines in the vacuum
\begin{equation}
\mathcal{S}_{1}^{(0)}(\epsilon_L) =  \tr\Big[ \mathcal{T}\Big[U_{\bar n}(0)U_n(0)\Big]\mathcal{\bar T}\Big[U_n^{\dagger}(0)^{\dagger}_{\bar n}(0)\Big]\mathcal{M}'\Big].
\end{equation}
At the tree level, this evaluates to $\mathcal{S}_{1}^{(0)}(\epsilon_L)=\delta(\epsilon_L) $. Compared to the vacuum result of the stage I jet function, this gives the following result for the matching coefficient in Eq.~\eqref{eq:mdfactorization}
\bea 
\mathcal{C}_{q\rightarrow 1}(\omega_J',z', \mu) =  \omega_J'\delta(1-z').
\eea
The one-loop real and virtual vacuum diagrams are shown in Fig.~\ref{vac2}. The virtual diagrams are scaleless and hence vanish in dimensional regularization. Since the full result is IR finite, it is sufficient to consider the real emission diagrams. For a gluon emission inside the jet, we get
\begin{equation}
\mathcal{S}^{(0)}_{qg} =\delta(\epsilon_L) \frac{\alpha_s}{\pi}\frac{(\mu^2e^{\gamma_E})^{\epsilon}}{\Gamma[1-\epsilon]}\int_0^{\infty}\frac{dq^-}{q^-}\int \frac{d|\bfq|}{|\bfq|^{1+2\epsilon}}\Theta\left(\frac{q^- R}{2}-|\bfq|\right).   
\end{equation}
The integral over $q^-$ in scaleless, and therefore the contribution vanishes in dimensional regularization. For the gluon emission outside the jet we obtain
\begin{equation}
\mathcal{S}^{(0)}_{q} = \frac{\alpha_s}{\pi}\frac{(\mu^2e^{\gamma_E})^{\epsilon}}{\Gamma[1-\epsilon]}\frac{1}{\epsilon_L}\int \frac{d|\bfq|}{\bfq^{1+2\epsilon}}\Theta\left(|\bfq|-\frac{ \epsilon_L R}{2}\right),
\end{equation}
which evaluates to 
\begin{equation}
\mathcal{S}^{(0)}_{q}= \frac{\alpha_s\,C_F}{2\pi \omega_J'}\delta(1- \tilde z)\Big[-\frac{1}{\epsilon^2}-\frac{l}{\epsilon}-\frac{l^2}{2}+\frac{\pi^2}{12}\Big]+\frac{\alpha_s\,C_F}{2\pi \omega_J'}\Big[\left(\frac{1}{\epsilon}+l\right)\frac{2}{(1-\tilde z)_+}-4\left(\frac{\ln(1- \tilde z)}{1-\tilde z}\right)_+\Big], 
\end{equation}
where we have written $\epsilon_L= (1-\tilde z) \omega_J'$. 
Here, we introduced $l= \ln (\mu^2/(\omega_JR/2)^2)$, which leads to a double logarithm in Mellin space of form $\sim \ln^2 (\mu^2 N^2/(\omega_JR/2)^2)$. We can read off the natural scale for this function, which is given by $(1-\tilde z)\omega_JR/2$. 
Comparing the one-loop result with the Stage I vacuum jet function in Appendix \ref{app:JVOneLoop}, we obtain the $\mathcal{O}(\alpha_s)$ correction to the matching coefficient $\mathcal{C}_{q\to 1}$. For the anti-k$_T$ jet algorithm, this leads to 
\bea
\mathcal{C}_{q\to 1}(z',\omega_J',\mu) &=&  \omega_J'\Bigg[\delta(1-z')+ \delta(1-z')\frac{\alpha_s}{2\pi}C_F\Bigg[\frac{l^2}{2}+\frac{3l}{2}+\frac{13}{2}-\frac{3\pi^2}{4}\Bigg] \nonumber\\
&+& \frac{\alpha_s}{2\pi}C_F\Big[-(1+z')l-2(1+z')\ln(1-z')-(1-z')]\nonumber\\
&+& \frac{\alpha_s\, l}{2\pi}P_{gq}(z')-\frac{\alpha_s}{2\pi}\Big[P_{gq}(z)2 \ln(1-z')+C_Fz'\Big]\Bigg],
\eea
where $l= \ln (\mu^2/(\omega_JR/2)^2)$. Note that in Mellin transform space, this result has no logarithms in $N$ at leading power. Hence, the natural scale for this function is  $\omega_JR/2$.
The running of the single subjet function between the scales $(1-z)\omega_JR/2$ to $\omega_J/R$ will allow us to resum the leading threshold logarithms. Further, the RG equation for this function is given by
\bea
\mu \frac{d}{d\mu}\mathcal{C}_{i\to 1}(z, \omega_J,\mu) = \sum_j\int _z^1 \frac{dz'}{z'} \gamma^{ji}_{\mathcal{C}_{i\to 1}}\left(\frac{z}{z'},\mu\right)\mathcal{C}_{j\to 1}(z',\omega_J,\mu).
\eea
We can read off the anomalous dimensions
\bea
\gamma^{qq}_{\mathcal{C}_{q\to 1}}(z) &=& \delta(1-z) \frac{\alpha_s C_F}{2\pi}\Big[4 l +3\Big] -\frac{\alpha_sC_F}{\pi}(1+z),\nn\\
\gamma^{qg}_{\mathcal{C}_{q\to 1}}(z)&=&\frac{\alpha_sC_F}{\pi}P_{gq}(z).
\label{eq:CRG}
\eea
Correspondingly, the anomalous dimension for the single subjet function reads 
\bea
\gamma^{qq}_{\mathcal{S}_1^{(0)}}(z)= -\delta(1-z)\frac{4\alpha_sC_F}{2\pi}l +\frac{\alpha_sC_F}{2\pi}\frac{4}{(1-z)_+}\,.
\label{eq:VacRG}
\eea
We note that as anticipated the addition of above two anomalous dimensions
\bea
\gamma^{qq}_{\mathcal{C}_{q\to 1}}(z)+\gamma^{qq}_{\mathcal{S}_1^{(0)}}(z) = \frac{\alpha_s}{\pi}P_{qq}(z), 
\eea
is simply the DGLAP evolution kernel for a quark jet.

%%%%%%%%%%%%%%%%%%%%%%%%%%%%%%%%%%%%%%%%%%%%%%%%%%%%%%%%%%%%%%%%%%%%%%%%%%%%%%%%%%%%%%%%%%%%%%%%%%%%%%%%%%%%%%%%%%%%%%%%%%%%%%%%%%%%%%%%%%%%%%%%%%%%%%%%%%%%%%%%%%%%%%%%%%%
\subsubsection{Matching the single medium interaction}

The second term in Eq.~\eqref{eq:seriess} contributes at order $\mathcal{O}(H^{\rm int}_{G,{\rm I}})^2$ and describes medium-induced radiation with a single interaction between collinear, collinear-soft and soft medium partons. The factorization formula, following the same series of steps that lead to Eq.~\eqref{eq:realminusvirtual} derived in the Stage I EFT, is given by
\begin{equation}
\mathcal{S}^{(1)}_{1}(\epsilon_L,R)=|C_G|^2\, \int d \bar x^-\int \frac{d^2\bfk}{(2\pi)^3 } \left(\textbf{F}_{r,1}^{(1)}-\textbf{F}_{v,1}^{(1)}\right)(\epsilon_L, R;  \bfk, \bar x^-)\varphi(\bfk, \bar x^-).
\end{equation}
where we have again introduced a real and virtual part for the collinear-soft function $\textbf{F}_1^{(1)}$. The real part corresponding to Glauber insertions on opposite sides of the cut, is given by
\begin{align}
&\textbf{F}^{(1)}_{r,1}(\epsilon_L, \bfk ,\bar x^-) = \frac{1}{2N_c}\frac{e^{-i\frac{\bar x^-}{2}(\mathcal{P}^{a}_{+}-\mathcal{P}^{b}_{+})}}{\bfk^2}\nonumber\\ 
& \sum_{X}\tr\Big[\langle 0|\bar{\mathcal{T}}\Big\{e^{-i\int dt H_{n}(t)}\Big[\delta(\mathcal{P}^-)\delta^2(\mathcal{P}_{\perp}-\bfk)\mathcal{O}_{n}^{qa}(0)+ \frac{1}{({\mathcal{P}}_{\perp}-\bfk)^2}\mathcal{O}_{\cs}^{ca}(0)t^c\Big]U^{\dagger}(n)U^{\dagger}(\bar n) \Big\} \mathcal{M}'|X\rangle \nonumber\\
&\langle X|\mathcal{T}\Big\{e^{-i\int dt H_{n}(t)} \Big[\delta(\mathcal{P}^-)\delta^2(\mathcal{P}_{\perp}+\bfk)\mathcal{O}_{n}^{qb}(0)+ \frac{1}{({\mathcal{P}}_{\perp}-\bfk)^2}\mathcal{O}_{\cs}^{cb}(0)t^c\Big]U(n)U(\bar n)\Big\}|0\rangle \Big]\delta^{ab}  \,.
\end{align}
The function $\varphi$ is identical to the medium correlator defined in Eq.~\eqref{eq:Medcorr3}. Similarly, we can define $\textbf{F}_{v,1}^{(1)}$, which corresponds to the insertion of Glauber operators on the same side of the cut. Below, we compute this jet function to the one-loop order. There are two sources of the collinear-soft radiation -- the Lipatov vertex operator $ \mathcal{O}_{\cs}^{ba}$ in Eq.~\eqref{eq:csop} and the Wilson line $U(n)$.
\begin{figure}
\centering
\includegraphics[width=0.9\linewidth]{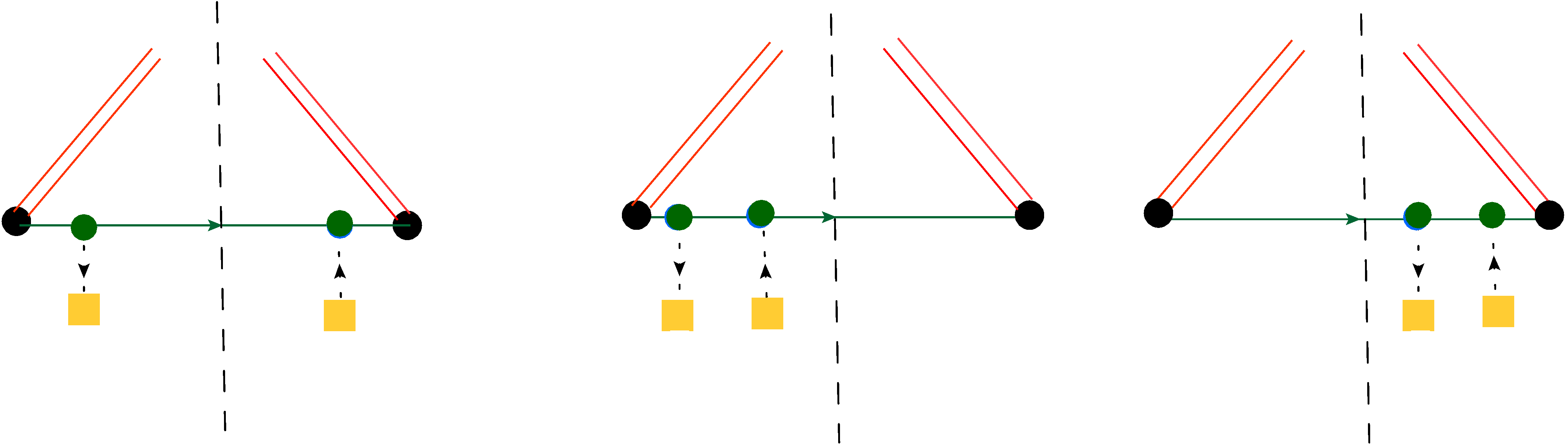}
\caption{Tree level diagrams for the single subjet operator. The green line represents the direction of the collinear quark, which has been integrated out, and the yellow squares are medium scattering centers. The green vertex is the Glauber operator $\mathcal{O}_{\cs}^{ba}$ insertion.~\label{LT}}
\end{figure}
The tree-level result is given by the interaction of the collinear mode with the medium which in our operator is just the leading order in g term of the operator $\mathcal{O}_{\cs}^{ba}$  and the corresponding diagrams are shown in Fig.~\ref{LT}\footnote{The $U(\bar n)$ Wilson line is not shown as it does not contribute to the interactions with the medium.}. Since the measurements on the jet are not affected by this interaction, the real and virtual contributions cancel each other and the diagrams add up to zero.

%%%%%%%%%%%%%%%%%%%%%%%%%%%%%%%%%%%%%%%%%%%%%%%%%%%%%%%%%%%%%%%

At one loop, there are four sets of diagrams. The relevant Feynman rules derived from the operator in Eq.~\eqref{eq:csop} for medium-induced collinear-soft radiation through the Lipatov vertex and the interaction of a collinear-soft gluon with the medium are given by
\begin{align}
\begin{minipage}{2cm}\includegraphics[width=\textwidth]{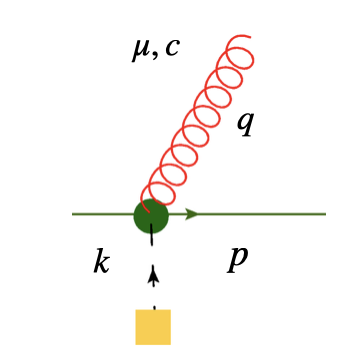} \end{minipage}=&8\pi \alpha_s igf^{abc}\Bigg(\!\!2\bfk^{\mu}-\bfq^{\mu}-\frac{\bar{n}^{\mu}}{2}n\cdot q+\bar{n}\cdot q \frac{n^{\mu}}{2}-n^{\mu}\frac{(\bfk-\bfq)^2}{n\cdot q}+\bar{n}^{\mu}\frac{\bfk^2}{\bar{n}\cdot q}\!\Bigg) \\
\begin{minipage}{2cm}\includegraphics[width=\textwidth]{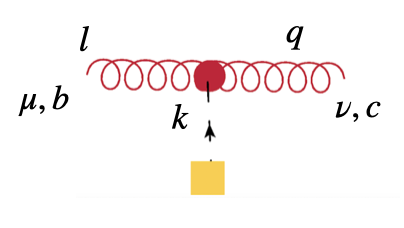} \end{minipage}=&-8\pi \alpha_s if^{abc}\Big(\bar{n}\cdot l\, g_{\perp}^{\mu \nu}-\bar{n}^{\mu}\bfl^{\nu}-\bar{n}^{\nu}\bfq^{\mu}+\frac{\bfq\cdot\bfl\, \bar{n}^{\mu}\bar{n}^{\nu}}{\bar{n}\cdot q} \Big). \,
\end{align}

\begin{figure}
\centering
\includegraphics[width=0.75\linewidth]{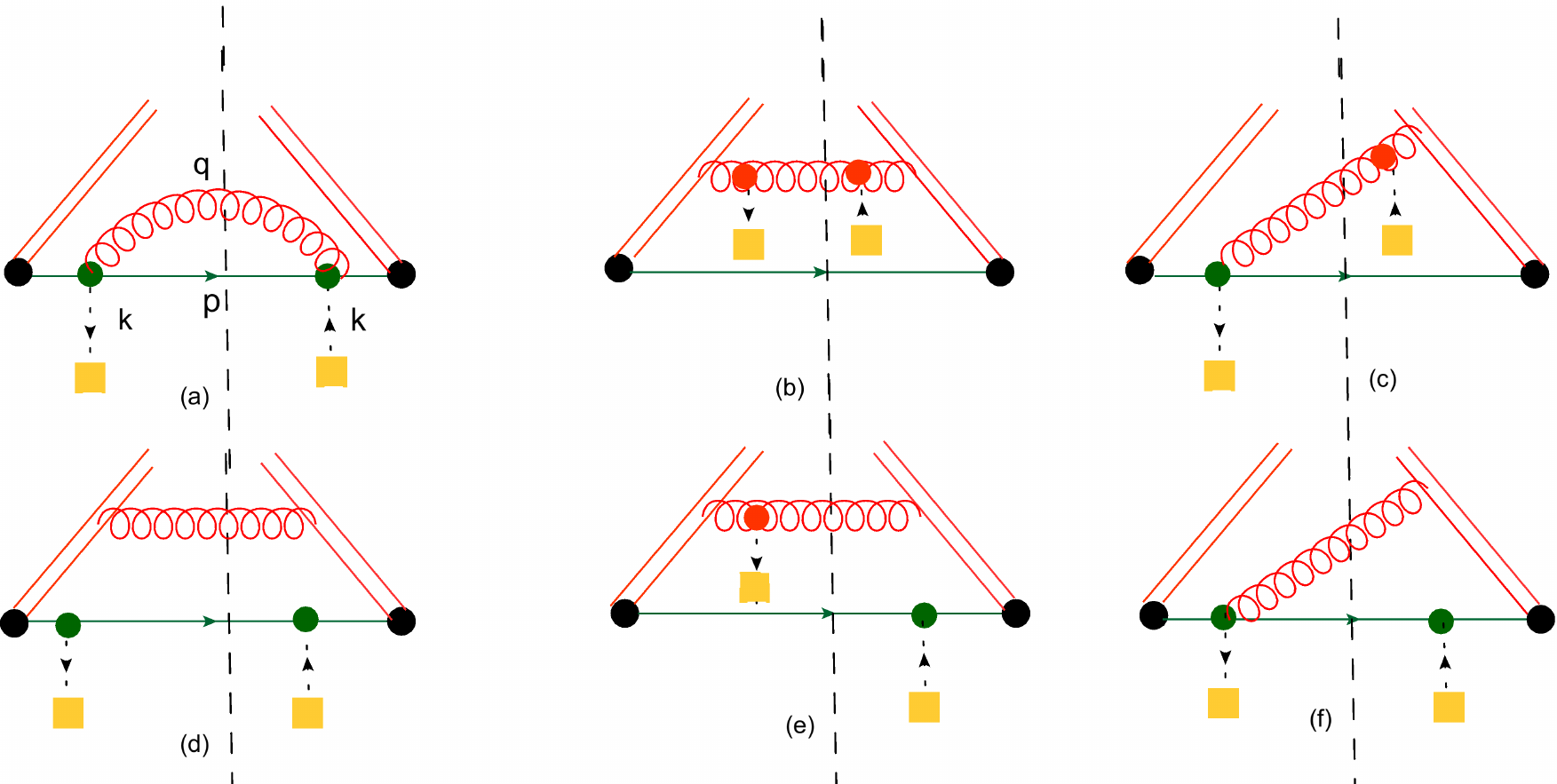}
\caption{Real emission diagrams at the one-loop level with Glauber insertions on opposite sides of the cut. The parallel red lines represent $U_{n}$ Wilson lines. The red vertex corresponds to the rescattering of cs gluon off the medium.}
\label{Loop2RR}
\end{figure}

We first consider the real emission diagrams with Glauber insertions on opposite sides of the cut as shown in Figure~\ref{Loop2RR}. First, we compute the one-loop contribution from the Lipatov vertex squared shown in Figure~\ref{Loop2RR}(a). This corresponds to medium-induced collinear-soft radiation. 
At one loop, Figure~\ref{Loop2RR}(a) results in the following expression
To carry out the matching to Stage I, we can conveniently divide the diagrams into two parts: First those that survive in the limit $L \rightarrow \infty$, i.e., do not contribute to any interference between the vacuum jet production and medium evolution. Second, diagrams that explicitly contribute to the interference and hence are proportional to the $F(x) = \cos[x] $ factor obtained in stage I Eq.~\ref{eq:verdiag}. \begin{align}
7a&=\frac{ g^2N_c(N_c^2-1)}{\bfk^2}\delta(\epsilon_L)\int \frac{d^4q \,\delta(q^2)}{(2\pi)^3(\bfq-\bfk)^4}\Theta_{\text{alg}} \nn\\
&\times  \Big[ (\bfk^{\mu}-\bfq^{\mu})+\bfk^{\mu}-\frac{ \bar n^{\mu}}{2} n \cdot q +\bar{n} \cdot q \frac{n^{\mu}}{2}-n^{\mu}\frac{(\bfk-\bfq)^2}{n \cdot q}+\bar n^{\mu} \frac{\bfk^2}{\bar n \cdot q}\Big]^2.  
\end{align}
where $\Theta_{\text{alg}} = \Theta\left( \frac{R}{2} -\frac{|\bfq|}{q^-}\right)$ for gluon emission inside the jet and $\Theta\left( \frac{|\bfq|}{q^-}-\frac{R}{2} \right)$ for cs radiation outside the jet. With some algebra, we can simplify this expression and we obtain
\begin{equation}
7a= \frac{g^2(N_c^2-1)}{2\bfk^2}\delta(\epsilon_L)\int \frac{d^2\bfq}{(2\pi)^3}\int\frac{dq^-}{q^-}\, \frac{\bfk^2}{\bfq^2 (\bfq-\bfk)^2}\Theta_{\text{alg}}    \,,
\end{equation}
which reproduces the square of the Lipatov vertex. Next, we consider the diagrams where the collinear-soft mode is sourced by $U(n)$ Wilson line which leads to 
\begin{equation}
7b = \frac{g^2(N_c^2-1)}{\bfk^2}\delta(\epsilon_L)\int \frac{d^2\bfq}{(2\pi)^3}\int\frac{dq^-}{q^-}\frac{1}{(\bfq-\bfk)^2}\Theta_{\text{alg}}. \label{eq:beforesubtract}
\end{equation}
In all these diagrams, we also need to be careful to avoid double counting arising from the Glauber limit of the collinear-soft modes. Therefore, we need to subtract the Glauber limit of the collinear-soft loop integrals from the $U(n)$ Wilson line diagrams. This is necessary for the three-sector EFT due to possible overlap between the collinear-soft and Glauber modes\footnote{The Glauber limit of diagrams not involving collinear-soft Wilson lines evaluates to zero.}. The possible overlaps and their zero-bin subtractions for the EFT involving all three modes have been discussed in Ref.~\cite{Rothstein:2016bsq}. The zero-bin subtraction here ensures that we only get the contribution from the diagrams where the collinear-emission emission goes on shell before it interacts with the medium so that it is genuinely sourced by the hard interaction that creates the jet and not from medium-induced radiation. Therefore, it removes the overlap with the Lipatov vertex in 8(a), which accounts for all the medium-induced radiation. As a result, after the proper zero-bin subtraction, the contribution in Eq.~(\ref{eq:beforesubtract}) reduces to half its value. Therefore, we can write the Glauber-subtracted result as  
\begin{equation}
7b_{\rm GS} = \frac{g^2(N_c^2-1)}{2\bfk^2}\delta(\epsilon_L)\int \frac{d^2\bfq}{(2\pi)^3}\int \frac{dq^-}{q^-}\frac{1}{(\bfq-\bfk)^2}\Theta_{\text{alg}}. \end{equation}
Further, we note that diagram 7(d) evaluates to zero due to Wilson line vertex contractions. Next, we consider the contributions from diagrams 7(c),  7(e), and 7(f) along with their complex conjugate. 
The contribution of diagram 7(c) proportional to $\bfk\cdot\bfq/(\bfq^2(\bfq -\bfk)^2)F[\bar x^-(\bfq-\bfk)^2/(2q^-)]$.
The diagram 7(e) is proportional to $F[\bar x^-(\bfl^2-\bfq^2)/2q^-]$ where $\bfl=\bfq-\bfk$ and gets canceled against the same term arising from diagram 8(d). The diagram 7(f) cancels out with diagrams 8(e) and 8(f) and reads as $F[\bar x^-\bfq^2/(2q^-)](-1+2\bfk^2/\bfq^2)/\bfl^2$. After performing the Glauber zero-bin subtraction, therefore only 7(c) along with the complex conjugate contributes and leads to the result 
\bea
 7c_{\text{GS}} +c.c.= \frac{\alpha_s(N_c^2-1)}{4\pi^2 \bfk^2}\, \int d^2\bfq\int\frac{dq^- }{q^-} \frac{2\bfk \cdot \bfq}{\bfq^2(\bfq-\bfk)^2}\delta(\epsilon_L)\Theta_{\text{alg}}F\Big[\frac{(\bfq-\bfk)^2 \bar x^-}{2q^-}\Big].
\eea
This term is the interference between the hard vertex and medium-induced emission and vanishes in the limit $L \rightarrow \infty$. 
\begin{figure}
\centering
\includegraphics[width=0.75\linewidth]{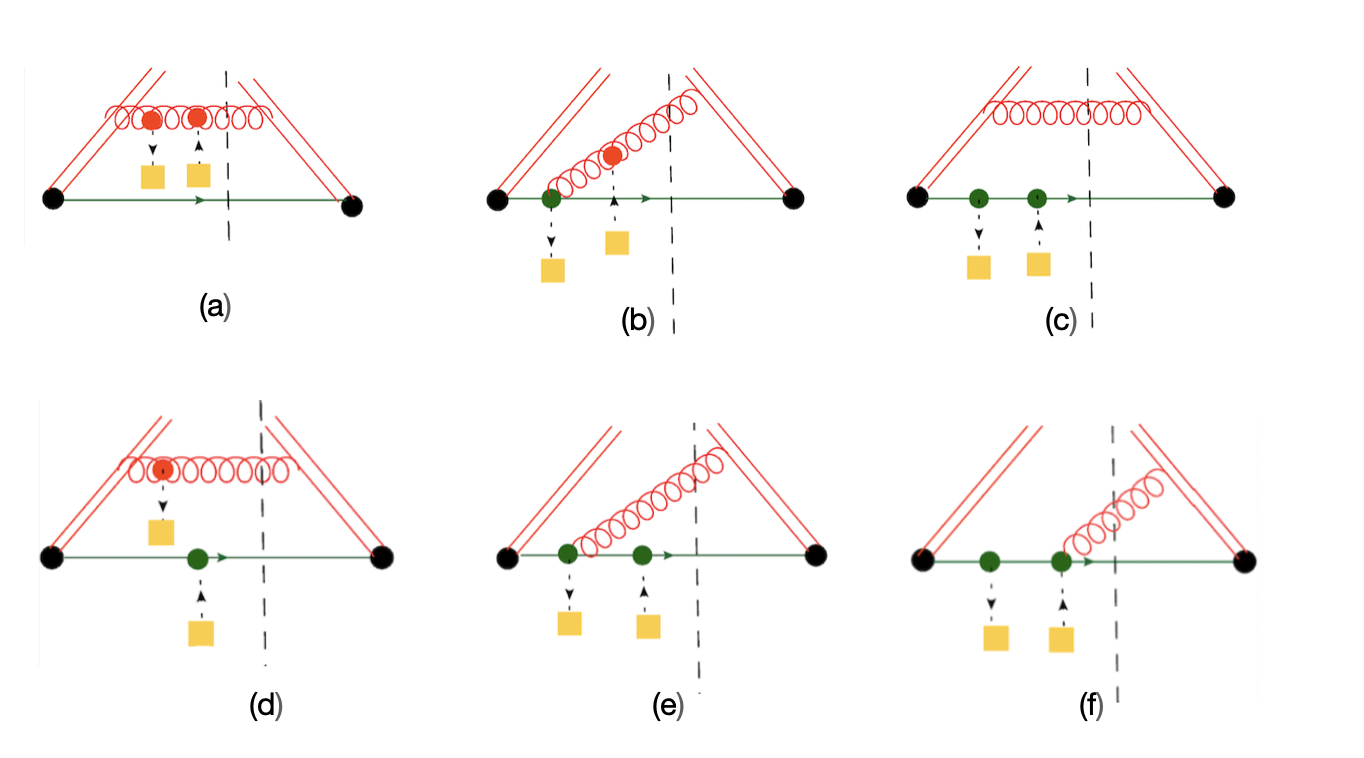}
\caption{One-loop diagrams for real emission and Glauber insertions on the same side of the cut.~\label{Loop2VR}}
\end{figure} 
Next, we consider the diagrams with Glauber insertions on the same side of the cut for a real gluon emission. The relevant diagrams are shown in Figure~\ref{Loop2VR}. The contribution from diagram 8(a) is
\begin{equation}
8a = \frac{g^2(N_c^2-1)}
{2\bfk^2}\delta(\epsilon_L)\int\frac{dq^-}{q^-}\int \frac{d^2\bfq}{(2\pi)^3}\frac{1}{\bfq^2}\Theta_{\text{alg}}  \,,
\end{equation}
which is another contribution to the broadening term. We find that diagrams 8(b) through 8(f) only give the interference term proportional to function $F(\bar x^-/t_f)$. The contribution from 8(c) vanishes due to Wilson line contractions. As mentioned earlier, 8(e) and 8(f) get canceled with 7(f) while 8(d) cancels out with 7(e). The contribution from figure 8(b) goes as $2F[\bar x^- \bfq^2/(2q^-)](\bfq\cdot\bfk+\bfk^2)/\bfl^2$. 
\begin{figure}
\centering
\includegraphics[width=0.75\linewidth]{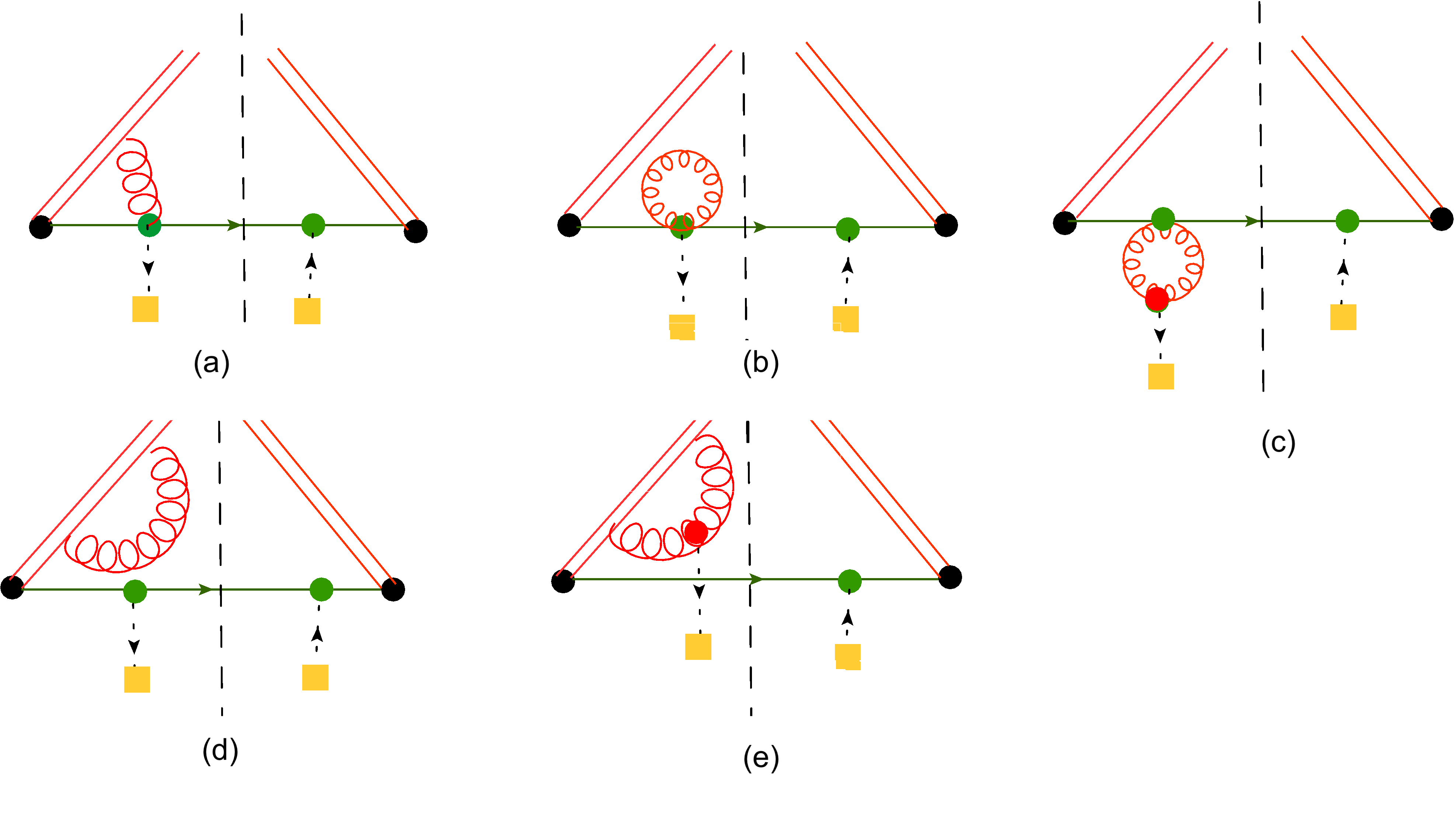}
\caption{Virtual emission diagrams at the one-loop level with Glauber insertions on opposite sides of the cut.~\label{Loop2RV}}
\end{figure} 
Next, we consider the diagrams with Glauber insertions on the opposite sides of the cut with a virtual gluon emission. The relevant diagrams are shown in Figure~\ref{Loop2RV}. After the Glauber zero bin subtraction, diagram 9(a) gives a result proportional to $F$. Further, diagram 9(d) evaluates to zero. All other diagrams cancel out against the corresponding virtual gluon diagrams with Glauber insertions on the same side of the cut. Diagram 9(a) gets canceled with diagrams 10(b) and 10(c), which are virtual contributions with Glauber insertions on the same side of the cut. Diagram 10(h) generates the same term as Diagram 7(c) and diagram 9(c) cancels against diagrams 10(f) and 10(g). The contribution of diagram 9(e) is $F[\bar x^-(\bfl^2-\bfq^2)/2q^-]\bfq\cdot\bfl/\bfl^2\bfq^2$.

%%%%%%%%%%%%%%%%%%%%%%%%%%%%%%%%%%%%%%%%%%%%%%%%%%%%%%%%%%%%%

\begin{figure}
\centering
\includegraphics[width=0.75\linewidth]{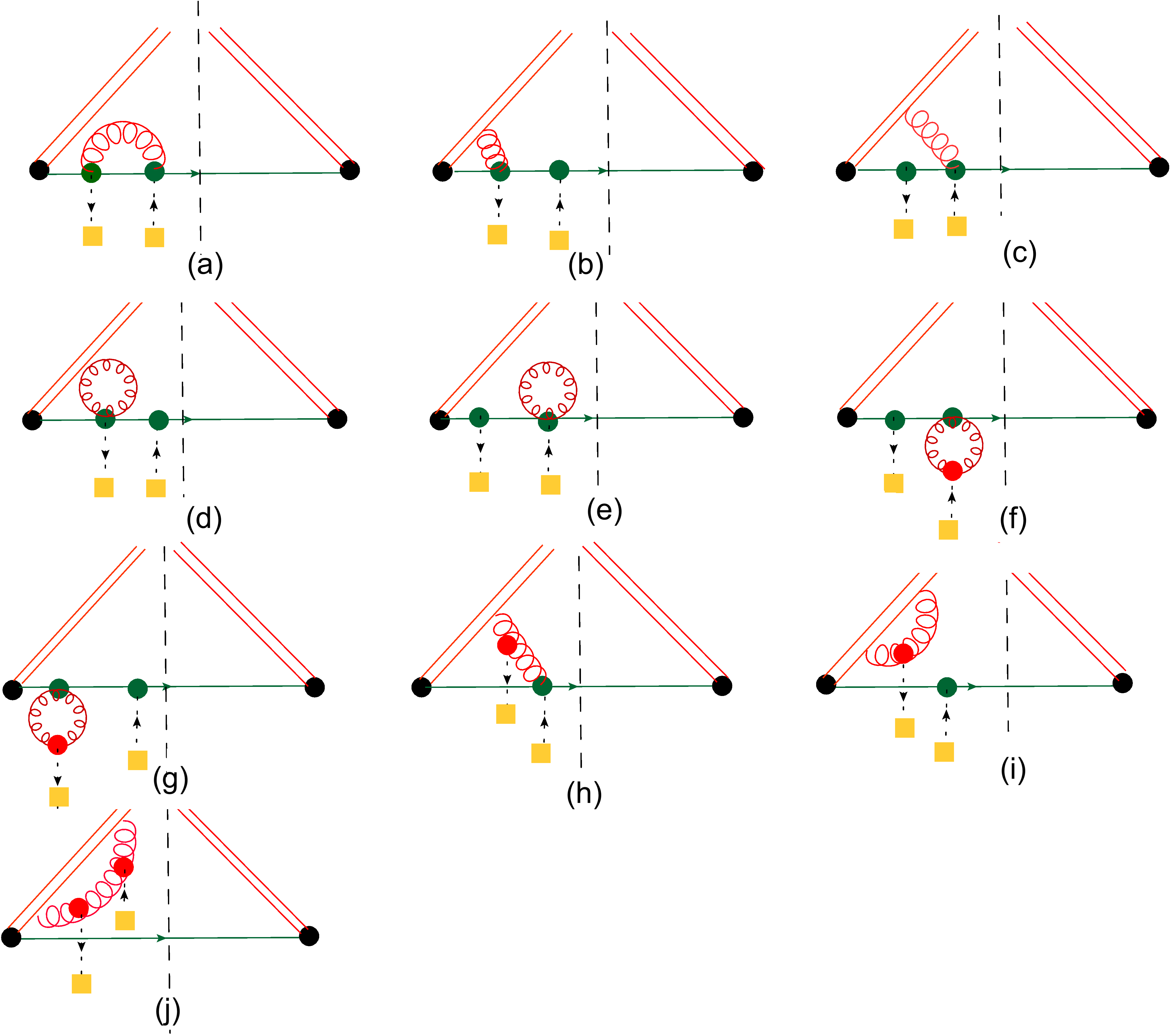}
\caption{Virtual correction diagrams at the one-loop level with Glauber insertions on the same side of the cut.~\label{Loop2VV}}
\end{figure} 
Lastly, we consider the diagrams with Glauber insertions on the same side of the cut and with a virtual gluon emission. The relevant diagrams for this case are shown in Fig.~\ref{Loop2VV}. Diagram 10(a) is again the square of the Lipatov vertex, hence, generates the contribution 
\begin{equation}
10a= \frac{g^2(N_c^2-1)}{2\bfk^2}\delta(\epsilon_L) \int\frac{dq^-}{q^-}\int \frac{d^2\bfq}{(2\pi)^3} \frac{\bfk^2}{\bfq^2 (\bfq-\bfk)^2}   \,.
\end{equation}
It correctly reproduces the virtual contribution that we had in the previous section. Further, diagrams b, c, h, i, and j  give terms proportional to function $F(L/t_f)$. Diagrams 10(d) through 10(g) cancel with the corresponding real diagrams. Contribution from diagram 10(h) is proportional to $F[\bar x^-\bfq^2/(2q^-)]\bfq\cdot\bfk/\bfq^4$ and from 10(i) $F[\bar x^-(\bfl^2-\bfq^2)/(2q^-)]\bfq\cdot\bfl/\bfl^2\bfq^2$.
Combining the contribution from all the diagrams we have 
\bea 
{\bf F}_{1}^{(1)} &=& \frac{\alpha_s(N_c^2-1)}{4\pi^2 \bfk^2}\, \int d^2\bfq\int\frac{dq^- }{q^-} \frac{2\bfk \cdot \bfq}{\bfq^2(\bfq+\bfk)^2}\Theta\left( \frac{2|\bfq|}{q^-}-R\right)\left( \delta(\epsilon_L) -\delta(\epsilon_L-q^-)\right)\nn\\
&\qquad\qquad\, \times & \left(1- F\Big[\frac{(\bfq+\bfk)^2\bar x^-}{2q^-}\Big]\right)\,.
\eea
We therefore find that the single interaction contribution from a single subjet completely reproduces the one-loop medium-induced contribution discussed in Eq.~\ref{eq:StageInloop} and Eq.~\ref{eq:StageOloop}. This is also consistent with the tree level matching coefficient $\mathcal{C}_{q\rightarrow 1} = \omega_J'\delta(1-z')$. 

%%%%%%%%%%%%%%%%%%%%%%%%%%%%%%%%%%%%%%%%%%%%%%%%%%%%%%%%%%%%%%%%%%%%%%%%%%%%%%%%%%%%

\section{Jet production in a dilute medium}
\label{sec:Dilute}

\begin{figure}
\centering
\includegraphics[width=0.75\linewidth]{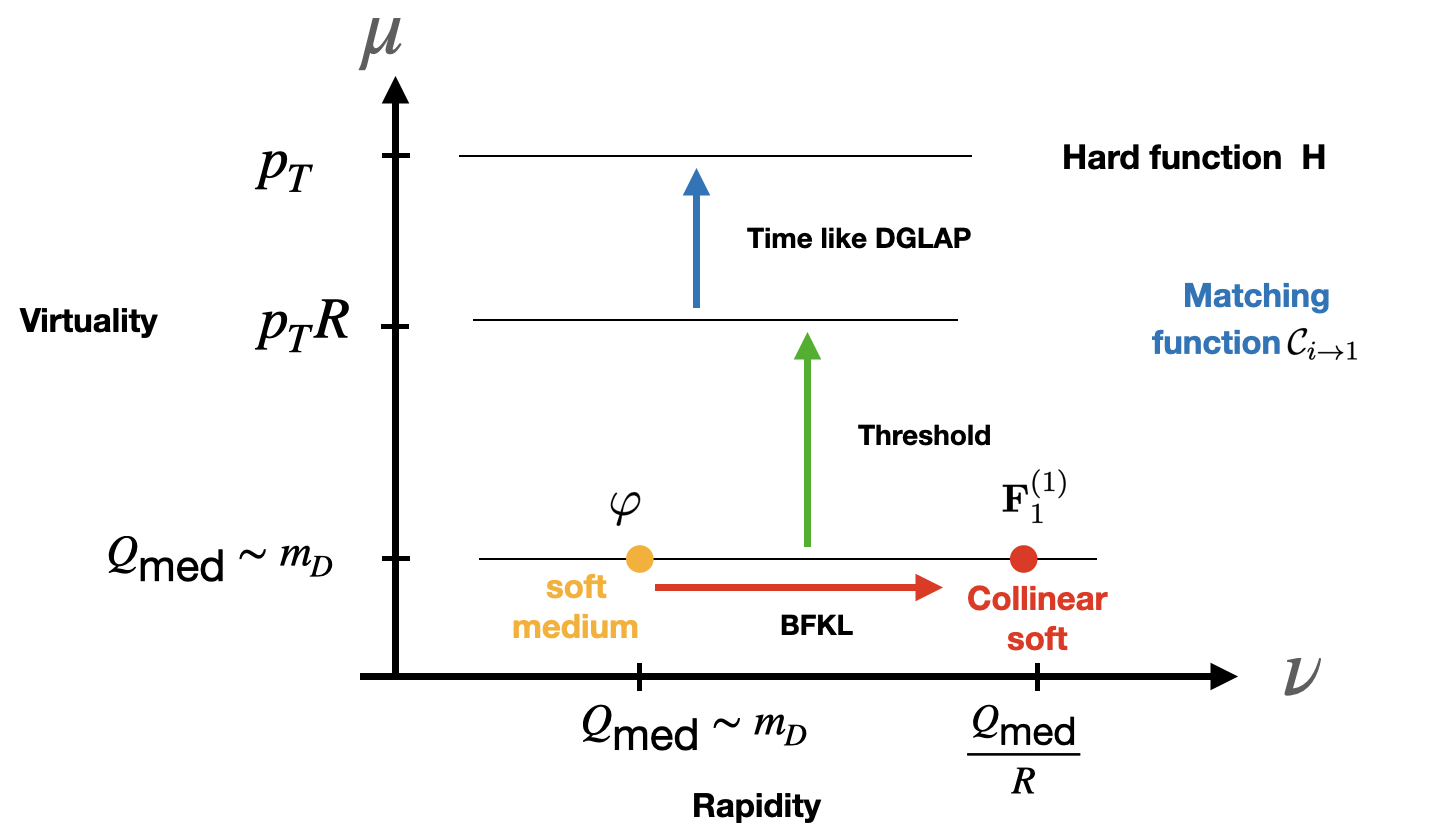}
\caption{The renormalization group flow for jet production in a dilute medium.~\label{RGD} }
\end{figure} 

In a dilute medium, we expect the interaction cross-section of the jet with the medium to be small.
Consequently, it suffices to keep terms only up to a single interaction with the medium, which is captured by the function ${\cal S}^{(1)}_1$. The complete factorization formula for a dilute medium for the case of an unresolved jet reads as
\bea 
\frac{d\sigma}{dp_T d\eta }&=&\int \frac{dz}{z}H_q(\omega;\mu)\int dz' \int d\epsilon_L\delta(\omega_J'-\omega_J-\epsilon_L)\,\mathcal{C}_{i\rightarrow 1}(\omega_J',z'; \mu){\cal S}_1(\epsilon_L, R; \mu),\nonumber \\
& \stackrel{\mbox{\tiny dilute}}{=} & \int \frac{dz}{z}H_q(\omega;\mu)\int dz' \int d\epsilon_L\delta(\omega_J'-\omega_J-\epsilon_L)\,\mathcal{C}_{i\rightarrow 1}(\omega_J',z'; \mu)\nonumber \\
&&\hspace{-1cm}\times \left(S_1^{(0)}(\epsilon_L, R;\mu)+ |C_G|^2 \int d\bar x^-\int \frac{d^2\bfk}{(2\pi)^2} \varphi(\bfk, \bar x^-; \mu,\nu)\,\textbf{F}_1^{(1)}(\epsilon_L,R,\bfk, \bar x^-;\mu,\nu)\right).
\eea
In this formula, we note that the hard function sits at the virtuality $\omega \sim p_T$, the matching coefficient at $p_TR$, and the collinear-soft and medium physics is at virtuality $\epsilon_LR$. This scale in turn is set by the transverse momentum scale $\bfk$ exchanged between the jet partons and the medium. For the single interaction considered here, this scale is set by $m_D \sim gT$. In a strongly coupled medium, this is a non-perturbative scale. Hence, for a \textit{dilute} medium, the entire object ${\cal S}_1(\epsilon_L, R; \mu)$ is non-perturbative. We therefore draw an important conclusion here that the non-perturbative physics, while independent of the jet $p_T$, depends on the jet radius $R$ and the path length $L$ in the medium. 

The analysis of the functions that appear in this formula also allows us to present a complete picture of the possible resummation of large logarithms that arise as a result of the factorization shown in Fig. \ref{RGD}. We have already established that the RG equation for the vacuum function ${\cal S}_1^{(0)}$ corresponds to threshold resummation given Eq.~\ref{eq:VacRG} with the corresponding RG for the matching coefficient in Eq.~\ref{eq:CRG}. Likewise, the medium correlator $\varphi$ obeys the BFKL equation in the rapidity cutoff $\nu$. By RG invariance of the cross section, we can then infer that the collinear-soft function $\textbf{F}_1^{(1)}$ obeys an RG equation in both $\mu$ and $\nu$, which are given by 
\begin{align}
\nu\frac{d}{d\nu}\textbf{F}_1^{(1)}(\epsilon_L, \bfk;\mu,\nu )= -\int d^2\bfu\,\mathcal{K}_{\text{BFKL}}(\bfk, \bfu)\textbf{F}_1^{(1)}(\epsilon_L, \bfu;\mu,\nu)\,,  
\label{eq:RGF1Nu}
\end{align}
and 
\bea
\mu \frac{d}{d\mu}\textbf{F}_1^{(1)}(z,\omega_J,\bfk;\mu,\nu )= \sum_j\int _z^1 \frac{dz'}{z'} \gamma_{\textbf{F}_1^{(1)}}\left(\frac{z}{z'},\mu\right)\textbf{F}_1^{(1)}(z',\omega_J, \bfk;\mu, \nu)\,.
\eea
Here the anomalous dimension is
\bea
\label{eq:RGF1Mu}
\gamma_{\textbf{F}_1^{(1)}}(z)= -\delta(1-z)\frac{4\alpha_sC_F}{2\pi}l +\frac{\alpha_sC_F}{2\pi}\frac{4}{(1-z)_+}\,,
\eea
where we have set $\epsilon_L= \omega_J'(1-z)$ and suppressed other arguments for simplicity. To verify these relations explicitly, we need to do an explicit two-loop computation for $\textbf{F}_1^{(1)}$ since the tree-level result vanishes. 
This also demonstrates the power of the factorization framework, which allows us to infer higher-order radiative corrections without explicit calculations.
%%%%%%%%%%%%%%%%%%%%%%%%%%%%%%%%%%%%%%%%%%%%%%%%%%%%%%%%%%%%%%%%%

\section{Factorization in a dense medium~\label{sec:Multi}}

So far, our discussion has been restricted to the case of a single interaction with the medium. In this section, we consider the factorization for an arbitrary number of interactions with the medium. As mentioned above, in this paper, we limit ourselves to the case $R\leq\theta_c $, i.e., a single subjet now undergoes multiple interactions with the medium. We begin again with the all-order definition of the subjet function ${\cal S}_1$, see Eq.~(\ref{eq:1subjet}),
\begin{align}
& \mathcal{S}_{1}(\epsilon_L,R) = \tr\Big[e^{-iH't} \mathcal{T}\Big[e^{-i\int dt H^{\text{int}}_{G,{\rm I}}}U_{n}U_{\bar n }\Big] \rho_M \mathcal{\bar T}\Big[e^{-i\int dt H^{\text{int}}_{G,{\rm I}}}U_{\bar n}^{\dagger}U_{n}^{\dagger}\Big]e^{i H't}\mathcal{M}'\Big],
\end{align}
which, as described earlier, can be expanded in terms of the number of interactions with the medium $ \mathcal{S}_{1} = \sum_{i=0}^{\infty}\mathcal{S}_{1}^{(i)} $. 
In the previous section, we derived the factorization for the first two terms in this series, which suffices for a dilute medium.
We can generalize the factorization to an arbitrary number of interactions $n$ as follows. The only assumption we make to derive the corresponding factorization is that successive interactions of the jet with the medium occur via independent scattering centers in the medium, i.e. the soft medium partons that the jet partons scatter off are color uncorrelated. This is a reasonable approximation if the average distance between the successive interactions of the jet with the medium, i.e. the mean free path $\ell_{\rm mfp}$, is larger than the inverse Debye screening mass $\ell_{\rm mfp} \gg 1/m_D$. This allows for multiple interactions since the length of the medium $L$ is much larger than the screening length scale. The scale $\ell_{\rm mfp}$ is itself an emergent scale. We leave a detailed derivation of this scale within the EFT framework for future work. At ${\cal O}(n)$, we can write the factorization formula for the subjet function as 
\begin{align}
\mathcal{S}_{1}^{(n)}(\epsilon_L,R) &= |C_{G}|^{2n}\Bigg[\prod_{i=1}^{n}\int dx^-_i\Theta(x^-_i-x_{i+1}^-)\int \frac{d^2\bfk_{i}}{(2\pi)^3}\varphi(\bfk_i, x^-_i, \mu,\nu)\Bigg]\nn\\
& \times \textbf{F}_1^{(n)}(\epsilon_L,R,x^-_1 \ldots ,x^-_n; \bfk_1,\bfk_2, \ldots, \bfk_n;\mu, \nu ). 
\end{align}
Here, we have $n$ copies of the medium correlator $\varphi$ and a single collinear-soft jet function  $\textbf{F}_1^{(n)}$ indicating that the interaction of the jet with the medium is fully coherent. The successive interactions of the jet with the medium are path-ordered along the direction of the jet propagation $x^-$. We will not write out the full operator definition for $\textbf{F}_1^{(n)}$ here since it is cumbersome but straightforward. It involves $2n$ insertion of the collinear sector Glauber operator with all possible permutations of these operators on each side of the cut. The function $\varphi$ obeys RG equations in $\mu$  and $\nu$ as defined in Eq.~(\ref{eq:BRG}). Using RG consistency, we can generalize the RG equation for $\textbf{F}_1^{(1)}$ in  Eqs.~\eqref{eq:RGF1Mu} and~\eqref{eq:RGF1Nu}. We find that the function $\textbf{F}_1^{(n)}$ has to obey the following equation
\begin{eqnarray}
\nu\frac{d}{d\nu}\textbf{F}_1^{(n)}(\epsilon_L, \bfk_1, \ldots, \bfk_n;\mu,\nu )= -\sum_{i=1}^n\int d^2\bfu\,\mathcal{K}_{\text{BFKL}}(\bfu,\bfk_{i})\textbf{F}_1^{(n)}(\epsilon_L, \bfk_{1}, \ldots,\bfk_{i-1},\bfu, \ldots ,\bfk_{n};\mu,\nu ) \nonumber  \\
\end{eqnarray}
where we have suppressed the dependence of the jet function on $R$ and $L$. Given the anomalous dimension of the matching coefficient $\mathcal{C}_{q \rightarrow 1}$ and RG consistency, we obtain 
\bea
\mu \frac{d}{d\mu}\textbf{F}_1^{(n)}(\bfk_1, \ldots, \bfk_n;\mu,\nu )= \sum_j\int _z^1 \frac{dz'}{z'} \gamma_{\textbf{F}_1^{(n)}}\left(\frac{z}{z'},\mu\right)\textbf{F}_1^{(n)}(z',\omega_J,\mu),
\eea
with 
\bea
\gamma_{\textbf{F}_1^{(n)}}(z)= -\delta(1-z)\frac{4\alpha_sC_F}{2\pi}l +\frac{\alpha_sC_F}{2\pi}\frac{4}{(1-z)_+}.
\eea
Ostensibly, the RG structure for the dense medium is similar to the dilute case with the only change being the multiple copies of the BFKL equation. Naively, the virtuality of the subjet function ${\cal S}_1$ is determined entirely by the scale $m_D$, which is the typical transverse momentum exchanged in a \textit{single} interaction. However, in a dense medium, we have to contend with a new scale, namely the \text{total} typical transverse momentum exchange between a jet parton with the medium. We have referred to this scale earlier as $Q_{\text{med}}$ but have been glib about its exact definition. In a dense medium, one may expect that this scale can be much larger than $m_D$ thereby dynamically introducing a new scale in our EFT hierarchy. A rough estimate of this scale may be obtained by comparing the leading order computation of multiple scatterings in a medium with data. In the literature, this scale is typically been written as $Q_{\text{med}} \sim \sqrt{ \hat q L}$ and leading order and it is in the range of $1-3$~GeV. 

The first important consequence of this is that the energy of the collinear-soft mode scales as $\epsilon_L \sim Q_{\text{med}}/R$.  Hence, the energy loss is now larger compared to the dilute case. If we choose the extreme value in this range, we obtain a scale separation between the three scales $p_TR \gg Q_{\text{med}} \gg m_D \sim T$. This would therefore require a further matching step where we expand out the result $m_D/Q_{\text{med}}$. A rigorous formulation of this matching is crucial to completely isolate the non-perturbative physics and establish its universality or lack thereof across distinct jet observables. This is beyond the scope of the current paper and will be addressed in the future.

%%%%%%%%%%%%%%%%%%%%%%%%%%%%%%%%%%%%%%%%%%%%%%%%%%%%%%%%%%%%%

\section{The $R \rightarrow 0$ limit}

The limit $R \rightarrow 0$ corresponds to the case where each jet consists of a single hadron, i.e. we recover the inclusive hadron production cross section in this limit~\cite{Aversa:1988vb}. In the following, we discuss how the medium EFT framework changes in this limit. The limit $R \rightarrow 0$ corresponds to $p_TR \rightarrow \Lambda_{\rm QCD}$. In the case of a strongly coupled medium, the following three scales are parametrically of the same order $p_TR \sim T \sim m_D$. In this limit, the two curves shown in Figure~\ref{Mm}, are no longer distinct but they merge into a single hyperbola. A further consequence of this scaling is that we no longer have distinct collinear and collinear-soft modes but only a single hard-collinear mode with momentum scaling
\bea 
p_{\rm hc} = p_T(1, R^2, R).
\eea
Further, the medium modes only consist of a soft mode with scaling
\bea
p_{\rm s} =(T, T, T) \sim p_T(R, R, R).
\eea
This is identical to the case described in section~\ref{sec:StageI} and was thoroughly discussed in the context of the transverse momentum imbalance between narrow jets in Refs.~\cite{Vaidya:2020lih, Vaidya:2021mly, Vaidya:2021vxu}.
In this case, the medium effects contribute to leading power in $R$, i.e. they are not suppressed compared to the vacuum. However, the formation time of the medium-induced radiation $t_f \sim 1/(p_TR^2) \sim p_T/\Lambda_{QCD}^2$ may become very large compared to the typical medium size $L$ in this regime. Therefore, there may be strong suppression of medium-induced radiation due to the LPM effect as was noted in \cite{Vaidya:2021vxu}\footnote{In principle, the limit $t_f \gg L$ would induce large logarithms in $t_f/L$. To resum these logarithms, we would need to introduce a collinear-soft mode with the scaling $p_{cs} \sim ( m_D^2L , 1/L, m_D)$. The EFT structure in that case is an interesting open problem.}. The final factorized result is then identical to the one in Eqs.~\eqref{eq:factI} and \eqref{eq:Sigma} in terms of a jet function evaluated in the medium background. We can do a further factorization written as a series in the number of jet-medium interactions as in Eq.~\eqref{eq:Jexp}. Each term in the series can be written in terms of multiple copies of the universal medium correlator and a medium jet function, which obeys both the DGLAP and BFKL evolution equations. For instance, the first two terms correspond to the vacuum evolution (section \ref{sec:vacI}) and single jet-medium interaction (section \ref{sec:JMedI}), respectively. 

Based on our analysis in the previous section, we conclude that in a dilute medium, the full jet function (now with a specific final state hadron $h$ instead of a jet) defined in Eq.~\eqref{eq:fact} sits at a virtuality of $m_D \sim \Lambda_{\rm QCD}$, i.e. it is non-perturbative. It depends on the hadron $p_T$ and the path length in the medium $L$. It denotes the probability of creating a hadron from a quark in a dilute medium of length $L$ and can therefore be interpreted as the hadron fragmentation function in the presence of the medium. In a dense medium, we would again expect the emergence of a perturbative scale $Q_{\text{med}}$, which would require further steps of matching. We leave a more detailed exploration of this case for future work.

%%%%%%%%%%%%%%%%%%%%%%%%%%%%%%%%%%%%%%%%%%%%%%%%%%%%%%%%%%%%%%%

\section{Discussion and outlook ~\label{sec:JetNature}} 

In this paper, we have taken the first step toward developing a comprehensive Effective Field Theory (EFT) framework for jet propagation in heavy-ion collisions. Based on this EFT, we show here a way to separate the physics at distinct scales \textit{at all orders in perturbation theory} in terms of gauge invariant operators. This crucially allows us to cleanly separate the perturbative from the non-perturbative physics in a systematically improvable manner. With the factorization formula, improving the accuracy requires computing to higher orders in perturbation theory in the factorized functions.
We presented the EFT for the most basic observable - the inclusive production cross section of high-energy jets with (parametrically) small jet radius $R$. The factorization formula presented in this paper is valid to leading power in two expansion parameters, the jet radius $R$ and $m_D/(p_TR)$. 

The open quantum system EFT framework developed in this paper allows us to make several important statements about the nature of jet propagation in a strongly coupled medium. These statements are valid within the domain of the hierarchy of scales described above
\begin{itemize}
\item The vacuum evolution of the jet above the scale $p_TR$ can be explicitly factorized from the evolution below this scale. This vacuum-like evolution above the jet scale follows DGLAP evolution, allowing us to resum the single logarithmic series in terms of the jet radius $\ln R$ similar to the vacuum case. 
\item Jet evolution below the scale $p_TR$ is governed by the relative hierarchy between the jet radius $R$ and the coherence angle $\theta_c$ in the medium. For $R \leq \theta_c$, the jet interacts with the medium coherently and behaves as a single source of energy loss. 
\item For $R > \theta_c$, the medium can resolve several subjets inside the jet, each of which interacts again coherently with the medium. The vacuum splittings at pairwise angles $\geq \theta_c$ act as the seeds around which each subjet evolves in the medium.
\item The dominant interaction of each subjet with the medium occurs via a collinear-soft mode, leading to jet energy loss. This mode is sourced either by vacuum splittings or by medium-induced effects. These modes span the radius of the jet and, hence, can lead to interactions and interference between subjets.
\item The factorization below the jet scale $p_TR$ therefore takes the form of a series expansion in the number of subjets. Each term in the series with increasing numbers of subjets has a perturbative matching coefficient at the scale $p_TR, p_T\theta_c$  convolved with a subjet function. Due to the possibility of interference between different subjets facilitated by the collinear-soft mode,  the subjet functions are distinct for each term in the series. These functions encode both the LPM interference between successive jet-medium interactions as well as color decoherence between distinct sources of medium-induced radiation. These functions depend on both the properties of the jet, namely its radius $R$, and also the properties of the medium, i.e., its temperature and size.
\item Computing the single subjet function perturbatively, reveals two types of contributions at one loop: broadening of the collinear-soft mode sourced by the vacuum splittings and medium-induced collinear-soft radiation. The leading contribution to jet energy loss is power suppressed by a factor $R$ but multiple interactions in a dense medium can lead to an enhancement.
\item For the regime $R \leq \theta_c$, the jet remains unresolved by the medium and only a single subjet function contributes. This can be expanded out order-by-order in the opacity expansion. At arbitrary order $n$, the contribution can be explicitly factorized in terms of a medium-induced jet function and $n$ copies of a \textit{universal} medium correlator. The medium correlator obeys two RG equations; one is in rapidity, which is the  BFKL equation, and the second is in terms of virtuality, which gives the QCD beta function. The correlator is sensitive only to the local microscopic properties of the medium and hence is a \textit{process independent} probe of medium properties. The appearance of an identical correlator in other jet hierarchies and observables \cite{Vaidya:2021vxu} confirms the universal nature of this object. The medium-induced jet function also obeys two RG equations, a BFKL equation in the rapidity scale with opposite sign and threshold evolution in virtuality. 
\item In a dilute medium, the virtuality of the subjet function is set by the scale $m_D$. Hence, in a strongly coupled medium, the subjet function is non-perturbative with a dependence on the jet radius $R$, the path length $L$ in the medium, and the temperature.  For other jet substructure observables, there will always be a dependence on $L$ and any angular measurement made on the jet partons. It is independent of the jet $p_T$ and therefore is universal across the $p_T$ bins. 
\end{itemize}
%%%%%%%%%%%%%%%%%%%%%%%%%%%%%%%%%%%%%%%%%%%%%%%%%%%%%%%%%%%%%%%%%%%%%%%

Several open questions naturally arise from these results. So far we have only addressed the case of one unresolved jet in detail. For $\theta_c< R$, there is phase space available for several subjets, with the $n^{th}$ term in this series being an $n$-subjet function whose matching coefficient starts at $O(\alpha_s^n)$ and therefore requires a corresponding $n$-loop calculation. We leave explicit calculations for this regime for the future. Calculations beyond the one-loop order have the potential to reveal new logarithms in the ratio $\theta_c/R$. The explicit computations in this paper are restricted to a single interaction with the medium which is valid for a dilute medium. In a dense medium, we expect multiple interactions to become important. This will enable us to answer several questions. First, we would like to obtain an estimate of the intrinsic medium scale $Q_{\text{med}}$, which will inform changes in the EFT structure. Based on this, we will be able to establish the universality (or lack thereof) of the non-perturbative physics in a dense medium. This in turn will determine the predictive power across different jet observables. 
  
An explicit operator definition for non-perturbative physics opens up the possibility of computing this object using numerical techniques such as Lattice/ Quantum computers depending on the nature of the correlator. Isolating the non-perturbative physics allows us to significantly reduce the complexity of the problem, which in turn reduces the resource cost of computation. While we have already recovered and gone beyond the well-known GLV result at one loop using the developed EFT, we also would like to recover the BDMPS-Z result and provide a systematic procedure to go beyond this leading order calculation. Finally, while we have focused on perhaps the simplest jet observable, the framework we have developed in this paper can be extended to more exclusive observables that probe the substructure of the jet. 

\section*{Acknowledgements}
V.V. and B.S.  are supported by startup funds from the University of South Dakota and by the U.S. Department of Energy, EPSCoR program under contract No. DE-SC0025545. Y. M.-T. was supported by the U.S. Department of Energy under Contract No. DE-SC0012704 and by Laboratory Directed Research and Development (LDRD) funds from Brookhaven Science Associates. FR was supported by the U.S. Department of Energy, Office of Science, Office of Nuclear Physics, Early Career Program under contract No. DE-SC0024358. 

\appendix
\section*{Appendix}
%%%%%%%%%%%%%%%%%%%%%%%%%%%%%%%%%%%%%%%%%%

%%%%%%%%%%%%%%%%%%%%%%%%%%%%%%%%%%%%%%%%%%

\section{Vacuum result } 
\label{app:JVOneLoop}

The vacuum cross section is factorized in terms of a hard and a jet function. Throughout this work, we consider only the case of a quark jet. The hard function can be found in Ref.~\cite{Kang:2016mcy}. Here, we give the one-loop results for the quark jet function.

\subsection{Jet function}

At the tree level, the vacuum jet function is given by
\bea 
J^0_{q}(z, \omega_J) = \delta(1-z)\,.
\eea
At one loop, when both quark and gluon are inside the jet, we have
\bea
J^0_{qg}(z, \omega_J)= \delta(1-z) \frac{\alpha_s  C_F}{2\pi}
\left[\frac{1}{\epsilon^2} + \frac{3}{2\epsilon} + \frac{1}{\epsilon} L+ \frac{1}{2}L^2+\frac{3}{2}L + \frac{13}{2} - \frac{3\pi^2}{4}\right],
\eea
where $L = \ln(4\mu^2/(\omega_J^2 R^2))$. Similarly, when only the quark is inside the jet, we obtain
\bea
J^{0}_{q}(z, \omega_J)& =&\frac{\alpha_s C_F}{2\pi}  \delta(1-z)\left[-\frac{1}{\epsilon^2} -\frac{1}{\epsilon} L - \frac{1}{2}L^2 + \frac{\pi^2}{12}\right]\nonumber \\
&+&\frac{\alpha_s C_F}{2\pi}  \left[\left(\frac{1}{\epsilon}+L\right) \frac{1+z^2}{(1-z)_+} -2(1+z^2)\left(\frac{\ln(1-z)}{1-z}\right)_+ - (1-z)\right],
\eea
and the case where only the gluon is inside the jet results in
\begin{align}
J^0_{gq}=\frac{\alpha_s}{2\pi}\Big(\frac{1}{\epsilon}+L\Big) P_{gq}(z)-\frac{\alpha_s}{2\pi}\Big[2\,P_{gq}(z)\ln(1-z)+C_F z \Big].  
\end{align}
Adding up all contributions, we obtain the following result up to the one-loop level
\begin{align}
J^0_{q}+J^0_{qg}+J^0_{gq}&=\delta(1-z)\Bigg[1+ \frac{\alpha_s  C_F}{2\pi}\Big(\frac{3}{2\epsilon}+\frac{3}{2}L+\frac{13}{2}-\frac{2\pi^2}{3} \Big) \Bigg] \nn\\
&+\frac{\alpha_s C_F}{2\pi}  \left[\left(\frac{1}{\epsilon}+L\right) \frac{1+z^2}{(1-z)_+} -2(1+z^2)\left(\frac{\ln(1-z)}{1-z}\right)_+ - (1-z)\right]\nn\\
&+\frac{\alpha_s}{2\pi}\Big(\frac{1}{\epsilon}+L\Big) P_{gq}(z)-\frac{\alpha_s}{2\pi}\Big[2\,P_{gq}(z)\ln(1-z)+C_F z \Big].
\end{align}
%%%%%%%%%%%%%%%%%%%%%%%%%%%%%%%%%%%%%%%%%%%%%%%%%%%%%%%%%%%%%%%%%%%%%

\section{Medium-induced jet function for hard to hard-collinear matching~\label{app:Loop}}

The medium-induced jet function is given in Eq. \eqref{eq:realminusvirtual}, 
\bea
 J_{n}^{(1)}(\omega, R, L, \bfk)= \left( J_{r,n}^{(1)}-J_{v,n}^{(1)}\right)(z,\omega, R; \bar x^-, \bfk). 
\eea
In the following, we evaluate both $J_{r,n}^{(1)}$ and $J_{v,n}^{(1)}$ at NLO. For the matching procedure discussed in Section~\ref{sec:StageII}, the diagrams at one loop contribute when the momentum $q$ of the emitted gluon scales as collinear-soft mode and, hence, we compute the result only in this limit. Below we evaluate both real and virtual diagrams in detail. The diagrams that contribute to this limit are the same as those that appear in \cite{Vaidya:2021mly} albeit with a modified measurement $\mathcal{M}$ imposed on the final state. Here we only show the non-vanishing diagrams and their contributions to the medium-induced jet function.
%%%%%%%%%%%%%%%%%%%%%%%%%%%%%

\subsection{Real contribution $J_{r,n}$}

As already noted in Section \ref{sssec:treeJ}, the tree-level result for the medium-induced jet function cancels out with the contribution from  $J_{v,n}$. Therefore, we focus on the one-loop diagrams. 

\begin{figure}
\centering
\includegraphics[width=0.85\linewidth]{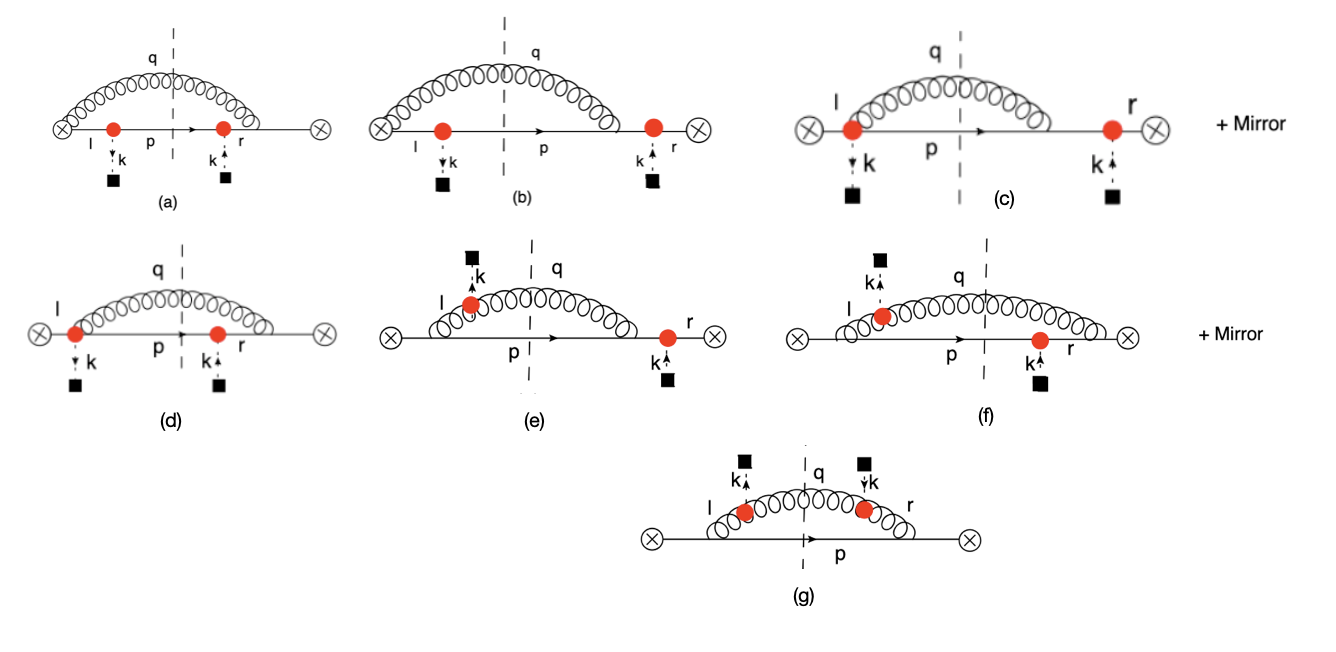}
\caption{The non-vanishing real gluon emission diagrams for Glauber insertions on opposite sides of the cut.}
\label{JR}
\end{figure}

\textbf{Real gluon emission $J_{r,n}^r$:} We first consider the set of diagrams, which consist of real gluon emission with Glauber insertions on opposite sides of the cut as shown in Figure~\ref{JR}.
Adding the corresponding mirror diagram and expanding around $q^-=0$ to retain the dominant contribution, we obtain
\bea
12(a)+12(b)&=&   g^2C_F \int \frac{dq^-}{q^-} \int \frac{d^{2}\bfq}{(2\pi)^{3}}\frac{1}{\bfq^2}\Bigg\{C_F-\frac{N_c}{2}\cos\Big[\frac{\bfq^2 \bar x^-}{2q^-}\Big]\Bigg\}\Theta_{\text{alg}}\delta(1-z),
\eea
where $\bfq^2/q^-$ can be identified as the formation time for the collinear-soft radiation. Next, we consider the diagrams with the Wilson line coming from the Glauber vertex. The diagrams (c) and (d) along with their mirrored analogs add up to   
\bea
12(c)+12(d)=\frac{g^2(N_c^2-1)}{2}\int \frac{dq^-}{q^-}\int \frac{d^{2}\bfq}{(2\pi)^3}\frac{1}{\bfq^2}\Bigg\{1-\frac{1}{2}\cos\Big[\frac{\bfq^2 \bar x^-}{2q^-}\Big]\Bigg\}\Theta_{\text{alg}}\delta(1-z).
\eea 
Next, we consider the set of diagrams with a collinear gluon-Glauber interaction along with a Lagrangian vertex insertion. Here, the diagrams (e) and (f) add up to 
\begin{align}
12(e)+12(f) = & -2\frac{g^2(N_c^2-1)}{2}\int \frac{dq^-}{q^-} \int \frac{d^{2}\bfq}{(2\pi)^3}\frac{\bfq\cdot (\bfq+\bfk)}{\bfq^2(\bfq+\bfk)^2} \nonumber\\
&\hspace{-2.5cm}\times\Bigg\{1-\cos\Big[\frac{(\bfq+\bfk)^2 \bar x^-}{2q^-}\Big]-\frac{1}{2}\cos\Big[\frac{\bfq^2 \bar x^-}{2q^-}\Big]+\frac{1}{2}\cos\Big[\frac{((\bfq+\bfk)^2-\bfq^2)\bar x^-}{2q^-}\Big]
\Bigg\}\Theta_{\text{alg}}\delta(1-z)\,.   
\end{align}
Finally, diagram 12(g) leads to
\bea
12(g) &=& 2\frac{g^2(N_c^2-1)}{2}\int \frac{dq^-}{q^-} \int \frac{d^{2}\bfq}{(2\pi)^3}\frac{1}{(\bfq+\bfk)^2} \Bigg\{1-\cos\Big[\frac{(\bfq+\bfk)^2\bar x^-}{2q^-}\Big]
\Bigg\}\Theta_{\text{alg}}\delta(1-z).
\eea
Adding up all the diagrams evaluated above that contribute to real gluon emissions, we obtain 
\begin{align}
\mathcal{J}_{n,R}^R&= \frac{g^2(N_c^2-1)}{2}\int \frac{dq^-}{q^-}\int \frac{d^{2}\bfq}{(2\pi)^3}\frac{1}{\bfq^2}\Bigg[\frac{\bfk^2}{(\bfq+\bfk)^2} \Bigg\{1-\cos\Big[\frac{(\bfq+\bfk)^2 \bar x^-}{2q^-}\Big]\Bigg\}+\cos\Big[\frac{(\bfq+\bfk)^2 \bar x^-}{2q^-}\Big]\nn\\
&-F\Big[\frac{\bfq^2 \bar x^-}{2q^-}\Big]-\frac{\bfq\cdot (\bfq+\bfk)}{(\bfq+\bfk)^2} \Bigg\{-\cos\Big[\frac{\bfq^2 \bar x^-}{2q^-}\Big]+\cos\Big[\frac{((\bfq+\bfk)^2-\bfq^2)\bar x^-}{2q^-}\Big]
\Bigg\}\Bigg]\Theta_{\text{alg}}\nn\\
&+\frac{g^2(N_c^2-1)}{2}\int \frac{dq^-}{q^-}\int \frac{d^{2}\bfq}{(2\pi)^3}\frac{1}{(\bfq+\bfk)^2} \Bigg\{1-\cos\Big[\frac{(\bfq+\bfk)^2 \bar x^-}{2q^-}\Big]
\Bigg\}\Theta_{\text{alg}}\,,    
\end{align}
where the term proportional to $C_F^2$ gets canceled out so we have ignored it in the expression above. In the limit $ L \rightarrow \infty$, all the interference terms (those that involve the function $F$) drop out. We obtain
\begin{align}
\mathcal{J}_{n,R}^R(L \rightarrow \infty)&=\frac{g^2(N_c^2-1)}{2}\int \frac{dq^-}{q^-} \int \frac{d^{2}\bfq}{(2\pi)^3}\bigg[\frac{\bfk^2}{\bfq^2(\bfq+\bfk)^2}+\frac{1}{(\bfq+\bfk)^2}\bigg]\Theta_{\text{alg}}\,.    
\end{align}

\begin{figure}
\centering
\includegraphics[width=0.75\linewidth]{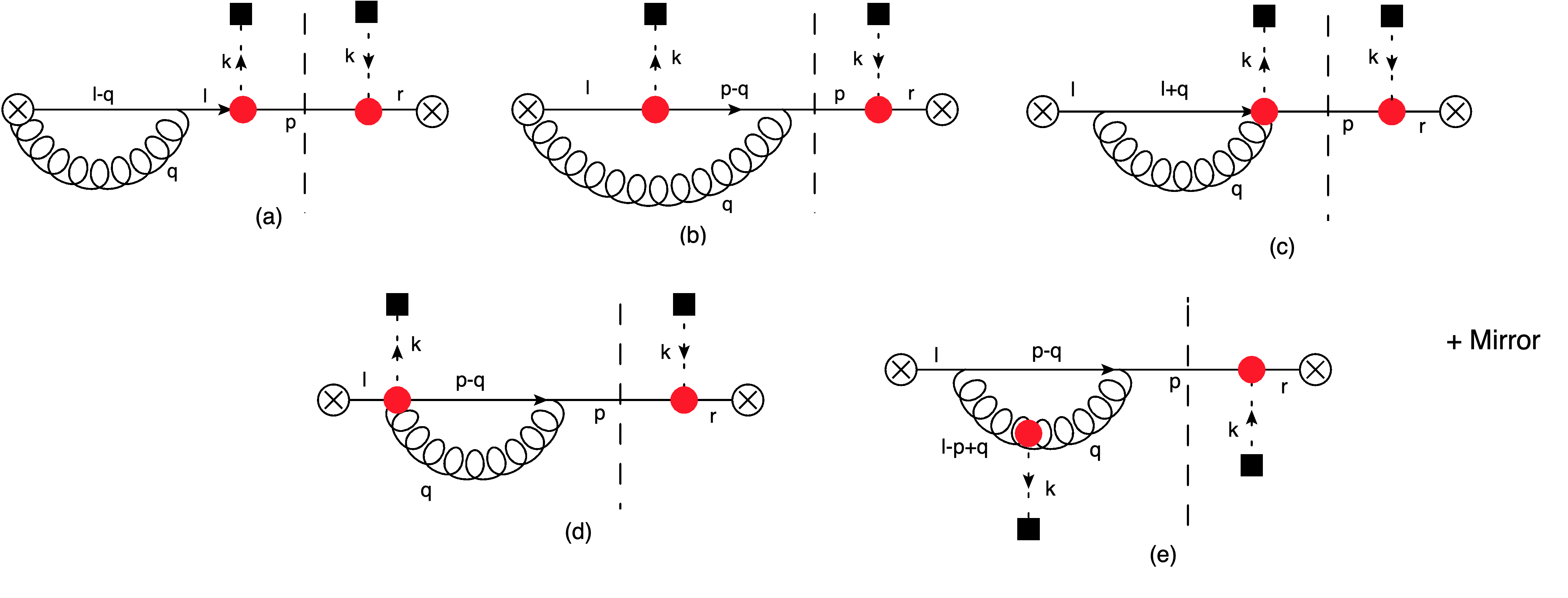}
\caption{Virtual Gluon emission diagrams with Wilson line emissions and Lagrangian insertion.}
\label{VGW}
\end{figure}

%%%%%%%%%%%%%%%%%%%%%%%%%%%%%%%%%%%%%%%%%%%%%%%%%%%%%%%%%%%%%%%%%%%%%
\textbf{Virtual gluon diagrams $\mathcal{J}_{n,R}^V$:} Next, we consider the diagrams contributing to virtual gluon corrections. We classify the diagrams in the same manner as the real gluon emissions. All the relevant diagrams in the collinear-soft limit of the loop momentum are shown in Figure~\ref{VGW}.
Here, diagrams (a) and (b) along with their mirror diagrams  add up to 
\bea
13(a)+13(b)&=&-g^2C_F\int_0^{\omega}\frac{dq^-}{q^-}\int \frac{d^2\bfq}{(2\pi)^{3}} \frac{ 1}{\bfq^2}\Bigg\{C_F-\frac{N_c}{2}\cos\Big[\frac{\bfq^2 \bar x^-}{2q^-}\Big]
\Bigg\}\,.
\eea

Again the terms proportional to $C_F$ cancel out with another diagram with an emission from the collinear Wilson line, leaving behind only the $N_c$ dependent term.  Next, we consider the diagrams with Wilson lines originating from the Glauber vertex as shown in diagrams (c) and (d). Adding up both contributions, we obtain 
\bea
13(c)+13(d)&=&-\frac{g^2(N_c^2-1)}{2}\int \frac{dq^-}{q^-}\int \frac{d^{2}\bfq}{(2\pi)^{3}} \frac{1}{\bfq^2}\Bigg\{1-\frac{1}{2}\cos\Big[\frac{\bfq^2 \bar x^-}{2q^-}\Big]\Bigg\}\delta(1-z)\,.
\eea
Finally, the contribution from the diagrams with the collinear gluon and the Glauber vertex shown in diagram (e) is
\bea
13(e)&=&\frac{g^2(N_c^2-1)}{4}\int \frac{dq^-}{q^-}\int \frac{d^{2}\bfq}{(2\pi)^{3}} \frac{2\bfq\cdot (\bfq+\bfk)}{(\bfq+\bfk)^2}\frac{1}{\bfq^2}\Bigg\{1
-\cos\Big[\frac{(\bfq+\bfk)^2 \bar x^-}{2q^-}\Big]\Bigg\}\delta(1-z).
\eea   
Therefore, adding all the diagrams shown in Figure~\ref{VGW}, we find
\bea
\mathcal{J}_{n,R}^V&=&-\frac{g^2(N_c^2-1)}{4}\int \frac{dq^-}{q^-} \int \frac{d^{2}\bfq}{(2\pi)^{3}} \Bigg[\frac{\bfk^2}{\bfq^2(\bfq+\bfk)^2}\Bigg\{1
-\cos\Big[\frac{(\bfq+\bfk)^2 \bar x^-}{2q^-}\Big]\Bigg\}\nn\\
&+& \frac{1}{\bfq^2}\Bigg\{\cos\Big[\frac{(\bfq+\bfk)^2 \bar x^-}{2q^-}\Big]-\cos\Big[\frac{\bfq^2\bar x^-}{2q^-}\Big]\Bigg\}\Bigg]\delta(1-z),
\eea 
which, in the limit $L \rightarrow \infty$, leads to 
\bea 
 \mathcal{J}_{n,R}^V(L \rightarrow \infty)&=&-\frac{g^2(N_c^2-1)}{4}\int \frac{dq^-}{q^-} \int \frac{d^{2}\bfq}{(2\pi)^{3}} \frac{\bfk^2}{\bfq^2(\bfq+\bfk)^2}\delta(1-z)\,.
\eea
\subsection{Virtual contribution $J_{v,n}$}	

Next, we evaluate the diagrams contributing to the jet function from  Glauber insertions on the same side of the cut.  Let us first consider real gluon emission diagrams.
 
\begin{figure}
\centering
\includegraphics[width=0.75\linewidth]{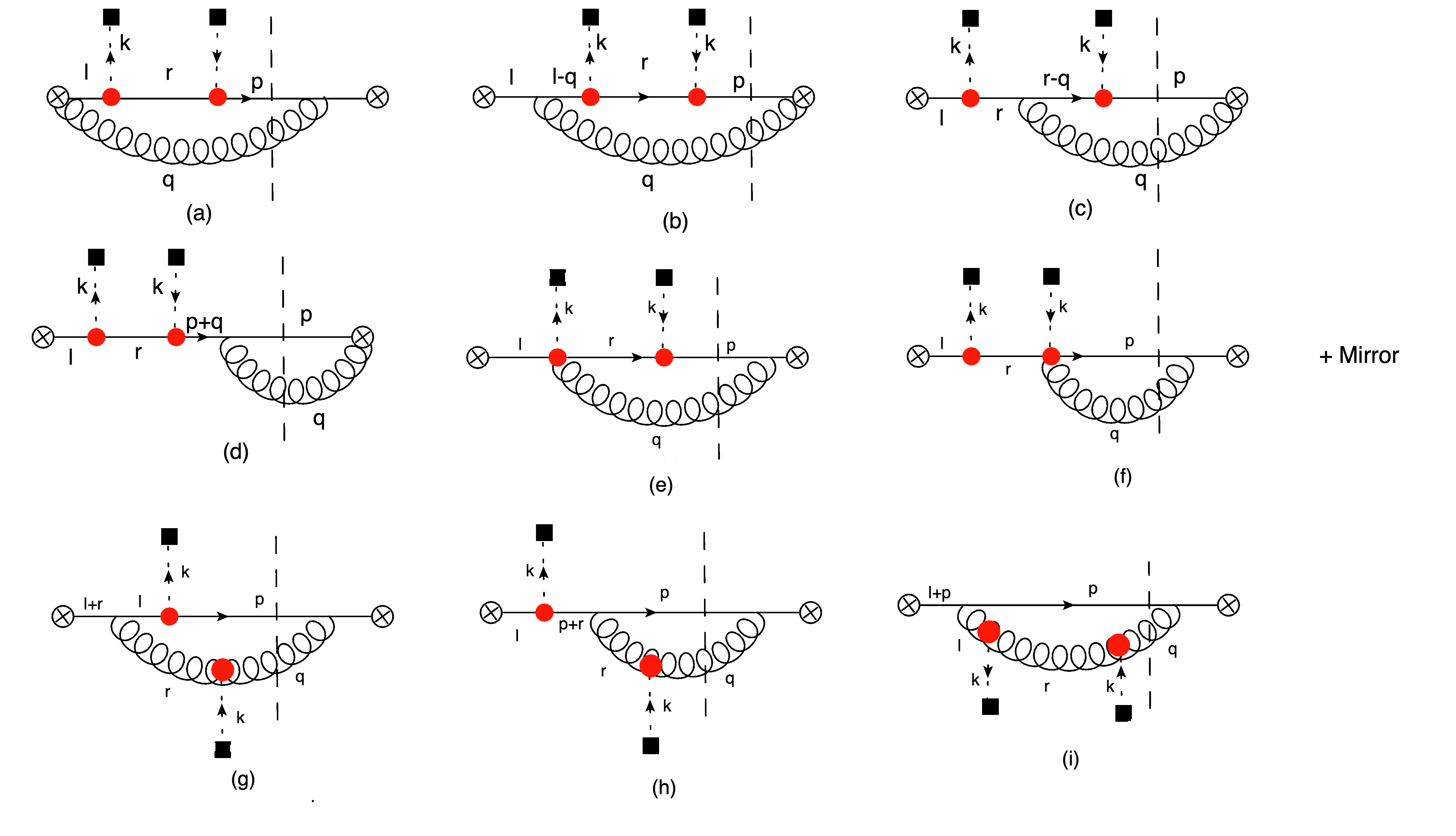}
\caption{Real Gluon emission for Glauber insertion on the same side of the cut.}
\label{WJV}
\end{figure}

\textbf{Real gluon emission $\mathcal{J}_{n,V}^R$:} We first consider the diagrams containing  Wilson line contribution from the hard vertex as shown in Figure~\ref{WJV}.  Their contributions, along with their mirror diagrams add up to 
\bea
14(a)+14(b)+14(c)+14(d)&=&\frac{g^2C_F^2}{2}\int \frac{dq^-}{q^-}\int \frac{d^2\bfq}{(2\pi)^{3}} \frac{1}{\bfq^2}\Theta_{\text{alg}}\delta(1-z).
\eea
This contribution gets canceled again and we therefore will not consider it further. The next set of diagrams are ones with Wilson lines from the Glauber vertex as shown in (e) and (f). Their sum vanishes
\bea 
14(e)+14(f) = 0\,. 
\eea
Next, we consider diagrams with one collinear gluon and a Glauber along with one quark Glauber vertex. These are shown in diagrams (g) and (h) and the corresponding contribution is
\bea
14(g)+14(h) &=&  -\frac{g^2(N_c^2-1)}{2}\int \frac{dq^-}{q^-}\int \frac{d^2\bfq}{(2\pi)^3}\frac{\bfq \cdot (\bfq+\bfk)}{\bfq^2(\bfq+\bfk)^2}\nn\\
&&\Bigg\{-\cos\Big[\frac{\bfq^2 \bar x^-}{2q^-}\Big]+\cos\Big[\frac{((\bfq+\bfk)^2-\bfq^2)\bar x^-}{2q^-}\Big]
\Bigg\}\Theta_{\text{alg}}\delta(1-z)\,.
\eea 
The last diagram with a double collinear gluon and Glauber vertex (i), results in the following contribution 
\bea
14(i)&=&   \frac{g^2(N_c^2-1)}{2}\int \frac{dq^-}{q^-}\int \frac{d^2\bfq}{(2\pi)^3}\frac{1}{\bfq^2}\Bigg\{1-\cos\Big[\frac{\bfq^2 \bar x^-}{2q^-}\Big]\Bigg\}\Theta_{\text{alg}}\delta(1-z)\,.
\eea
Adding up all contributions, we obtain
\bea 
J_{n,V}^R&=&  -\frac{g^2(N_c^2-1)}{2}\int \frac{dq^-}{q^-}\int \frac{d^2\bfq}{(2\pi)^3}\frac{\bfq \cdot (\bfq+\bfk)}{\bfq^2 (\bfq+\bfk)^2}\Bigg\{-\cos\Big[\frac{\bfq^2 \bar x^-}{2q^-}\Big]+\cos\Big[\frac{((\bfq+\bfk)^2-\bfq^2)\bar x^-}{2q^-}\Big]
\Bigg\}\Theta_{\text{alg}}\nn\\
&+&\frac{g^2(N_c^2-1)}{2}\int \frac{dq^-}{q^-}\int \frac{d^2\bfq}{(2\pi)^3}\frac{1}{\bfq^2}\Bigg\{1-\cos\Big[\frac{\bfq^2 \bar x^-}{2q^-}\Big]\Bigg\}\Theta_{\text{alg}}\delta(1-z),
\eea
which, in the limit $L \rightarrow \infty$, leads to 
\bea 
J_{n,V}^R(L \rightarrow \infty)&=&  \frac{g^2(N_c^2-1)}{2}\int \frac{dq^-}{q^-} \int \frac{d^2\bfq}{(2\pi)^3}\frac{1}{\bfq^2}\Theta_{\text{alg}}\delta(1-z).
\eea

\begin{figure}
\centering
\includegraphics[width=0.75\linewidth]{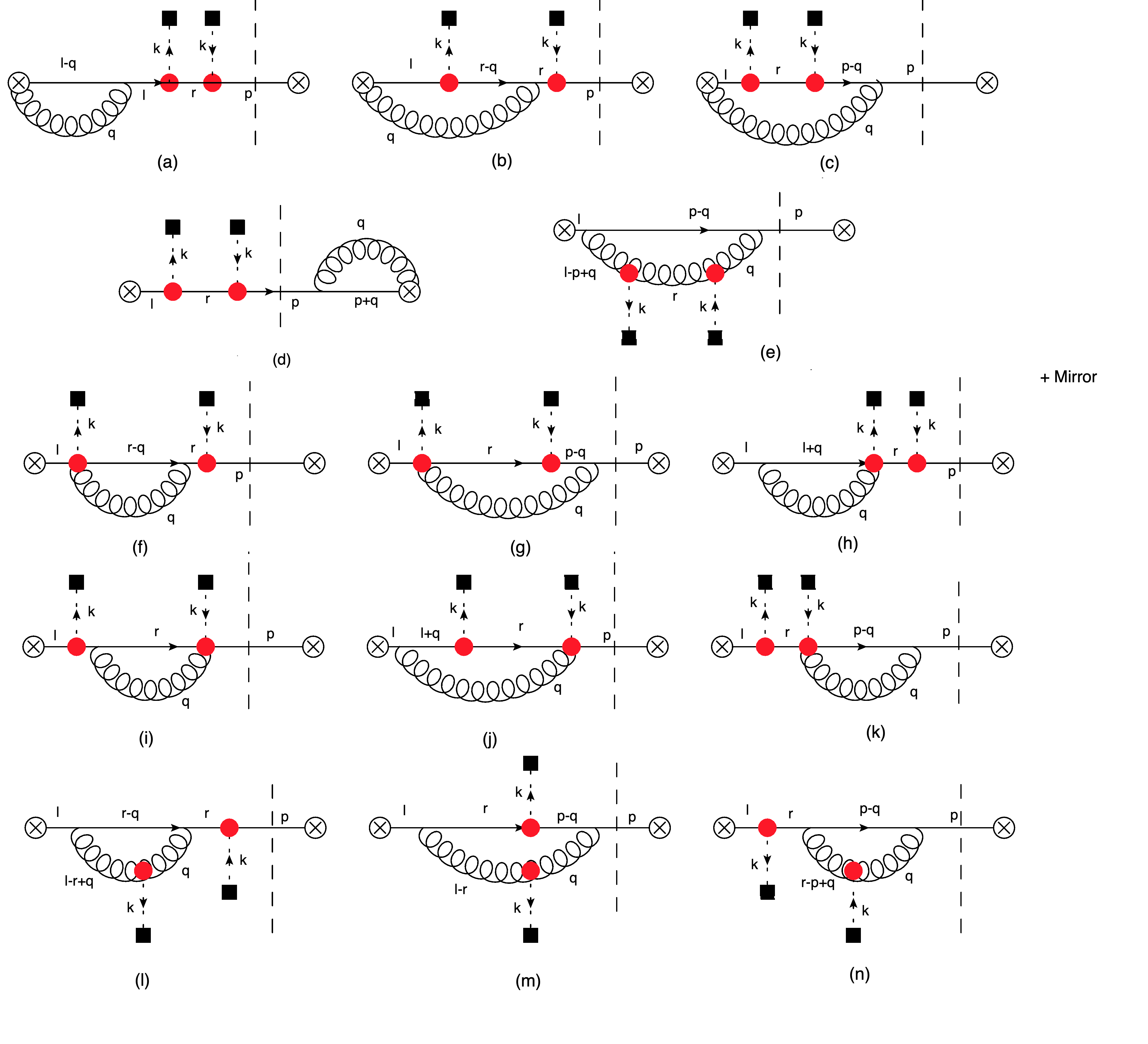}
\caption{Virtual gluon emission diagrams with rapidity divergence for Glauber insertion on the same side of the cut}
\label{WG2V}
\end{figure}

\textbf{Virtual gluon emission $\mathcal{J}_{n,V}^V$:}  We now consider the final set of diagrams that involve a virtual gluon contribution from Glauber insertions on the same side of the cut.  The diagrams with Wilson lines from the hard vertex shown in Figure~\ref{WG2V} along with the complex conjugates add up to 
\bea
15(a)+15(b)+15(c)+15(d)&=&-g^2C_F^2\int \frac{dq^-}{q^-} \int \frac{d^2\bfq}{(2\pi)^{3}} \frac{1}{\bfq^2}\delta(1-z).
\eea
Diagram (e) evaluates to    
\bea
15(e)&=&\frac{g^2(N_c^2-1)}{2}\int \frac{dq^-}{q^-} \int \frac{d^2\bfq}{(2\pi)^3}\frac{1}{\bfq^2}\Bigg\{1-\cos\Big[\frac{\bfq^2 \bar x^-}{2q^-}\Big]\Bigg\}\delta(1-z).
\eea       
The contribution from all the diagrams with the Wilson lines from the Glauber vertex vanishes
\bea 
15(f)+15(g)+.. +15(k) = 0\,. 
\eea
Next, we consider the diagrams with one Glauber and collinear gluon vertex. These add up to 
\begin{equation}
15(l)+15(m)+15(n)= -\frac{g^2(N_c^2-1)}{2}\int \frac{dq^-}{q^-} \int \frac{d^2\bfq}{(2\pi)^3}\frac{\bfq\cdot (\bfq+\bfk)}{\bfq^2(\bfq+\bfk)^2}\Bigg\{1-\cos\Big[\frac{(\bfq+\bfk)^2\bar x^-}{2q^-}\Big]\Bigg\}\delta(1-z)\,.    
\end{equation}
Adding all the contributions, we obtain
\begin{align}
\mathcal{J}_{n,V}^{V}= &\, \frac{g^2(N_c^2-1)}{4}\int \frac{dq^-}{q^-}\int \frac{d^2\bfq}{(2\pi)^3}\Bigg[\frac{\bfk^2}{\bfq^2(\bfq+\bfk)^2}\Bigg\{1-\cos\Big[\frac{(\bfq+\bfk)^2 \bar x^-}{2q^-}\Big]\Bigg\}\nn\\
&+\frac{1}{\bfq^2}\Bigg\{\cos\Big[\frac{(\bfq+\bfk)^2\bar x^-}{2q^-}\Big]-\cos\Big[\frac{\bfq^2 \bar x^-}{2q^-}\Big]\Bigg\}\Bigg]\,,  
\end{align}
which, in the limit $L \rightarrow \infty$, becomes 
\begin{align}
\mathcal{J}_{n,V}^{V}(L \rightarrow \infty)&= \frac{g^2(N_c^2-1)}{4}\int \frac{dq^-}{q^-} \int \frac{d^2\bfq}{(2\pi)^3}\frac{\bfk^2}{\bfq^2(\bfq+\bfk)^2}\delta(1-z)  \,.  
\end{align} 
Finally, combining all contributions with Glauber insertions on the same and opposite side of the cut, we obtain the quark jet function where both the quark and gluon are inside the jet
\begin{align}
J^{(1)}_{qg}(\omega,z,R;\bfk)&= \frac{g^2(N_c^2-1)}{2}\int_0^1\frac{dx }{1-x}\int \frac{d^2\bfq}{(2\pi)^3}\Bigg[    \Big\{\frac{\bfk^2}{\bfq^2(\bfq+\bfk)^2}+\frac{1}{(\bfq+\bfk)^2}-\frac{1}{\bfq^2}\Big\}\nn\\
& \times \Big\{1- \cos\Big[\frac{(\bfq+\bfk)^2\bar x^-}{2\omega(1-x)}\Big] \Big\}\left(\Theta_{\rm alg}-1 \right)\Bigg] \delta(1-z) \,.
\end{align}
where we have set $q^- = \omega(1-x)$
Here, $\Theta_{\rm alg}=\Theta(\omega(1-x)R/2 -|\bfq|)$. For the case where the gluon emission is outside the cone, the quark jet function is
\begin{align}
J^{(1)}_{q}(z,\omega,R;\bfk)&=\frac{g^2(N_c^2-1)}{2}\frac{z}{1-z} \int \frac{d^2\bfq}{(2\pi)^3 }\Bigg[  \frac{\bfk^2}{\bfq^2(\bfq+\bfk)^2}+\frac{1}{(\bfq+\bfk)^2}-\frac{1}{\bfq^2}\Bigg] \nn\\
&\times \Big\{1- \cos\Big[\frac{(\bfq+\bfk)^2\bar x^-}{2\omega(1-z)}\Big] \Big\}\Theta'_{\rm alg} \,,
\end{align}
where $\Theta'_{\rm alg}=\Theta(|\bfq|-R(1-z)\omega/2)$. We also note that the contribution when the quark is outside the jet vanishes.

%%%%%%%%%%%%%%%%%%%%%%%%%%%%%%%%%%%%%%%%%%%%%%%%%%%%%%%%%%%%%%%%
\section{Integrating out the collinear sector}\label{sec:collint} 

In this section, we derive the final form of the factorization in terms of the hard-collinear to collinear-soft matching discussed in Section~\ref{sec:StageII}, which is obtained by integrating out the collinear mode. We start from the single subjet function in Eq.~\eqref{eq:subjet1}, which is given here again for convenience
\begin{align}
\mathcal{S}_{1}(\epsilon_L)& = \int d^2\bfr \int dr^+e^{-i\omega r^+}\tr\Big[ \mathcal{T}\Big[e^{-i\int dt \mathcal{H}}[U(n)U(\bar n)\bar{\chi}_{cn}](0)\Big]\frac{\slashed{\bar n}}{2} \rho_M\nn\\
&\quad\quad\, \mathcal{\bar T}\Big[e^{-i\int dt \mathcal{H}}[U(n)^{\dagger}U(\bar n)^{\dagger}\chi_{cn}](\bfr, r^+)\Big]\mathcal{M}'\Big]\,,
\label{eq:SJ1}
\end{align}
where $\mathcal{M'} = \delta( \epsilon_L- \bar n\cdot \mathcal{P}_{\text{out}})\Theta_{\text{alg}}$
and $\bar n\cdot \mathcal{P}_{\text{out}}$ is the longitudinal momentum component of the collinear-soft radiation that flows out of the jet.  Here, $H=H_c+H_{\cs}+H_{\s}+H_G^{\rm int}$ is the three sector Hamiltonian, and $H_G^{\rm int}$ corresponds to the Glauber interaction given in Eq.~(\ref{eq:LG}). The Hamiltonians $H_{c}$ and $H_{\cs}$ are two copies of the collinear SCET Hamiltonian, while $H_{\s}$ is soft SCET Hamiltonian. Further, $U_{n}$  is a collinear-soft Wilson line. As discussed in section~\ref{sec:cal1loop}, we note that the contribution to the energy loss from collinear radiation is suppressed compared to collinear-soft emissions. Therefore, in the collinear-soft jet function, we can treat the collinear Hamiltonian $H_c$ as the free theory Hamiltonian.  
The physical process here is the propagation of a high-energy quark that interacts with the medium and emits medium-induced collinear-soft radiation. Therefore, in the Hamiltonian $\mathcal{H}$, the field $\chi_{cn}$ appears either through the free theory quark Hamiltonian $H_{c}$, or in the Glauber interaction operator $H_{G}^{\rm int}$ through the operators in Eq.~\eqref{eq:LG}
\begin{align}
H_G^{\text{int}}&= H_{\cs\text{-}\s} + H_{n\text{-}\cs\text{-}\s} \,,\nn\\
 H_{n\text{-}\cs\text{-}\s} &=\sum_{i,j}\mathcal{O}_n^{ib}\frac{1}{\mathcal{P}_{\perp}^2}\mathcal{O}_{\cs}^{ba} \frac{1}{\mathcal{P}_{\perp}^2} \mathcal{O}_{\s}^{ja} \equiv \sum_{i,j}\mathcal{O}_n^{ib} \mathcal{O}_{\cs\text{-}\s}^{jb}  \,.
\end{align}
For a quark-initiated jet, ${\cal O}_n^{qb}$ is the collinear quark current, which is given 
\bea
\mathcal{O}_n^{qb} =  \bar \chi_n \frac{\slashed{\bar n}}{2} t^b\chi_n\,.
\eea
The Hamiltonian in this case reads as
\bea
H = H_{n}^{(0)}+ H_{\cs}+  H_{\s} + H_{\cs\text{-}\s}+\mathcal{O}_n^{qb}\frac{1}{\mathcal{P}_{\perp}^2}\mathcal{O}_{\cs\text{-}\s}^{b}\equiv H' + \mathcal{O}_n^{qb}\frac{1}{\mathcal{P}_{\perp}^2}\mathcal{O}_{\cs\text{-}\s}^{b}\,.
\eea
We now expand out the second interaction term order by order. Due to the non-commutativity of the Hamiltonian, we first need to rewrite the action by dressing this operator with the remaining terms of the Hamiltonian, so that we obtain
\begin{align}
\mathcal{S}_{1}(z)& = \tr\Big[\mathcal{T}\Big[e^{-i H' t}e^{-i\int dt \mathcal{O}_{n,{\rm I}}^{qb}\frac{1}{\mathcal{P}_{\perp}^2}\mathcal{O}_{\cs\text{-}\s,{\rm I}}^{b}}[U_{n}U_{\bar n}\bar{\chi}_{cn}]_{\rm I}(0)\Big]\frac{\slashed{\bar n}}{2}\nn\\
 &\times \rho_M \mathcal{\bar T}\Big[e^{-i\int dt H'}e^{-i\int dt \mathcal{O}_{n,{\rm I}}^{qb}\frac{1}{\mathcal{P}_{\perp}^2}\mathcal{O}_{\cs\text{-}\s,{\rm I}}^{b}}\delta^2(\mathcal{P}_{\perp})\delta(\omega- \bar n \cdot P)[U_{\bar n}^{\dagger}U_{n}^{\dagger}\chi_{cn}]_{\rm I}(0)\Big]\mathcal{M}'\Big]  \,,
\end{align}
where the subscript I indicates that the operator is dressed with the Hamiltonian $H'$. We have also explicitly performed the $r$ coordinate space integrals.
We continue by considering one side of the cut where the result at the amplitude level can be written as
\begin{align}
A=\mathcal{\bar T}\Big[e^{-i\int dt \mathcal{H}'}e^{-i\int dt\, \mathcal{O}_{n,\rm I}^{qb}\frac{1}{\mathcal{P}_{\perp}^2}\mathcal{O}_{\cs\text{-}\s,\rm I}^{b} }\delta(\omega -\bar{n}\cdot\mathcal{P})\delta^2(\mathcal{P}_{\perp})[U_{n}^{\dagger}U_{\bar n}^{\dagger}\chi_{cn}]_{\rm I}(0)\Big]\mathcal{M}'|X \rangle \,.    
\end{align}
We can now expand out the exponent and simplify the result order by order in the collinear sector. Since there are no radiative corrections in the collinear sector, the quark propagators are free theory propagators. We find the following results
\begin{figure}
\centering
\includegraphics[width=\linewidth]{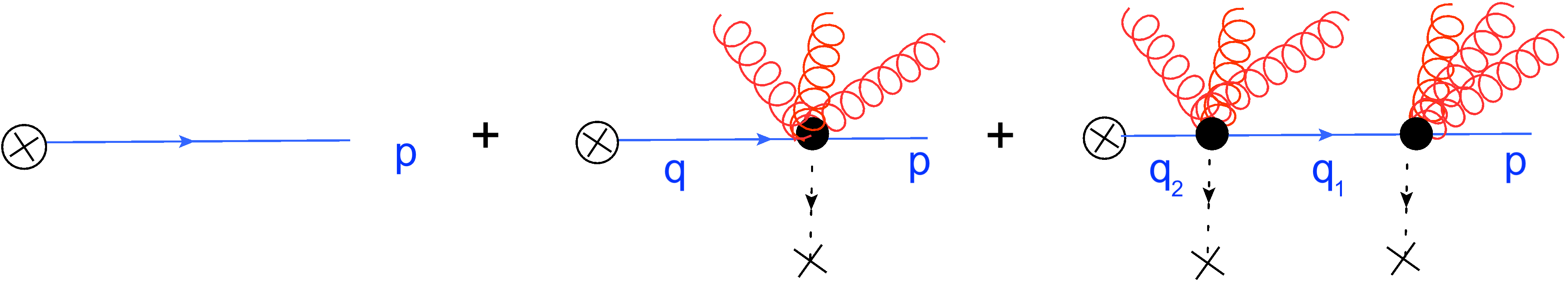}
\caption{A single collinear quark (blue) with insertions of Glauber interactions with the medium (black vertex) inducing collinear-soft radiation (red).}
\label{proof}
\end{figure}      
\begin{itemize}
\item{The result with no insertions is}
\begin{align}
A^{(0)}&=  \mathcal{\bar T}\Big[e^{-i\int dt H'}\delta(\omega-\bar{n}\cdot\mathcal{P})\delta^2(\mathcal{P}_{\perp})[U_{n}^{\dagger}U_{\bar n}^{\dagger}\chi_{cn}]_{\rm I}(0)\Big]\mathcal{M}|X\rangle \nonumber \\
&= \mathcal{\bar T}\Big[e^{-i\int dt\, H'}\delta^2(\bfp+\mathcal{P}_{\perp})[U_{\bar n,{\rm I}}^{\dagger}U_{n,{\rm I}}^{\dagger}]\Big]\mathcal{M}'|X\rangle \delta(\omega-\bar{n}\cdot p)u(p)\,,    
\end{align}
where $p$ is the momentum of the final state collinear quark as shown in Fig.~\ref{proof}. 
\item{For one insertion, we obtain} 
\begin{align}
A^{(1)}&= \int d^4x \mathcal{\bar T}\Big[e^{-i\int dt H'}\delta^2(\mathcal{P}_{\perp})\delta(\omega-\bar{n}\cdot\mathcal{P})[U_{n}^{\dagger}U_{\bar n}^{\dagger}]_{\rm I} \frac{1}{\mathcal{P}_{\perp}^2}\mathcal{O}^b_{\cs\text{-}\s,{\rm I}}(x)\nonumber \\
& \Big(\chi_{cn} \Big[\bar \chi_n(0) \frac{\slashed {\bar n}}{2} t^b \chi_n\Big](x)\Big)\Big]\mathcal{M}|p X\rangle \,.   
\end{align}
We can drop the $x^+$ dependence of the operator ${\cal O}_{\rm cs-s}$ since the contribution to the $x^-$ component of momentum is power suppressed compared to the collinear mode.
One of the $\chi$ fields at $x$ annihilates the final quark with momentum $p$ and the other two give us a quark propagator. In momentum space, we get 
\begin{align}
A^{(1)}= &\,\int d^4x e^{-ip \cdot x} \int d^2 \bfk e^{-i\bfk \cdot \bfx}  \int d^4q\, e^{i q\cdot x}\mathcal{\bar T}\Big[e^{-i\int dt H'}\delta^2(\bfq+ \mathcal{P}_{\perp})[U_{n}^{\dagger}U_{\bar n}^{\dagger}]_{\rm I} \frac{1}{\bfk^2}\mathcal{O}^b_{\cs\text{-}\s,{\rm I}}(\bfk, x^-)\Big]\nn\\
&\delta(\omega-\bar{n}\cdot p)\frac{i \bar{n}\cdot q}{q^2-i\epsilon} \frac{\slashed { n}}{2} \frac{\slashed {\bar n}}{2} t^b u(p)\mathcal{M}|X\rangle \,,
\end{align}
where $q$ is the collinear quark momentum. With this simplification, we can now do the integrals over $\bs x$, $x^+$, which leads to 
\begin{align}
A^{(1)}= &\,  \int d^2\bfk \int dx^- e^{-i(p^+-q^+)x^-}  \int d^4 q \mathcal{\bar T}\Big[e^{-i\int dt H'}\delta^2(\bfq+ \mathcal{P}_{\perp})[U_{n}^{\dagger}U_{\bar n}^{\dagger}]_{\rm I} \frac{1}{\bfk^2}\mathcal{O}^b_{\cs\text{-}\s,{\rm I}}(x^-,\bfk)\Big]   \nn \\
& \frac{i \bar{n}\cdot q}{q^2-i\epsilon} \frac{\slashed {\bar n}}{2} t^b u(p)\mathcal{M}|X\rangle \delta(p^--q^-)\delta^2(\bfp+\bfk-\bfq)\delta(\omega-\bar{n}\cdot p)\,.   
\end{align}
Using the delta function that enforces transverse momentum conservation, we can perform the $\bfq$ integration to obtain
\begin{align}
A^{(1)}&=  \int d^2\bfk  \int dq^+ \int dx^- e^{-i(p^+-q^+)x^-} \mathcal{\bar T}\Big[e^{-i\int dt H'}\delta^2(\bfp+ \bfk +\mathcal{P}_{\perp})[U_{n}^{\dagger}U_{\bar n}^{\dagger}]_{\rm I}\frac{1}{\bfk^2}\mathcal{O}^b_{\cs\text{-}\s,{\rm I}}(x^-,\bfk)\Big]  \nn \\
&\delta(\omega-\bar{n}\cdot p) \frac{i }{q^+-(\bfp+\bfk)^2/p^--i\epsilon}\frac{\slashed {n}}{2} \frac{\slashed {\bar n}}{2} t^b u(p)\mathcal{M}'|X\rangle\,.    
\end{align}
Next, we can perform the integral over $q^+$ by closing the contour in the upper half plane. We obtain
\begin{align}
A^{(1)} = &\, \int d^2\bfk   \int dx^- e^{-i(p^+-(\bfp+\bfk)^2/p^-)x^-}\mathcal{\bar T}\Big[e^{-i\int dt H'}\delta^2(\bfp+\bfk+ \mathcal{P}_{\perp})[U_{n}^{\dagger}U_{\bar n}^{\dagger}]_{\rm I} \frac{1}{\bfk^2}\mathcal{O}^b_{\cs\text{-}\s,{\rm I}}(x^-,\bfk)\Big]\nn\\
&\frac{\slashed {n}}{2} \frac{\slashed {\bar n}}{2} t^b \delta(\omega-\bar{n}\cdot p)u(p)\mathcal{M}'|X\rangle \,.
\end{align}
Note that we can now drop the phase factor $e^{-i(p^+-(\bfp+\bfk)^2/p^-)x^-}$ since the contribution from the plus component of the momentum is subleading compared to the one from the $\mathcal{O}_{\cs- \s}$ operator. This can be seen from the plus component scaling of the collinear-soft and soft modes compared to the collinear mode. 
At the level of the amplitude squared, there is a phase space integral over the final state quark $\int d^2 \bfp \int dp^-/p^-$. Since no other term depends on $\bfp$, we will eventually perform the phase space integral over $\bfp $ to eliminate $\delta^2(\bfp+ \bfk+ \mathcal{P}_{\perp})$. Therefore, we ignore this delta function from now on. We obtain
\begin{align}
A^{(1)}&=\int dx^- \int d^2\bfk  \frac{\slashed {\bar n}}{2}\frac{\slashed {n}}{2} t^b \delta(\omega-\bar{n}\cdot p)u(p)\mathcal{\bar T}\Big[e^{-i\int dt H'}[U_{n}^{\dagger}U_{\bar n}^{\dagger}]_{\rm I}\frac{1}{\bfk^2}\mathcal{O}^b_{\cs\text{-}\s,{\rm I}}(x^-,\bfk) \Big]\mathcal{M}'|X\rangle  \nn \\
&=\frac{\slashed {n}}{2} \frac{\slashed {\bar n}}{2} \delta(\omega-\bar{n}\cdot p)u(p)\mathcal{\bar T}\Big[e^{-i\int dt \mathcal{H}}[U_{n}^{\dagger}U_{\bar n}^{\dagger}]_{\rm I}\int dx^-{\bf O}^b_{\cs\text{-}\s,{\rm I}}(x^-)  t^b\Big]\mathcal{M}'|X \rangle \,,
\end{align}
where we have defined the operator ${\bf O}^b_{\cs\text{-}\s,{\rm I}}$ as
\bea
{\bf O}^b_{\cs\text{-}\s,{\rm I}}(x^-) = \int d^2\bfk \frac{1}{\bfk^2}\mathcal{O}^b_{\cs\text{-}\s,{\rm I}}(x^-,\bfk)\,.
\eea
We can follow the same procedure for higher-order Glauber insertions.   
\item{} For two insertions, we find 
\begin{align}
A^{(2)}&=\frac{1}{2}\int d^4x_1 \int d^4x_2 \mathcal{\bar T}\Big[e^{-i\int dt H'}\delta(\omega-\bar{n}\cdot\mathcal{P})\delta^2(\mathcal{P}_{\perp})[U_{n}^{\dagger}U_{\bar n}^{\dagger}]_{\rm I} \mathcal{O}^b_{\cs\text{-}\s,{\rm I}}(x_1) \mathcal{O}^c_{\cs\text{-}\s,{\rm I}}(x_2)\nonumber \\
&\chi_{cn}(0) \Big(\bar \chi_n \frac{\slashed {\bar n}}{2} t^b \chi_n\Big)(x_1) \Big(\bar \chi_n \frac{\slashed {\bar n}}{2} t^c \chi_n\Big)(x_2)\Big]\mathcal{M}|p X\rangle    \,.
\end{align}
As before, we can drop the $x_i^+$ dependence of the $\mathcal{O}_{\cs\text{-}\s}$ operators and, in momentum space, we get two quark propagators. We can then perform the integrations over $x_i^+$, $\bfx_i$, which yields momentum conservation in the minus and transverse directions.  Further, we can perform the contour integration similar to the previous case and drop the phase factor term to obtain
\begin{align}
A^{(2)} = &\,\frac{\slashed {n}}{2} \frac{\slashed {\bar n}}{2} \delta(\omega-\bar{n}\cdot p)u(p)\frac{1}{2}\mathcal{\bar T}\Big[e^{-i\int dt \mathcal{H}'}[U_{n}^{\dagger}U_{\bar n}^{\dagger}]_{\rm I}\int dx_1^-{\bf O}^b_{\cs\text{-}\s,{\rm I}}(x_1^-)  t^b\nn\\
&\int dx_2^-{\bf O}^c_{\cs\text{-}\s,{\rm I}}(x_2^-)  t^c\Big]\mathcal{M}'|X \rangle \,.    
\end{align}
The time ordering between the two $\mathcal{O}_{\cs\text{-}\s}$ operators can be written as a path ordering along the $x^-$ direction since $\theta( x_1^0 -x_2^0) $ simplifies to $\theta( x_1^- -x_2^-)$. Therefore, we get
\begin{align}
A^{(2)}&= \frac{\slashed {n}}{2} \frac{\slashed {\bar n}}{2} \delta(\omega-\bar{n}\cdot p)u(p)\mathcal{\bar T}\Big[e^{-i\int dt H'}[U_{n}^{\dagger}U_{\bar n}^{\dagger}]_{\rm I}\Big\{\int dx_1^-{\bf O}^b_{\cs\text{-}\s,{\rm I}}(x_1^-)  t^b\nn\\
&\int_0^{x_1^-} dx_2^-{\bf O}^c_{\cs\text{-}\s,{\rm I}}(x_2^-)  t^c\Big\}\Big]\mathcal{M}'|X \rangle \,.
\end{align}
\end{itemize}
We can now generalize this to an arbitrary number of insertions. Adding up all orders, we find
\begin{align}
A&= \frac{\slashed {n}}{2} \frac{\slashed {\bar n}}{2} \delta(\omega-\bar{n}\cdot p)u(p)\mathcal{\bar T}\Big[e^{-i\int dt H'}[U_{n}^{\dagger}U_{\bar n}^{\dagger}]_{\rm I}{\bf P}e^{i\int dx_1^-{\bf O}^b_{\cs\text{-}\s,{\rm I}}(x_1^-)  t^b}\Big]\mathcal{M}'|X \rangle \,.
\end{align}
As discussed earlier, at the level of the amplitude squared, we get a phase space integral over $p$, the final state quark, along with a Dirac trace, which will eliminate $u(p)$ and the $\gamma$ matrices. We note that this term is a common factor for any order in the Glauber interaction expansion.
Further, note that all the ${\bf O}_{\cs\text{-}\s}$ operators still have the superscript I, which means also at this stage they are all dressed with the Hamiltonian $H'$. Therefore, we can invert the dressing, reversing the process we started with, to capture the operator term in the Hamiltonian. As a result, the effective Hamiltonian reads as  
\bea
\int dt\, H'= \int dt (H_{\cs}+  H_{\s} + H_{\cs\text{-}\s}) + \int dx^- {\bf O}_{\cs\text{-}\s}^a(x^-)t^a. 
\eea
The last term in the above equation can be understood as an operator for medium-induced radiation along the world line of the quark.

\newpage

\bibliographystyle{utphys.bst}
\bibliography{SemiJet.bib}

\end{document}